\documentclass[10pt, a4paper, twocolumn]{article} 

\usepackage[english]{babel} 

\usepackage{microtype} 

\usepackage{amsmath,amsfonts,amsthm} 

\usepackage[svgnames]{xcolor} 

\usepackage[small, labelfont=bf, up]{caption} 

\usepackage{booktabs} 

\usepackage{lastpage} 

\usepackage{graphicx} 

\usepackage{enumitem} 
\setlist{noitemsep} 

\usepackage{sectsty} 
\allsectionsfont{\usefont{OT1}{phv}{b}{n}} 

\usepackage{geometry} 

\usepackage[T1]{fontenc} 
\usepackage[utf8]{inputenc} 

\usepackage{XCharter} 

\usepackage{fancyhdr} 
\fancypagestyle{firstpage}{ 
	\fancyhf{}
}

\newcommand{\authorstyle}[1]{{\large\usefont{OT1}{phv}{b}{n}\color{DarkRed}#1}} 

\newcommand{\institution}[1]{{\footnotesize\usefont{OT1}{phv}{m}{sl}\color{Black}#1}} 

\usepackage{titling} 

\newcommand{\HorRule}{\color{DarkGoldenrod}\rule{\linewidth}{1pt}} 

\pretitle{
	\vspace{-30pt} 
	\HorRule\vspace{10pt} 
	\fontsize{19}{21}\usefont{OT1}{phv}{b}{n}\selectfont 
	\color{DarkRed} 
}

\posttitle{\par\vskip 15pt} 

\preauthor{} 

\postauthor{ 
	\vspace{10pt} 
	\par\HorRule 
	\vspace{20pt} 
}

\usepackage{lettrine} 
\usepackage{fix-cm}	

\usepackage{xstring} 

\usepackage[autostyle=true]{csquotes} 

\usepackage[T1]{fontenc}
\usepackage[utf8]{inputenc}

\DeclareUnicodeCharacter{0301}{*************************************}

\usepackage{graphics}
\usepackage[margin=1.5cm, font=small]{caption}
\usepackage{graphicx}
\usepackage{amsmath}
\usepackage{amssymb}
\usepackage{setspace}
\usepackage{nicefrac}
\usepackage{bm}
\usepackage{xr}
\usepackage{relsize}
\usepackage{caption}
\usepackage[scr]{rsfso}
\usepackage{ulem}
\usepackage{comment}
\usepackage[left]{lineno}

\usepackage[backend=biber,style=numeric,sorting=none]{biblatex}
\usepackage{pifont}   
\newcommand{\cmark}{\ding{51}} 
\newcommand{\xmark}{\ding{55}} 

\usepackage[colorlinks = true, linkcolor = purple, urlcolor  = purple, citecolor = blue, anchorcolor = purple]{hyperref}

\usepackage{xcolor}
\usepackage{cuted}
\usepackage{xurl}
\usepackage{hyperref}
\PassOptionsToPackage{hyphens}{url}
\usepackage{siunitx}

\RequirePackage[utf8]{inputenc}
\RequirePackage{calc}
\RequirePackage{indentfirst}
\RequirePackage{fancyhdr}
\RequirePackage{graphicx,epstopdf}
\RequirePackage{lastpage}
\RequirePackage{ifthen}
\RequirePackage{lineno}
\RequirePackage{float}
\RequirePackage{amsmath}
\RequirePackage{setspace}
\RequirePackage{enumitem}
\RequirePackage{mathpazo}

\usepackage{ulem}
\usepackage{mathtools}

\newtagform{brackets}{[}{]}
\usetagform{brackets}

\renewcommand{\labelenumi}{\theenumi}
\renewcommand{\theenumi}{(\roman{enumi})}

\definecolor{ao}{rgb}{0.0, 0.5, 0.0}

\usepackage[autostyle=true]{csquotes}

\title{Unifying non-Markovian Dynamics and Agent Heterogeneity in Scalable Stochastic Networks}

\author{\authorstyle{Aur\'elien P\'elissier$^{1,2,3}$, Miroslav Phan$^{1,2}$, Didier Le Bail$^{4}$, Niko Beerenwinkel$^{2}$ and Mar\'ia Rodr\'iguez Mart\'inez$^{1,5,\dagger}$}\newline\newline
\textsuperscript{1}\institution{IBM Research Europe, 8803 Rüschlikon, Switzerland}\\
\textsuperscript{2}\institution{Department of Biosystems Science and Engineering, ETH Zurich, 4058 Basel, Switzerland}\\
\textsuperscript{3}\institution{Institute of Computational Life Sciences, Zürich University of Applied Sciences (ZHAW), 8820, Wädenswil, Switzerland}\\
\textsuperscript{4}\institution{Centre de Physique Th\'eorique (CPT), Aix-Marseille University, CNRS, 13009 Marseille, France.}\\
\textsuperscript{5}\institution{Department of Biomedical Informatics \& Data Science, Yale School of Medicine, New Haven, CT, United States.}\\
\textsuperscript{$\dagger$}\institution{Corresponding author \href{maria.rodriguezmartinez@yale.edu}{maria.rodriguezmartinez@yale.edu}}}

\date{\vspace{-5ex}}

\begin{document}

\maketitle

\begin{strip}

\begin{center}
\begin{minipage}{14.6cm}
\vspace{-2cm}
\begin{center}
{\large \textbf{Abstract}}
\end{center}
\vspace{-0.2cm}

Stochastic processes underpin dynamics across biology, physics, epidemiology, and finance, yet accurately simulating them remains a major challenge. Classical approaches such as the Gillespie algorithm are exact for Markovian, time-independent systems, where propensities depend only on the current state and agents of a given type are statistically identical. While efficient, this framework misses a defining feature of many real systems: heterogeneity and memory at the level of individual agents. Cells may divide or differentiate on distinct intrinsic timescales, individuals may preferentially interact with specific partners, and inter-event-time distributions can deviate strongly from the exponential.
We introduce MOSAIC (Modeling of Stochastic Agents with Individual Complexity), a general and scalable framework that embeds agent-specific properties directly into the dynamics. MOSAIC unifies heterogeneous rates, dynamic interaction preferences, and both Markovian and non-Markovian waiting-time distributions within a single stochastic formalism, while retaining Gillespie-like computational cost.
Applications to delayed biochemical reactions, competitive immune-cell dynamics, and temporal social networks show that MOSAIC reproduces empirical features that existing methods either miss or capture only at prohibitive computational cost, establishing it as a practical tool for simulating heterogeneous stochastic systems.

\end{minipage}

\end{center}

\end{strip}

\vfill

\normalsize
\onecolumn

Stochastic processes provide a powerful framework for describing the random dynamics of complex systems across biology, physics, chemistry, epidemiology, and the social sciences~\cite{ross1996stochastic, van1992stochastic, cao2009discrete, yan2008distribution, kijima2002stochastic, tseng2020forecasting}. Classical approaches typically rely on the Markovian assumption, where the probability of future events depends only on the present state of the system. This assumption has enabled elegant mathematical analysis and efficient algorithms, most notably the Gillespie stochastic simulation algorithm (SSA)~\cite{gillespie1976general,gillespie1977exact}, which generates statistically exact trajectories for continuous-time Markov processes whose reaction propensities depend only on the current state. It has become a cornerstone of stochastic modeling~\cite{arkin1998stochastic, thomas2019probabilistic, pelissier2020computational}, as well as hybrid solvers such as COPASI~\cite{hoops2006copasi} that combine deterministic and stochastic descriptions within this framework.

Yet, the simplicity of the Markovian paradigm comes at a cost. Real systems rarely behave in a memoryless way, and most cannot be reduced to propensities that depend only on the instantaneous state. Molecular, cellular, and social processes often exhibit memory and temporal correlations: molecules may undergo delays before reacting, neurons exhibit refractory periods between firings, and human activity often follows bursty patterns separated by extended intervals of inactivity~\cite{barabasi2005origin, bratsun2005delay, stumpf2017stem, corral2004long}. Importantly, they also exhibit heterogeneity at the level of individual agents. Cells within the same population may divide or differentiate on different timescales, individuals may favor specific interaction partners, and inter-event time distributions (IEDs) may vary widely, from long-tailed to narrow or near normally distributed. While analytical approaches can capture certain classes of such non-Markovian effects, especially at steady state~\cite{zhang2019markovian}, simulating large, heterogeneous systems with general non-Markovian timing remains challenging. Ignoring this diversity systematically leads to simulations that deviate from empirical observations, while incorporating it markedly improves agreement with experimental and population-level data~\cite{gregg2021agent, grossmann2021heterogeneity, le2023modeling}.

Several approaches have sought to overcome the limitations of simple Markovian models. Hidden Markov Models, for example, capture long-range dependencies through latent states and have been applied in domains such as climate and finance~\cite{rabiner1986introduction, bracken2014hidden, nguyen2015hidden}, but they demand heavy parameterization, computationally intensive inference, and often lack a clear mechanistic link to the underlying process. Other strategies modify stochastic simulation directly. Delay-based algorithms and non-Markovian Gillespie variants incorporate memory by allowing rates to depend on elapsed time or mixtures of exponentials~\cite{anderson2007modified, boguna2014simulating, masuda2018gillespie}, while network-based constructions generate bursty temporal patterns~\cite{sheng2023constructing, ubaldi2017burstiness}. These methods extend the Gillespie framework to non-exponential dynamics, but typically assume homogeneous agents and incur higher computational costs, scaling as $O(\log N)$ or worse per event due to reaction-queue management or complex rate sampling.

Agent-based models (ABMs) provide a complementary path, enabling explicit heterogeneity in rules and interaction patterns~\cite{figueredo2014comparing, garcia2023understanding}. High-performance epidemic simulators such as GEMFsim, EpiFast, EpiSimdemics, and EpiHiper likewise support agent-level heterogeneity at large scale~\cite{sahneh2017gemfsim, bisset2009epifast, barrett2008episimdemics, chen2025epihiper}, but typically rely on discrete-time updates, large event queues, and distributed-memory HPC resources. A notable subclass of ABMs, activity-driven network models, highlights the importance of heterogeneous activity rates~\cite{perra2012activity}, but generally lacks rigorous stochastic guarantees and is not optimized for efficient, exact simulation. Collectively, these approaches underscore the value of heterogeneity (and, in some cases, memory), yet there is still no general framework that combines agent-specific dynamics and non-Markovian timing in continuous time while retaining Gillespie-like simplicity and single-node scalability.

Here we introduce MOSAIC (Modeling of Stochastic Agents with Individual Complexity), a rejection-based simulation framework that preserves the efficiency of Gillespie while allowing each agent to follow its own dynamics. Importantly, MOSAIC considers all instantaneous reaction rates at the start of each iteration, which permits dynamic, state-dependent changes in interaction preferences or individual inter-event distributions. In doing so, MOSAIC addresses both major sources of complexity, memory effects that generate non-exponential waiting times and dynamic agent-level heterogeneity in rates, preferences, and inter-event distributions, while maintaining update steps that scale independently of system size.
This represents a paradigm shift in stochastic modeling, enabling scalable simulations that link microscopic diversity and memory to emergent macroscopic behavior.

We demonstrate MOSAIC’s capabilities through three representative applications: clonal B-cell dynamics during immune responses, RNA transcription with feedback regulation, and temporal social networks with heterogeneous node activity. Across these domains, MOSAIC reproduces empirical features that existing methods either fail to capture or capture only at a heavy computational cost, establishing it as a general and scalable framework for the next generation of stochastic simulations.

\section{Results}

\subsection{Modeling of Stochastic Agents with Individual Complexity (MOSAIC).}

\noindent At its core, MOSAIC generalizes the Gillespie algorithm by representing each agent as its own renewal process with a time-dependent reaction rate~\cite{boguna2014simulating}. Unlike the standard Gillespie (SG) algorithm, which collapses identical particles into reaction channels with fixed exponential waiting times, MOSAIC allows every process to follow a distinct IED or to have a rate that depends on individual properties and interactions. This makes it possible to model systems where both memory effects and agent heterogeneity drive the dynamics.  
We denote by $N$ the total number of elementary stochastic processes (potential events) in the system. In a multi-agent model, each admissible agent--agent interaction defines one such process. Thus $N$ is fixed by the model structure and is not a tunable parameter.  
A central challenge is scalability. In many realistic systems $N$ is very large, either because every individual agent must be tracked separately or because interactions are pairwise, so that $N$ grows combinatorially with the population sizes (e.g., $\mathcal{O}(N_\text{A} N_\text{B})$ for bimolecular reactions). Treating each process as an independent channel, as in SG, would lead to prohibitive computational costs~\cite{thomas2019probabilistic}.  

MOSAIC overcomes this barrier by tracking only a single global maximum rate, $\lambda_\text{max}$, rather than updating all individual rates at every step. Candidate processes are drawn uniformly at random from the population, each with probability $1/N$, and then accepted with probability proportional to their instantaneous rate, $\lambda_j(t_j)$ (Eq.~\ref{eq:rejection}). Here $\lambda_j(t_j)$ is shorthand for a state-dependent rate, which may depend both on the elapsed time $t_j$ since the last event of process $j$ and on the current configuration of the system. This rejection sampling step ensures that the effective probability of firing is correctly weighted by the true internal rates of the processes, while avoiding the need to explicitly maintain and update all rates~\cite{thanh2014efficient,st2019efficient}. By relying on a global bound rather than explicit enumeration, each update requires constant-time operations (Methods~\ref{Methods_MOSAIC}), where the complexity per accepted event depends on the expected number of rejections, $r$, which we show later is constant and independent of $N$, unlike the $O(N)$ scaling of the standard Gillespie algorithm. Moreover, MOSAIC does not require maintaining a reaction queue, further reducing memory usage and implementation complexity. This design ensures that MOSAIC remains efficient even in populations with millions of heterogeneous agents or densely interacting networks, while still capturing the full diversity of waiting times and agent-specific interaction preferences.\\

Algorithmically, MOSAIC explicitly tracks a global time $T$ and, for each process $j$, the time $\tau_j$ of its last event. The elapsed time of process $j$ is then given implicitly by $t_j = T - \tau_j$. Let $\lambda_{j}(t_j)$ denote its time-dependent (and state-dependent) rate. Each iteration of MOSAIC proceeds in four steps:

\begin{enumerate}
    \item \textit{Set the global maximum rate.}  
    Choose $\lambda_\text{max}$ such that
    \begin{equation}
        \lambda_\text{max} \geq \max_{1 \leq j \leq N} \lambda_j(t_j).
    \end{equation}
    In practice, $\lambda_\text{max}$ is a user-chosen upper bound: it can be set to the true maximum rate in the system, or increased beyond this value to improve accuracy at the cost of additional rejections.\\

    \item \textit{Advance the time.}  
    Draw $u \sim \mathcal{U}[0,1]$, set $\Delta t$ and advance the global time
    \begin{equation}
        \Delta t = \frac{\ln (1/u)}{N \cdot \lambda_\text{max}}. \ \ \text{and} \ \ T \leftarrow T + \Delta t
        \label{eq:deltat}
    \end{equation}

    \item \textit{Select a candidate process.}  
    Choose $j$ uniformly at random:
    \begin{equation}
        p_j = \frac{1}{N}.
        \label{eq:choice}    
    \end{equation}

    \item \textit{Accept or reject the event.}  
    Compute the updated elapsed time of the candidate process as $t_j = T - \tau_j$ and accept with probability
    \begin{equation}
        p_\text{accept} = \frac{\lambda_j(t_j)}{\lambda_\text{max}}, 
        \label{eq:rejection}
    \end{equation}
    in which case the last-event time of the firing process is reset, $\tau_j \leftarrow T$. This state update may in principle change the rates of other processes, but MOSAIC does not require maintaining or updating the full vector $(\lambda_1,\dots,\lambda_N)$; only the global bound $\lambda_\text{max}$ must remain an upper bound on all instantaneous rates, which can be enforced in $O(1)$ time as described in Methods~\ref{Methods_MOSAIC}. If the event is rejected, the system state and all $\tau_k$ remain unchanged, and the only effect of the iteration is that the global time $T$ has advanced by $\Delta t$.
\end{enumerate}

MOSAIC recovers the Gillespie renewal theory, in the limit $\Delta t \rightarrow 0$ (equivalently when $N \to \infty$ or $\lambda_\text{max} \to \infty$), the simulated IED for process $j$ converges to the renewal-theory form:  
\begin{equation}
\psi_j(t_j) = \lambda_j(t_j) \exp \!\left( - \int_0^{t_j} \lambda_j(\tau) \, d\tau \right).
\label{eq_PDF_main}
\end{equation}
A full derivation is provided in Supplementary Section~\ref{MOSAIC_proof}. Although rejection introduces an approximation at finite step size, MOSAIC becomes exact in the small-$\Delta t$ limit and can reach arbitrarily high accuracy by increasing $\lambda_\text{max}$, even when $N$ is modest.\\  

In terms of accuracy, the expected error in the sampled inter-event times of a process satisfies (Supplementary Section~\ref{quant_error_MOSAIC}):
\begin{equation}
\mathbb{E}\!\left[ \text{error}_{\text{\tiny MOSAIC}} \right] 
   \sim \frac{\langle \lambda' \rangle}{N \cdot \lambda_\text{max}^2},
\label{Eq:MOSAIC_final_error}
\end{equation}
with $\langle \lambda' \rangle = \int_0^\infty \lambda'(t)\psi(t)\,dt$ the mean derivative rate. In practice, this error correlates to the Earth Mover’s Distance~\cite{rubner1998metric} between the true and simulated distribution of inter-event times for that process. Here $N$ is fixed by the model, whereas $\lambda_\text{max}$ is freely chosen and provides a direct accuracy–efficiency trade-off.

The computational complexity is affected by the number of rejected reactions, which scales as $r \propto \lambda_\text{max}/\lambda_0$ per accepted event, where $\lambda_0$ is the mean event rate (Supplementary Section~\ref{proof_complexity}). 
Thus, increasing $\lambda_\text{max}$ reduces the error quadratically, but raises the number of rejections only linearly.\\ 

Finally, MOSAIC generalizes the standard Gillespie algorithm by embedding heterogeneity directly at the level of individual agents. Each process can follow its own inter-event distribution, with rates defined as $\lambda_j(t) = \psi_j(t)/\Psi_j(t)$. This formulation allows Gamma, Weibull, Pareto, log-normal, and other biologically relevant waiting-time distributions to be simulated without approximation, enabling accurate representation of memory effects (Supplementary Section~\ref{MOSAIC_distributions}). In addition, reaction rates can depend on agent-specific traits $\mathcal{P}_j$ and on the current system state, such as susceptibility in epidemics, binding affinity in biochemistry, or sociability in networks. Unlike traditional approaches that require discretizing these properties into subchannels~\cite{thomas2019probabilistic}, MOSAIC evaluates $\lambda_j = \lambda_j(t_j, \mathcal{P}_j, \text{state})$ directly, thereby preserving the full variability of individual properties.

\begin{table}[htbp]
\centering
\captionsetup{width=1\linewidth}
\setstretch{1.3} 
\resizebox{\textwidth}{!}{
\begin{tabular}{lcccll}
\toprule
\textbf{Algorithm} & \textbf{Arbitrary inter-event} & \textbf{Dynamic} & \textbf{Accuracy} & \textbf{Complexity/} & \textbf{Complexity /} \\
                & \textbf{time distribution} & \textbf{delays} &  & \textbf{simulation} & \textbf{time step} \\
\midrule
Standard Gillespie & \xmark & \cmark & Exact & $O(N^2)$ & $O(N)$ \\
Tree-based Gillespie~\cite{gibson2000efficient} & \xmark & \cmark  & Exact & $O(N k \log N)$ & $O(k \log N)$ \\
DelaySSA~\cite{fu2022delayssatoolkit} & \cmark & \xmark & Exact & $O(N k \log N)$ & $O(k \log N)$ \\
nMGA~\cite{boguna2014simulating} & \cmark & \cmark & Only when $N \rightarrow \infty$ & $O(N^2)$ & $O(N)$ \\
\textbf{MOSAIC (ours)} & \cmark & \cmark & Arbitrary high & $O(rN)$ & $O(1)$ \\
\bottomrule
\end{tabular}}
\caption{Comparison of stochastic simulation algorithms for systems with heterogeneous agents. $N$ is the number of processes; $k$ is the number of propensity updates per iteration; $r$ is the rejection/acceptance ratio (Supplementary Section~\ref{proof_complexity}). Details on complexity derivations are provided in the Supplementary section~\ref{SI_complexity}.}
\label{table:algorithm_comparison_IP}
\end{table}
\renewcommand{\arraystretch}{1}

In summary, MOSAIC extends the Gillespie framework to non-Markovian systems with full agent-level heterogeneity. By tracking only a global maximum rate, it scales efficiently to large populations and multi-reactant interactions, while capturing diverse waiting-time distributions and individual-specific properties. This combination of accuracy and scalability establishes MOSAIC as a general tool for stochastic simulation across biology, physics, and networked systems.

\subsection{Application 1: Competitive selection in heterogeneous B-cell populations}

\begin{figure}[h!t]
    \centering
    \captionsetup{width=1\linewidth}
    \includegraphics[width=0.83\linewidth]{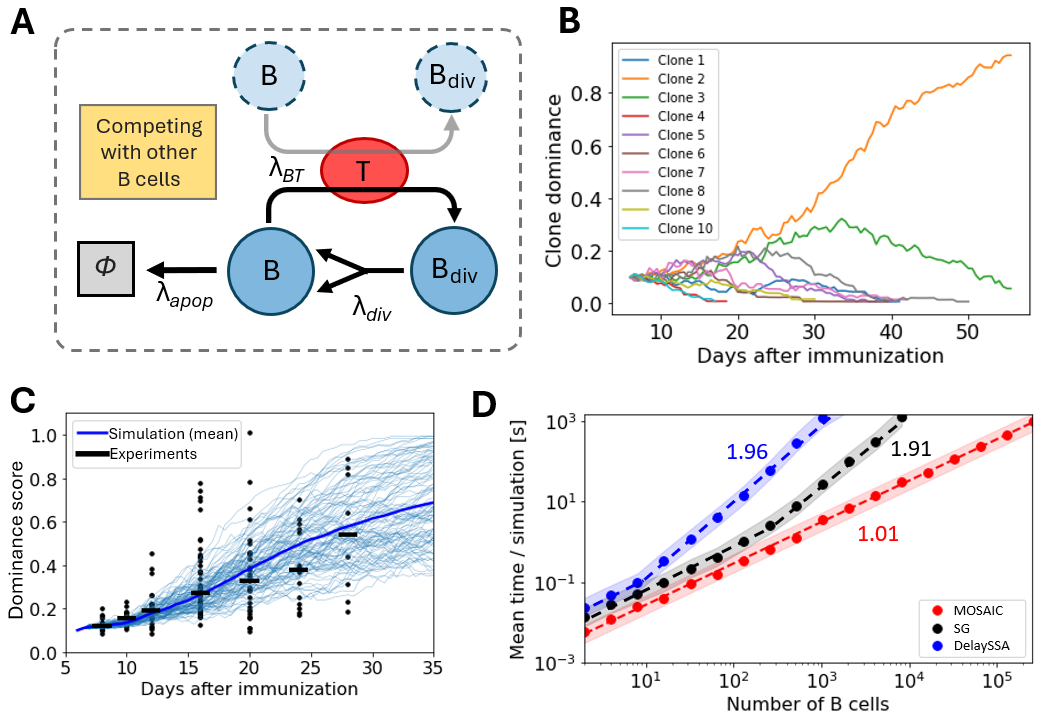}
    \caption{\textbf{Efficient simulation of affinity maturation with MOSAIC.}
    (A) Cartoon of affinity maturation. B cells compete for T-cell help: higher-affinity clones are rescued from apoptosis and expand, whereas lower-affinity clones undergo apoptosis. 
    (B) Evolution of 10 clonal families in a representative simulation. Family 2 dominates after several days.  
    (C) Dominance of the largest clone across 1000 simulations (thin blue lines = replicates, thick blue = mean). Experimental data from~\cite{tas2016visualizing} are shown as scatter points with median bars.  
    (D) Computational performance of different algorithms as $N_B$ increases, with slopes obtained from piecewise polynomial fits. Shaded regions denote standard deviations.}
    \label{fig:immune_response}
\end{figure}

\noindent B-cell affinity maturation provides a canonical example of a heterogeneous stochastic process shaped by competition. During an immune response, B cells expressing diverse antigen receptors compete for survival signals delivered by a limited pool of T cells. Clones, i.e. groups of B cells descended from the same ancestor, with higher antigen affinity are preferentially rescued from apoptosis, undergo further rounds of proliferation, and acquire affinity-enhancing mutations, ultimately giving rise to dominant clonal families (Figure~\ref{fig:immune_response}A). Although many cellular processes, such as apoptosis or division, are governed by intrinsic programs with characteristic timescales that deviate from exponential waiting-time distributions, we focus here on a simplified \textit{Markovian} setting in which all reactions are assumed exponential. This approximation, which has previously been shown to capture the essential features of germinal center dynamics~\cite{thomas2019probabilistic,pelissier2020computational}, enables us to isolate the role of clonal heterogeneity and to demonstrate the efficiency gains of MOSAIC over existing Markovian methods. Extensions to non-exponential dynamics will be addressed later in this article (Sections~\ref{result:application2}~\&~\ref{result:application3}).

We consider a minimal model consisting of five processes (Methods~\ref{GC-system}): (i) binding of B cells to T cells, (ii) displacement of bound B cells by higher-affinity competitors, (iii) apoptosis of B cells that did not receive survival signals from T cells, (iv) spontaneous unbinding after sufficient T-cell signaling has been received, and (v) subsequent division with affinity updates. Among these, competition (process ii), in which B cells with higher-affinity receptors preferentially displace lower-affinity counterparts from T-cell interactions, plays the central role in affinity maturation. We represent this process as
\begin{equation}
    [B_1T] + B_2 \xrightarrow{(IP)} [B_2T] + B_1,
\end{equation}
where \textit{(IP)} indicates that the rate depends on the individual properties of the cells, in this case, receptor affinity. In our model, a bound B cell $B_1$ can only be displaced if the incoming competitor $B_2$ has strictly higher affinity. This pairwise competition drives clonal selection but also introduces algorithmic challenges, as each event can alter the rates of many other competing pairs. MOSAIC overcomes this challenge by using a rejection-based scheme, in which all potential interactions are sampled, but only those that satisfy the strict affinity threshold defined by the existing interaction are accepted (Eq.~\ref{eq:rejection}). In this application the dynamics are purely Markovian, so the simulation is exact and we simply set 
\(\lambda_{\text{max}} = \max_{1 \le j \le N} \lambda_j\), i.e., equal to the true maximum rate over all processes.

Simulations with $N_B = 1000$ B cells and $N_T = 10$ T cells, modeling 50 days of affinity maturation, reproduce experimental observations.
Figure~\ref{fig:immune_response}B shows the trajectories of the ten most expanded clones in a representative run, where one clonal family (clone 2) becomes dominant after several days. When results are aggregated across replicates (Figure~\ref{fig:immune_response}C), the dominance of the leading clone increases monotonically, in agreement with experimental data~\cite{tas2016visualizing}.

Next, we examine computational scaling. Conventional Gillespie methods~\cite{gillespie1976general,gillespie1977exact} and its variant like COPASI~\cite{hoops2006copasi} face a combinatorial burden due to pairwise competition. After each event, it is necessary to check whether any unbound B cell has higher affinity than those currently bound to T cells, which requires maintaining and updating a list of competing pairs. Because each displacement event can change the rates of many potential competitors, this bookkeeping becomes computationally expensive. In a standard Gillespie-style implementation for this heterogeneous setting, one must treat each admissible B–T pair as a separate stochastic process with its own propensity, leading to $N = N_B \cdot N_T$ such processes and $O(N)$ complexity per step. Tree-based Gillespie~\cite{gibson2000efficient} and DelaySSA~\cite{fu2022delayssatoolkit} improve event selection to $O(\log N)$ through hierarchical data structures, but still require updating $k$ propensities after each event, leading to an overall cost of $O(k \log N)$ that grows superlinearly with system size. MOSAIC avoids this overhead entirely by relying on the rejection scheme, where all potential interactions are sampled but only those satisfying the affinity condition are accepted (Eq.~\ref{eq:rejection}). This reduces the per-iteration cost to constant time on average, independent of $N$ or $k$. Empirical benchmarks confirmed this scaling behavior (Figure~\ref{fig:immune_response}D; Table~\ref{table:algorithm_comparison_IP}), with MOSAIC maintaining near-linear runtime even in large, densely interacting populations.

These results demonstrate that even under purely Markovian assumptions, MOSAIC faithfully reproduces the qualitative and quantitative features of affinity maturation while maintaining computational efficiency at scales where conventional methods become prohibitive. This establishes its utility for modeling heterogeneous, competitive systems.

\subsection{Application II: Reactions with state-dependent non-exponential delays}

\label{result:application2}

\noindent Many biological processes display memory effects, where reaction times deviate from simple exponential behavior and instead reflect intrinsic programs or multi-step mechanisms. For instance, apoptosis and cell division unfold on characteristic timescales, while transcription elongation can pause intermittently, giving rise to broad inter-event time distributions. Classical extensions of Gillespie, such as DelaySSA~\cite{anderson2007modified,fu2022delayssatoolkit}, can simulate arbitrary IEDs by scheduling delays at the moment of initiation. However, once a delay has begun, it cannot be modified if system conditions change. This limitation prevents DelaySSA from capturing state-dependent waiting times, where effective delays vary dynamically with the molecular context. In contrast, MOSAIC continuously evaluates instantaneous rates, allowing each process to adapt in real time to changes in the system. This flexibility enables the modeling of feedback-regulated processes in which delays are coupled to evolving system states.

\begin{figure}[h!t]
    \centering
    \captionsetup{width=1\linewidth}
    \includegraphics[width=0.95\linewidth]{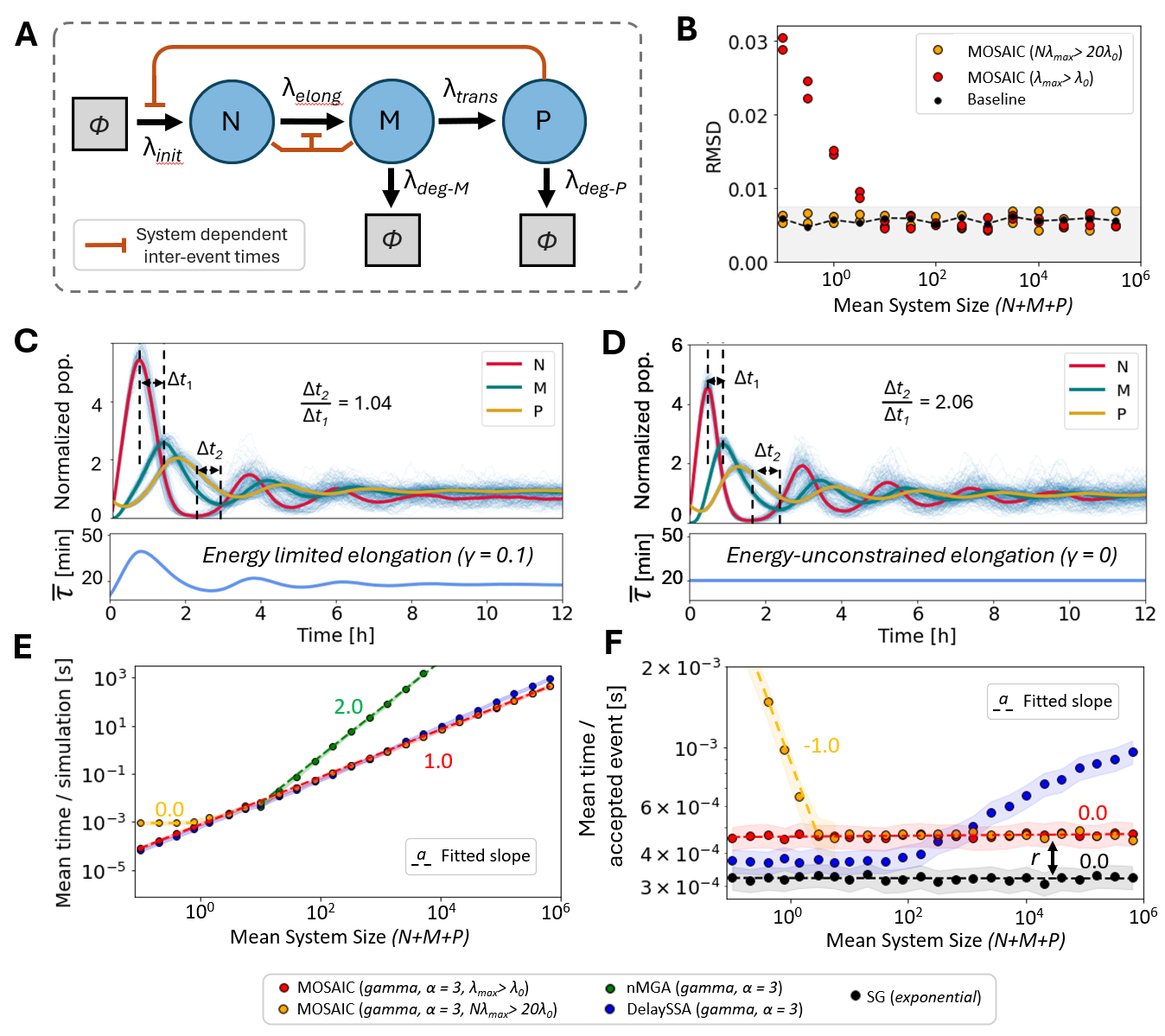}
    \caption{\textbf{MOSAIC accurately simulates Hes1 dynamics with state-dependent elongation delays} (A) Schematic of the Hes1 negative feedback model. Nascent RNA (N) elongates until it matures into mRNA (M), which is translated into protein (P). Both mRNAs and proteins degrade at constant rates. Hes1 protein inhibits its own transcription, introducing a feedback loop that depends on the current system state. (B) Root mean square deviation (RMSD) between MOSAIC and DelaySSA trajectories as a function of mean system size ($\approx 25\beta$). Black dots show the baseline RMSD between two DelaySSA ensembles, while red dots compare MOSAIC against DelaySSA. Orange dots show MOSAIC with stricter $\lambda_\text{max}$ settings, further reducing error. Shaded regions indicate the 95\% confidence interval expected between identical simulations. (C–D) Population oscillations in nascent RNA, mature mRNA, and protein for energy-limited ($\gamma > 0$, panel C) and energy-unconstrained ($\gamma = 0$, panel D) elongation. Thin blue lines show single realizations, while thick lines show the average over 1000 simulations. Altered ratios of peak-to-trough delays ($\Delta t_2 / \Delta t_1$) highlight the impact of state-dependent elongation times. (E) Average runtime per simulation as $\beta$ increases, showing near-linear scaling of MOSAIC compared to superlinear growth in alternative methods. (F) Runtime normalized by accepted events. MOSAIC maintains constant per-event cost, differing from standard Gillespie only by a rejection factor ($r \approx 1.6$ in this model).}
    \label{fig:hes1_model}
\end{figure}

We illustrate this capability with a classic model of the transcription factor Hes1, which represses its own transcription through a delayed negative feedback loop~\cite{hirata2002oscillatory, monk2003oscillatory, barrio2006oscillatory} (Methods~\ref{hes1-system}). In this system (Figure~\ref{fig:hes1_model}A), RNA polymerase initiates transcription, the nascent RNA undergoes a non-exponential elongation phase before maturing into mRNA, and the mRNA is then translated into protein. The resulting Hes1 protein feeds back to inhibit further initiation, producing oscillations in RNA and protein levels.

The elongation step provides a critical test case. Experiments indicate that elongation times are broadly distributed, with heavy tails arising from polymerase pausing~\cite{rajala2010effects,qian2021basic}. We model elongation as a Gamma-distributed waiting time, with the rate parameter further modulated by the number of nascent and mature RNAs ($N_N$ and $N_M$ respectively) to account for limited cellular resources:
\begin{equation}
    \lambda_0 = \tau_0^{-1} \left( 1 + \gamma \cdot \frac{N_N + N_M}{\beta} \right)^{-1}.
\end{equation}
Here, $\tau_0$ defines the baseline mean delay, $\gamma$ controls resource sensitivity, and $\beta$ scales with system size. When $\gamma > 0$, elongation slows as transcripts accumulate, dynamically reshaping the waiting-time distribution. Importantly, such state-dependent delays cannot be captured by DelaySSA, which fixes delays at initiation, but are naturally accommodated by MOSAIC.

To benchmark accuracy, we first examine the case $\gamma = 0$ (no resource coupling). In this regime, elongation delays are fixed, and DelaySSA is exact, providing a natural ground truth for evaluating MOSAIC. MOSAIC trajectories closely track those of DelaySSA across a wide range of system sizes (Figure~\ref{fig:hes1_model}B), with deviations vanishing as $\beta$ increases, consistent with theoretical error bounds (Eq.~\ref{Eq:MOSAIC_final_error}). Stricter choices of $\lambda_\text{max}$ further reduce errors in small systems, at only modest additional computational cost (Supplementary Section~\ref{arbitrary_accuracy}); we find here that increasing $\lambda_{\text{max}}$ above the true maximum rate is only beneficial for systems with fewer than about 20 processes.
When $\gamma > 0$, elongation delays become context-dependent, producing marked changes in oscillatory behavior. In particular, oscillations in mRNA and protein populations lengthen in period as elongation slows at high transcript abundance (Figure~\ref{fig:hes1_model}C–D). This effect reflects competition for limited cellular resources, which in practice constrains the energy and machinery available for transcription. MOSAIC naturally captures these shifts by continuously updating reaction rates with the system state. In contrast, DelaySSA schedules delays at initiation and keeps them fixed, making it unsuitable for systems with dynamically evolving inter-event times (Table~\ref{table:algorithm_comparison_IP}). These results demonstrate how MOSAIC extends Gillespie-like simulation to biologically realistic settings with state-dependent delays.
 
We next assess runtime scaling across $\beta$, which controls the average number of molecules (Figure~\ref{fig:hes1_model}E–F). For comparison, we include the non-Markovian Gillespie algorithm (nMGA)~\cite{boguna2014simulating}, which extends the Gillespie framework to non-exponential waiting times by explicitly tracking all pending reactions. As expected, nMGA exhibits quadratic scaling, while DelaySSA incurs $O(\log N)$ cost per accepted event due to maintaining a priority queue of delays (Table~\ref{table:algorithm_comparison_IP}). In contrast, MOSAIC achieves linear scaling with a constant per-event cost, differing from the standard Gillespie algorithm only by a multiplicative rejection factor $r$ (here $r \approx 1.6$). Thus, MOSAIC unites the flexibility to model dynamic, non-Markovian delays with the computational efficiency of classical stochastic simulation.

Together, these results demonstrate that MOSAIC enables realistic modeling of feedback-regulated processes, including transcription, signaling, and cell-cycle progression, where capturing non-exponential dynamics is critical for understanding system behavior.

\subsection{Application III: Non-Markovian temporal networks (MOSAIC-TN)}

\label{result:application3}

\begin{figure}[h!t]
    \centering
    \captionsetup{width=1\linewidth}
    \includegraphics[width=1\linewidth]{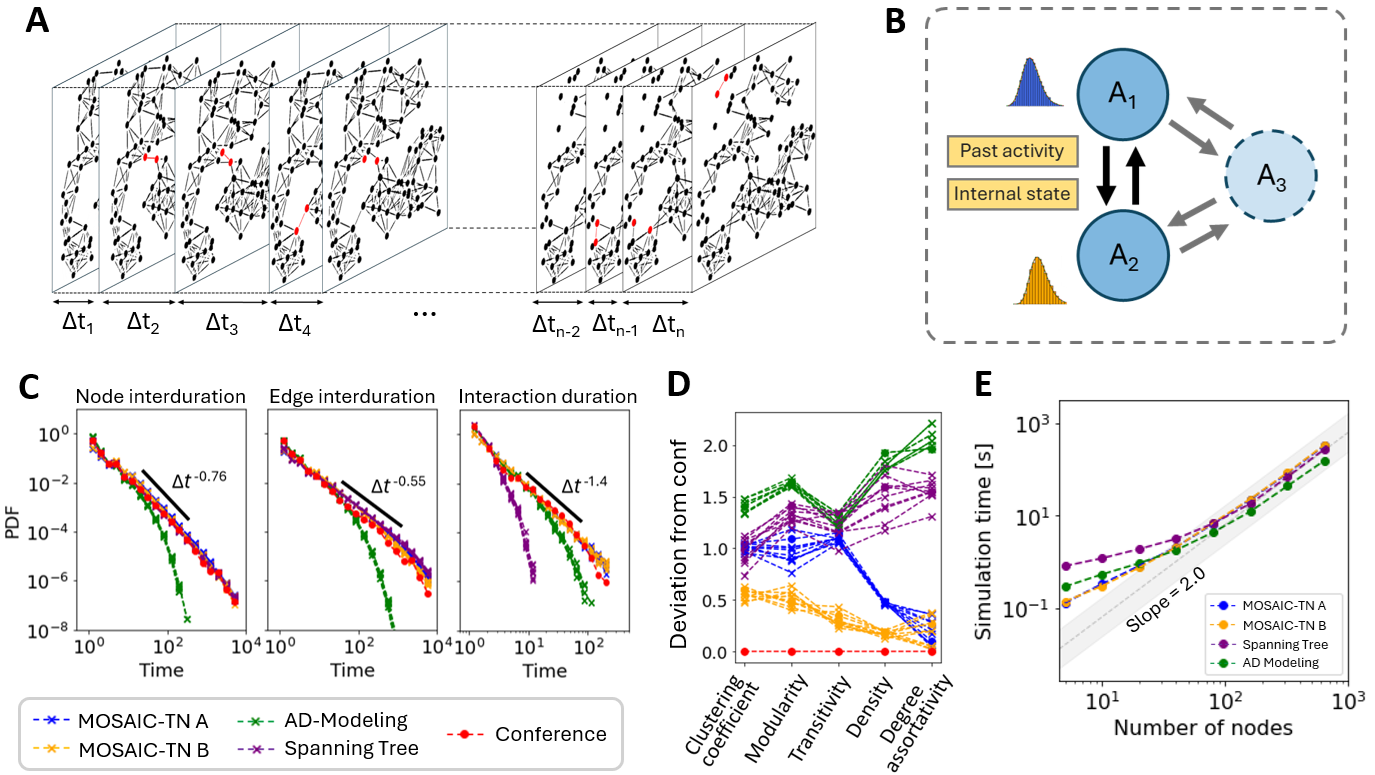}
    \caption{\textbf{MOSAIC-TN reproduces empirical social-temporal statistics with bursty node activity and interaction memory.}
    (A) Schematic of a temporal network simulated with MOSAIC-TN. At each time step, one link at most is updated; modified nodes and edges are shown in red. Rejected steps leave the network unchanged. The time increment $\Delta t$ is variable and computed according to Eq.~\ref{deltat_TN}.
    (B) Cartoon of the face-to-face interaction model. Each node tracks its own inter-event time distribution (IED), shown as blue and brown curves, and pairwise interaction probabilities depend on both intrinsic activity and past contacts.  
    (C) Probability densities for temporal interaction statistics in conference data~\cite{genois2022combining} and in simulated networks, averaged over 10 runs per model. Left: node interdurations, i.e. time intervals between successive interactions of the same node. Middle: edge interdurations, i.e. time intervals between successive interactions of the same pair of nodes. Right: interaction durations, i.e. edge lifetimes. Power-law exponents fitted to the tails of the conference distributions are annotated in the panels. Two MOSAIC-TN variants are shown: MOSAIC-TN~A (no partner preference) and MOSAIC-TN~B (reinforced past interactions), alongside other baselines.
    (D) Absolute deviations of aggregated network metrics (clustering, modularity, transitivity, density, assortativity) from conference data. Deviations are normalized by the standard deviation across models; values are averaged over 10 simulations per model.  
    (E) Computational runtime as a function of network size ($N_A$). For $N_A > 100$, all models exhibit $O(N_A^2)$ scaling (shaded region).
    }
    \label{fig:social_interactions}
\end{figure}

\noindent Temporal networks describe systems in which connections between nodes evolve over time, capturing both the timing and the duration of interactions (Figure~\ref{fig:social_interactions}A). They provide a natural framework for studying dynamical processes such as opinion spreading~\cite{li2006dynamics, wang2020public}, epidemic transmission~\cite{holme2016temporal, keeling2005implications}, brain activity~\cite{lynn2019physics}, and the diffusion of innovations~\cite{iacopini2018network}. Many such processes are inherently non-Markovian, characterized by bursty human activity, long inactive periods, and history-dependent dynamics that cannot be captured by memoryless exponential waiting times~\cite{scholtes2014causality, han2023non, williams2019effects}.
Temporal networks can be modeled according to two complementary perspectives: \textit{edge-driven} or \textit{node-driven}. In edge-driven models, each  link $(i, j)$ has its own clock, and an interaction occurs when that clock fires~\cite{vestergaard2015temporal, unicomb2021dynamics}. In node-driven models, each node carries an activity process that generates partner interactions. When all clocks are exponential, these two perspectives are mathematically equivalent: the superposition of exponential edge clocks yields exponential node activity, and vice versa. In this Markovian setting, standard Gillespie methods apply seamlessly to either formulation. 

For non-exponential IEDs, however, this equivalence breaks down (Supplementary Section~\ref{distribution_relationship}). Node activity statistics can no longer be obtained by simply combining edge clocks, and conversely, prescribing node-level IEDs does not translate into straightforward edge processes. Existing Gillespie extensions (DelaySSA, nMGA, Laplace Gillespie) remain fundamentally edge-centric, unable to explicitly capture heterogeneous node activity patterns or adapt to self-exciting and bursty dynamics. Yet many real-world temporal networks are inherently node-driven. For instance, in social systems, individuals alternate between long inactive periods and short bursts of intense interaction~\cite{barabasi2005origin, genois2022combining}. In neuroscience, neurons generate spike trains with refractory periods and history-dependent firing~\cite{kelly2012framework}. In molecular biology, receptors and enzymes cycle through inactive and active states that shape their binding opportunities~\cite{guevara2014use}. In all these examples, it is the \textit{intrinsic dynamics of nodes}, rather than the edges, that primarily govern interaction statistics.

In this final use case, we show that MOSAIC can be extended to node-driven systems to capture these behaviors (MOSAIC-TN; Methods~\ref{method:MOSAIC-TN}). Namely, we consider a temporal network where each node $A_i$ carries its own clock $t_i$, which records the time since the last interaction. Pairwise events are governed by a kernel rate $\Lambda_{ij}(t_i, t_j)$, which combines the activities of the two nodes and can incorporate history-dependent interaction terms. More generally, $\Lambda_{ij}(t_i, t_j)$ can be scaled by a global network-topology factor $H(i,j)$, allowing it to encode complex, non-independent relationships between nodes (e.g., higher-order structural effects beyond direct adjacency).
At each iteration, two nodes are sampled according to the desired interaction topology (e.g., by drawing an edge from the underlying static network), and their interaction is accepted with probability $\Lambda_{ij}(t_i, t_j)/\Lambda_\text{max}$, where $\Lambda_\text{max}$ is an upper bound on all pairwise rates. As in MOSAIC, rejection sampling ensures $O(1)$ complexity per candidate event while allowing arbitrary node-level IEDs. 

For a multiplicative kernel of the form $\Lambda_{ij} \propto \lambda_i(t_i) \cdot \lambda_j(t_j)$, interactions occur only when both nodes are concurrently available, since the product suppresses events whenever either node’s intrinsic rate is low. This probabilistic synchronization couples node and edge dynamics, and in the limit where the number of links $N \to \infty$, the activity of each node converges to its intrinsic IED (Supplementary Section~\ref{MOSAIC_paired_rate}). In this way, MOSAIC-TN generalizes Gillespie-style simulation to non-Markovian temporal networks, consistently linking individual node activity patterns to emergent interaction dynamics.\\


We next apply MOSAIC-TN to model face-to-face interactions recorded at the International Conference on Computational Social Science (IC2S2, 2017)~\cite{genois2022combining}. Conferences provide a nearly well-mixed setting in which any two participants can, in principle, interact, so we model the system as a fully connected underlying contact graph and let temporal dynamics determine which edges are active at any given time. Empirical analyses show that both the \textit{node interduration}, i.e. the time between successive interactions of the same individual, and the \textit{interaction duration}, i.e. the time an edge persists, follow heavy-tailed distributions, well approximated by Pareto laws~\cite{genois2022combining, le2023modeling}. Accordingly, we fit Pareto distributions to the IC2S2 data and use the resulting parameters to define two kinetic channels (Methods~\ref{param_optimization}):
(i) $A_i + A_j \to [A_i A_j]$ with $\psi(t) \sim t^{-\alpha_A}$ for node interdurations, and
(ii) $[A_i A_j] \to \varnothing$ with $\psi(t) \sim t^{-\alpha_\varnothing}$ for interaction durations (Figure~\ref{fig:social_interactions}B,C).

To test the role of interaction memory, we implement two variants. In MOSAIC-TN~A, all pairs interact with equal probability, independent of past history. In MOSAIC-TN~B, partner selection is history-dependent: once $A_i$ is chosen, nodes $A_j$ with prior interactions are $w$ times more likely to be selected. This mechanism captures first-order temporal neighborhood effects, where past contacts increase the likelihood of repeated interactions. Remarkably, despite introducing only a single additional parameter, $w$, this history-based reinforcement substantially improves model performance. Optimizing $w$ to minimize deviations in network clustering, modularity, transitivity, density, and assortativity (Methods~\ref{param_optimization}) yields markedly better agreement with empirical data on four of the five metrics compared to MOSAIC-TN~A (Figure~\ref{fig:social_interactions}D).

We then benchmark MOSAIC-TN against two established methods, the spanning-tree framework and activity-driven (AD) models. The spanning-tree approach~\cite{sheng2023constructing} imposes specified node and edge IEDs but lacks flexibility in modeling interaction durations, leading to systematic underestimation of edge lifetimes (Figure~\ref{fig:social_interactions}C). AD models~\cite{perra2012activity, le2023modeling} provide greater flexibility but require extensive parameter tuning and inherently generate exponential tails, failing to reproduce the heavy-tailed node interdurations observed in empirical data (Figure~\ref{fig:social_interactions}C). By contrast, MOSAIC-TN directly enforces arbitrary node-level IEDs while preserving exact event-driven timing, enabling it to simultaneously match both the interduration statistics and the distribution of interaction durations (Figure~\ref{fig:social_interactions}C). In our instantaneous-rate framework, node and edge interdurations are mathematically related (Supplementary Section~\ref{distribution_relationship}), and MOSAIC-TN reproduces the empirically observed burstiness at both levels. Previous models that incorporated bursty node IEDs~\cite{ubaldi2017burstiness, karsai2012universal} treated interactions independently: nodes could display bursty activation while edge IEDs remained effectively memoryless. The spanning-tree framework~\cite{sheng2023constructing} captures edge-level burstiness but at the cost of reduced flexibility. MOSAIC-TN unifies these advantages by enforcing bursty node IEDs, reproducing bursty edge interdurations, and retaining flexible, realistic interaction durations within a single coherent framework.

We next assess computational scaling with network size, where $N_A$ denotes the number of nodes in the social network (Figure~\ref{fig:social_interactions}E). In this fully connected setting, the number of processes scales as $N \sim N_A^2$, since each pair of nodes can potentially interact. At smaller $N_A$, runtimes differ across methods: MOSAIC-TN~A and B achieve the fastest execution, AD models followed by leveraging parallel updates, while the spanning-tree approach incur higher costs. As $N_A$ increased, all methods converge to the expected quadratic scaling $O(N_A^2)$, consistent with the combinatorial growth in interactions. Importantly, MOSAIC-TN preserve its constant per-event cost across the full range of network sizes, thereby matching the asymptotic efficiency of classical Gillespie-type approaches. Moreover, MOSAIC-TN~A and MOSAIC-TN~B remain indistinguishable, confirming that history-based reinforcement introduces no measurable computational overhead.

In summary, MOSAIC-TN combines the statistical precision of Gillespie-style simulation with the flexibility to impose arbitrary node-level IEDs and incorporate temporal memory. When applied to social contact data, it faithfully reproduces heavy-tailed interdurations, realistic edge lifetimes, and aggregated network statistics—outperforming both spanning-tree and AD models. Beyond social systems, this flexibility is essential for capturing complex real-world dynamics, including temporal neighborhood effects~\cite{le2024flow, le2025generalizing} and higher-order interactions~\cite{han2024probabilistic, lin2024higher, sheng2024strategy}, where nodes preferentially engage with neighbors of their neighbors~\cite{majhi2022dynamics, le2023modeling, iacopini2024temporal}. Thus, MOSAIC-TN establishes a principled and scalable framework for simulating non-Markovian temporal networks, with broad applications ranging from human interaction patterns to neuronal firing and molecular binding.

\section{Discussion}

\noindent Real-world systems often display memory effects and and heterogeneity among agents and nodes, from transcriptional regulation to epidemic spreading~\cite{barabasi2005origin, stumpf2017stem, corral2004long}. These features are closely connected, as diversity in agents and their interactions naturally produces memory effects and non-exponential dynamics.
MOSAIC introduces a unifying framework that bridges two frameworks long kept separate. Gillespie-style algorithms offer elegance, efficiency, and analytical rigor but assume statistical homogeneity among agents~\cite{zhang2019markovian, boguna2014simulating, masuda2018gillespie, aquino_chemical_2017}. Agent-based approaches, in contrast, can encode arbitrary heterogeneity but typically sacrifice scalability and mathematical tractability~\cite{gregg2021agent, garcia2023understanding, perra2012activity}. MOSAIC combines the strengths of both: it retains the rigor and scalability of Gillespie-type methods while extending their flexibility to heterogeneous agents with distinct distributions, rates, and interaction preferences. This synthesis establishes MOSAIC as a general framework for systems in which memory effects and heterogeneity are both essential.

We have demonstrated MOSAIC’s versatility across three distinct domains. First, in a system of B cell competition and evolution, MOSAIC efficiently models clonal selection driven by B cell receptor diversity. B cells with different antigen affinities compete for limited T cell help, where only the highest-affinity B cells are rescued from apoptosis, allowed to divide, and may accumulate beneficial mutations. This competition is central to affinity maturation but also introduces algorithmic challenges, since each B cell must be represented with its own receptor-specific affinity. In standard Gillespie implementations, such heterogeneity leads to a combinatorial explosion of possible interactions. By contrast, MOSAIC uses a rejection-based scheme in which candidate interactions are sampled uniformly and only those meeting the strict affinity threshold are accepted. This avoids maintaining or updating a full list of competing pairs, enabling constant-time updates per event even in large heterogeneous populations, while faithfully reproducing experimentally observed patterns of clonal dominance during affinity maturation. Second, in an RNA transcription system with delayed negative feedback, MOSAIC captures state-dependent, non-exponential waiting times that classical delay-based algorithms cannot dynamically adjust. In this system, transcription initiation is followed by broadly distributed elongation times shaped by polymerase pausing, and protein negative feedback that inhibits further initiation. By continuously updating reaction rates with the evolving system state, MOSAIC reproduces realistic oscillatory dynamics in both mRNA and protein abundances, consistent with experimental data, and does so without the restrictions of fixed-delay methods, which lock delays at initiation and prevent adaptation to changing cellular conditions. Third, in temporal social networks, MOSAIC-TN captures node-driven activity patterns in which individuals alternate between long inactive periods and short bursts of interactions. By directly enforcing arbitrary node-level inter-event distributions and incorporating memory in partner selection, it reproduces heavy-tailed inter-event times, realistic edge lifetimes, and higher-order network features. In doing so, it outperforms spanning-tree and activity-driven models, which either underestimate interaction durations or rely on exponential assumptions. Together, these use cases show that MOSAIC unifies heterogeneity and memory in a single scalable framework, recovering empirical features that existing methods approximate poorly or fail to capture altogether.

MOSAIC is most efficient for large systems with moderate heterogeneity and broad inter-event distributions. Its advantage diminishes when rejection rates are high or populations extremely small (Supplementary Section~\ref{proof_complexity}), where delay-based approaches may remain preferable~\cite{anderson2007modified}. A further limitation arises when inter-event time distributions become highly peaked or nearly discrete (e.g., concentrated on multiples of a fixed time step): in this regime, representing the dynamics via a continuous-time hazard requires very large upper bounds in the thinning procedure, which leads to heavy rejection and poor efficiency. For strictly time-discrete models with atomic event times, dedicated discrete-time or DelaySSA-style algorithms remain better suited, and MOSAIC should be viewed only as a continuous-time approximation rather than the method of choice.

More generally, all stochastic simulation frameworks face challenges when applied to systems with very large numbers of reactants and reaction channels. In such cases, analytical simplifications or surrogate approaches, such as training neural networks on limited simulations to generate large-scale ensembles~\cite{jiang2021neural}, can reduce computational cost. Another fundamental difficulty arises when reaction propensities span several orders of magnitude: the dominance of high-frequency events can make rare events prohibitively expensive to observe, as noted in RNA transcription studies~\cite{herbach2017inferring}. Piecewise-deterministic Markov process (PDMP) models provide an effective alternative by simulating rare events stochastically while handling frequent events deterministically, leading to major speedups in diverse applications~\cite{herbach2017inferring, boxma2005off, pakdaman2010fluid, koshkin2024stochastic}. Hybrid extensions of MOSAIC that combine these ideas represent a natural next step.

In our temporal-network examples, we focused on a fully connected topology and networks of up to $10^3$ agents, primarily to keep the exposition and comparison with SSA-based baselines as simple and transparent as possible. This choice is made for illustration rather than realism, and many applications will naturally involve constrained, sparse, but large contact networks~\cite{koher2019contact, valdano2015analytical}. Within MOSAIC-TN, such constrained topologies can be incorporated straightforwardly: candidate interactions can be restricted to edges of an arbitrary static network, and both the pair-selection mechanism and the pairwise kernel $\Lambda_{ij}(t_i,t_j)$ can encode dynamic topological structure. In that setting, the number of processes would scale with the number of admissible edges rather than the square of the population size, opening the door to larger systems on sparse graphs. Nonetheless, our current implementation is not an HPC framework and is not engineered for national-scale, distributed-memory simulations. Large-scale tools such as EpiFast~\cite{bisset2009epifast} and EpiHiper~\cite{chen2025epihiper} target a different regime, namely discrete-time dynamics on massive sparse networks. Systematically exploring sparse and application-specific topologies, and developing optimized implementations for that regime, is therefore an important direction for future work~\cite{cure2025fast}.

In modeling interactions, we focused here on the case where both nodes must be simultaneously available, which is natural for undirected contact networks such as face-to-face encounters or synaptic transmission. This multiplicative formulation coordinates node and edge activity and is well suited for systems where neither participant can sustain an interaction independently. Importantly, MOSAIC is not restricted to this choice. In settings where interactions are driven primarily by one participant, such as epidemic spreading, email exchanges, or message transmissions, an additive formulation of the pairwise rate, $\Lambda_{ij} \propto \lambda_i(t_i) + \lambda_j(t_j)$, can be used instead (Supplementary Section~\ref{MOSAIC_paired_rate}). The framework therefore accommodates both synchronous and sender-driven dynamics, underscoring its flexibility across different classes of networks.

Looking ahead, MOSAIC provides a foundation for extending stochastic simulation to increasingly complex and large real-world systems. For instance, hybrid deterministic–stochastic schemes can further improve efficiency in systems with widely separated timescales. Coupling MOSAIC with differential-equation models, i.e. ordinary differential equations for population averages or partial differential equations for spatially structured processes, can enable true multiscale integration, linking individual variability to emergent population-level outcomes. On the application side, MOSAIC provides a versatile framework for modeling patient-specific variability in personalized medicine, capturing behavioral heterogeneity in epidemic control, and simulating adaptive dynamics in synthetic biology, neuroscience, and financial markets.

In conclusion, MOSAIC demonstrates that mechanistic realism and computational efficiency need not be traded off. By unifying memory and heterogeneity in a scalable stochastic framework, it establishes a versatile foundation for modeling complex systems and opens powerful new avenues for understanding how diversity and memory drive collective dynamics.

\section{Methods}

\label{Methods}

\subsection{MOSAIC implementation}
\label{Methods_MOSAIC}

\noindent The MOSAIC implementation was optimized around two key aspects to reduce computational cost. First, instead of updating the elapsed time $t_j$ since the last reaction for every process at each iteration, we only store the last reaction timestamp $\tau_{j}$ for process $j$ and compute $t_j = T - \tau_{j}$ when required. Second, the calculation of $\lambda_\text{max}$ (the upper bound on instantaneous reaction rates) is avoided at every step by exploiting the structure of the inter-event time distributions. Four scenarios can be distinguished:

\begin{itemize}
    \item \textbf{Bounded rates.} For distributions with a bounded instantaneous rate (e.g., Pareto, log-normal, Cauchy, delayed exponential, Gamma with $\alpha > 1$), $\lambda_\text{max}$ can be set as the fixed upper bound.
    
    \item \textbf{Unbounded, increasing rates.} For distributions with an unbounded rate that increases monotonically with time (e.g., normal, Weibull with $\alpha > 1$), $\lambda_\text{max}$ is determined by the reactant that has not reacted for the longest time, $\lambda_\text{max} = \lambda(t_\text{max})$. The bound only needs updating when this reactant undergoes a reaction.
    
    \item \textbf{Unbounded, decreasing rates.} For distributions with an unbounded rate that decreases monotonically with time (e.g., Weibull with $\alpha < 1$, Gamma with $\alpha < 1$, or power-law forms), $\lambda_\text{max}$ is determined by the reactant with the shortest waiting time. Since this can yield excessively large $\lambda_\text{max}$, MOSAIC is not recommended for such cases, where the Laplace Gillespie algorithm~\cite{masuda2018gillespie} is more efficient.
    
    \item \textbf{Arbitrary rates.} For distributions with arbitrary shapes (e.g., numerical fits from experimental data), the user should precompute bounds on $\lambda(t)$ within intervals. This reduces the problem to one of the three categories above for each interval.
\end{itemize}

\subsection{B-cell affinity maturation model}

\label{GC-system}

\noindent During an immune response, B cells and T cells migrate within germinal centers (GCs)~\cite{mesin2016germinal}, where they encounter one another through random motion and transient contacts. B cells that successfully bind antigen present fragments to T helper cells, which in turn provide survival and proliferation signals. Because the number of available T cells is limited, B cells compete for this help, and clones with higher-affinity receptors are more likely to be rescued from apoptosis, to proliferate, and to accumulate affinity-enhancing mutations. This competitive selection process is the basis of affinity maturation, and we capture it here with a minimal stochastic model consisting of five reaction processes:  
\begin{itemize}
    \item \textbf{B cell binding to T~cells}: B + T $\xrightarrow{\lambda_\text{BT}}$ $[$BT$]$. A B~cell encounters and initiates an interaction with a T~cell. Here, we assume that T~cells are all identical.
   
    \item \textbf{B-cell competition for T-cell help}: $[$B$_1$T$]$ + B$_2$ $\xrightarrow{\lambda_\text{BT} \ (IP)}$ $[$B$_2$T$]$ + B$_1$. Even if a cell B$_1$  is already interacting with a T cell and receiving survival signals  (B$_1$T), it can be displaced by a new cell B$_2$ with a higher-affinity receptor, i.e.  $\text{affinity}(B_2) > \text{affinity}(B_1)$. This competition is central to driving affinity maturation in B~cells.  (IP) in the reaction equation denotes that the process incorporates the individual properties (IP) of each B cell, accounting for variations in receptor affinity, and resulting in different propensities, or rates, for each existing (B, T) pair.
    
    \item \textbf{B-cell apoptosis}: B $\xrightarrow{\lambda_\text{apop}} \varnothing$. B~cells undergo apoptosis if not rescued by T~cells. This mechanism eliminates non-competitive B~cells from the system.
    
    \item \textbf{B-cell spontaneous unbinding from T~cells}: $[$BT$]$ $\xrightarrow{\lambda_\text{unbind}}$ B$_\text{div}$ + T. After receiving sufficient survival signals, B~cells detach from their T~cell and prepare for further divisions.
    
    \item \textbf{B-cell division}: B$_\text{div}$ $\xrightarrow{\lambda_\text{div}}$ B + B. Selected B cells undergo division, producing two daughter cells. Let $B_p$ and $B_d$ represent a parent and a daughter B~cell, respectively. After each division event, the affinity change of the daughter cell  $B_d$  is computed as follows:
    \begin{equation}
        \label{affinity_change}
        \Delta_\text{aff} = 
        \text{affinity}(B_d) - \text{affinity}(B_p) =
        \frac{u - \text{affinity}(B_p)}{\beta},
    \end{equation}
    where $u$ is a uniformly distributed random variable in the range $[0, 1]$, and $\beta$ is a scale parameter that determines the magnitude of the affinity change. This formulation makes it more challenging for cells with higher affinity values to further improve their affinity (Supplementary Figure~\ref{fig:GC_affinity}A) 

    In our simulations, B-cell affinities start at 0 and evolve dynamically through successive divisions. After 50 days, affinity values reach approximately $0.7 \pm 0.05$ (Supplementary Figure~\ref{fig:GC_affinity}B), reflecting the gradual accumulation of affinity-enhancing mutations over time.
    
\end{itemize}
We initialized the system with $N_B = 1000$ B cells and $N_T = 10$ T cells. Simulations were run for 50 days using the following parameters (from~\cite{pelissier2020computational}): $\beta = 10$, $\lambda_\text{BT} = \SI{0.146}{h^{-1}}$, $\lambda_\text{unbind} = \SI{2}{h^{-1}}$, $\lambda_\text{apop} = \SI{0.084}{h^{-1}}$, and $\lambda_\text{div} = \SI{0.134}{h^{-1}}$. For the reaction involving B–T competition, we set $\lambda_\text{max} = \lambda_\text{BT}$ and kept this choice fixed throughout the simulation.

\subsection{Hes1 transcription model with delayed negative feedback} 

\label{hes1-system}

\noindent We modeled RNA transcription and protein synthesis in the \textit{hes1} system, which is known to exhibit oscillatory behavior driven by delayed auto-inhibition~\cite{hirata2002oscillatory, monk2003oscillatory, barrio2006oscillatory}. The protein Hes1 represses the transcription of \textit{hes1} RNA, generating a negative feedback loop that has been widely studied as a prototype for regulatory circuits with memory~\cite{cao2020analytical, park2018chemical}.  
The system was represented by three species: nascent RNA ($N$), mature mRNA ($M$), and protein ($P$), with populations denoted $N_N$, $N_M$, and $N_P$. The dynamics of \textit{hes1} transcription and translation were described by five stochastic reactions, each assigned an IED that specifies the timing of events. Building upon the work of Monk~\cite{monk2003oscillatory}, where parameters were fit from experimental data, we specified the reactions as follows:

\begin{itemize}

    \item \textbf{RNA transcription initiation:} $\varnothing \xrightarrow{\beta \cdot G(N_P)} \text{N}$.  
    The initiation of \textit{hes1} transcription is modeled with an instantaneous rate that decreases with the amount of Hes1 protein $N_P$~\cite{barrio2006oscillatory}. Here, $G(N_P)$ is a Hill function with coefficient $h = 4.1$:  
    \begin{equation}
        G(N_P) = \frac{1}{1 + \left(\frac{N_P}{\beta}\right)^h} \, ,
    \end{equation}
    where $\beta$ is the protein concentration required for half-maximal repression.  

    \item \textbf{nRNA elongation:} $\text{N} \xrightarrow{ \tau \sim \mathcal{G}(\lambda_0, \ \alpha_\text{elong})} \text{M}$.  
    Elongation of an RNA molecule typically takes 15–20 minutes. We modeled this delay using a Gamma distribution $\mathcal{G}$ with mean $\tau_0 = \SI{20}{min}$, shape parameter $\alpha_\text{elong} = 3$, and rate $\lambda_0 = 1/\tau_0$, as parametrized in Supplementary Section~\ref{SI-Gamma-params}. The Gamma distribution captures the variability of elongation times, which are often heavy-tailed due to pausing events~\cite{qian2021basic, rajala2010effects}.  
    To incorporate resource limitation, we used a generalized form~\cite{boguna2014simulating}:  
    \begin{equation}
        \lambda_0 = \tau_0^{-1} \left(1 + \gamma \cdot \frac{N_N + N_M}{\beta}\right)^{-1}, 
    \end{equation}
    where $\gamma$ controls the effect of limited resources, and $N_N$ and $N_M$ are normalized by $\beta$. Setting $\gamma = 0$ recovers a constant elongation rate.  

    \item \textbf{Protein synthesis:} $\text{M} \xrightarrow{\lambda_\text{trans}} \text{M} + \text{P}$.  
    Mature mRNA molecules are translated into protein with rate $\lambda_\text{trans} = \SI{0.01}{min^{-1}}$. Translation is treated as effectively instantaneous, since completion typically occurs within one minute~\cite{barrio2006oscillatory}. Each mRNA can be reused multiple times before degradation.  

    \item \textbf{RNA degradation:} $\text{M} \xrightarrow{\lambda_\text{deg-M}} \varnothing$.  
    mRNA molecules degrade with rate $\lambda_\text{deg-M} = \SI{0.029}{min^{-1}}$. This process was modeled as a Markovian exponential decay, consistent with experimental observations~\cite{wang2002precision}.  

    \item \textbf{Protein degradation:} $\text{P} \xrightarrow{\lambda_\text{deg-P}} \varnothing$.  
    Proteins degrade with rate $\lambda_\text{deg-P} = \SI{0.031}{min^{-1}}$, modeled as a Markovian exponential decay process.  

\end{itemize}
For the nRNA elongation reaction, the waiting time $\tau$ follows a Gamma distribution $\mathcal{G}(\lambda_0,\alpha_\text{elong})$ which we model in MOSAIC with the instantaneous rate
\begin{equation}
    \lambda_\text{elong}(t)
    = \frac{\psi_\text{elong}(t)}{1 - F_\text{elong}(t)}
    = \frac{(\alpha_\text{elong} \lambda_0)^{\alpha_\text{elong}} t^{\alpha_\text{elong}-1} 
            e^{-\alpha_\text{elong} \lambda_0 t}}
           {\Gamma\!\big(\alpha_\text{elong}, \alpha_\text{elong} \lambda_0 t\big)} \leq \lambda_\text{max}^\text{(elong)} = \alpha_\text{elong} \lambda_0,
\end{equation}
where $\Gamma(\cdot,\cdot)$ denotes the upper incomplete gamma function, and $\lambda_\text{max}^\text{(elong)}$ the global upper bound for the elongation channel in MOSAIC. For small systems, where $N = N_N + N_M + N_P$ is modest, we also evaluated a more conservative choice to ensure a smaller effective time step $\Delta t$ (Supplementary Section~\ref{arbitrary_accuracy}), by enforcing
\begin{equation}
    N \lambda_\text{max}^\text{(elong)} \geq 20 \lambda_0, \ \ \ \text{which implies an average time step} \ \ \mathbb{E}[\Delta t] \leq \frac{1}{20 \lambda_0}.
\end{equation}

We assessed MOSAIC's accuracy under the specific condition of $\gamma = 0$, where the distribution of RNA elongation inter-event times remains unaffected by the RNA molecule population (energy-unconstrained elongation). We executed MOSAIC and DelaySSA across $N_\text{sim} = 10^4 / \sqrt{\beta}$ simulations and compared the root mean square deviation (RMSD) of normalized averaged populations for different values of the scale parameter $\beta$. As a baseline, we used the mean RMSD between two independent realizations of DelaySSA simulations. The number of simulations was scaled by $1/\sqrt{\beta}$ to compensate for the increased stochasticity in smaller systems.

\subsection{The MOSAIC-TN Algorithm}

\label{method:MOSAIC-TN}

\noindent MOSAIC-TN extends the rejection-based Gillespie framework to node-driven temporal networks, where each node follows its own IED and carries an internal clock $t_i$ encoding its elapsed time since the last event. Interactions are constrained by a prescribed topology. Let $N_A$ denote the number of nodes, and let $\mathcal{E}$ denote the set of admissible node pairs $(i,j)$ consistent with the desired interaction graph. We write $N = |\mathcal{E}|$ for the number of such pairs; in the fully connected case, $N = N_A (N_A - 1)/2$. The pairwise rates $\Lambda_{ij}(t_i, t_j)$ depend on the internal times of nodes $i$ and $j$. The procedure is as follows:

\begin{enumerate}

    \item Set $\Lambda_\text{max}$ as an upper bound on all possible node--node interaction rates over admissible pairs:
    \begin{equation}
        \Lambda_\text{max} \;\geq\; \max_{(i,j) \in \mathcal{E}} \Lambda_{ij}(t_i, t_j).
    \end{equation}

    \item Compute the time increment to the next candidate event using $\Lambda_\text{max}$. Draw a uniform random variable $u \in \mathcal{U}[0,1]$ and advance time by
    \begin{equation}
        \Delta t \;=\; \frac{\ln (1/u)}{N \,\Lambda_\text{max}},
        \label{deltat_TN}
    \end{equation}

    \item Select a candidate pair $(i,j)$ for the next event by sampling uniformly from the admissible set $\mathcal{E}$. Thus, each pair has probability
    \begin{equation}
        p_{ij} \;=\; \frac{1}{N}, \qquad (i,j) \in \mathcal{E}.
    \end{equation}
    (More general proposal distributions over $\mathcal{E}$ can also be used, provided $\Lambda_\text{max}$ remains a valid global bound, as discussed in Supplementary Section \ref{MOSAICbin})\\

    \item Accept the event with probability
    \begin{equation}
        \label{Eq:MOSAIC_compute_pairwise_rate}
        p_\text{accept} \;=\; \frac{\Lambda_{ij}(t_i, t_j)}{\Lambda_\text{max}},
    \end{equation}
    and update the states and clocks of nodes $i$ and $j$ accordingly. If the event is rejected, the step is treated as a null event: time advances by $\Delta t$, but the node states remain unchanged.
    
\end{enumerate}

\subsection{Practical implementation for Pareto-distributed activity}

\label{Pareto-method}

\noindent For social interactions we model node activity with a Pareto law $\mathcal{P}(\lambda_A,\alpha_A)$, consistent with empirical human contact data~\cite{genois2022combining, le2023modeling}. The IED of node activity is parametrized as
\begin{equation}
\psi(t) = \frac{\alpha_A , t_{\min}^{\alpha_A}}{t^{\alpha_A+1}}, 
\qquad t \geq t_{\min}, \qquad \text{with instantatenious rate} \ \lambda(t) = \frac{\alpha_A}{t}.
\end{equation}
We define the characteristic interaction rate $\lambda_A$ as the reciprocal of the median inter-event time,
\begin{equation}
t_{\text{med}} = t_{\min}  2^{1/\alpha_A},
\qquad
\lambda_A = \frac{1}{t_{\text{med}}} = \frac{2^{-1/\alpha_A}}{t_{\min}}.
\end{equation}
This definition remains valid for all $\alpha_A > 0$, including heavy-tailed regimes where the mean diverges.

For MOSAIC-TN, we consider a fully connected underlying network and adopt a multiplicative pairwise rate,  
\begin{equation}
    \Lambda_{ij}(t_i,t_j) = \frac{\lambda_i(t_i)\,\lambda_j(t_j)}{(N_A-1)\,\lambda_A},
\end{equation}
which guarantees that as $N_A \to \infty$, each node’s IED converges to its prescribed Pareto form (Supplementary Section~\ref{MOSAIC_paired_rate}).  
Finally, because the Pareto hazard rate $\lambda(t)$ decreases monotonically after $t_{\min}$, we can set a constant global upper bound for the entire simulation as  
\begin{equation}
    \lambda_\text{max} = \frac{\alpha_A}{t_{\min}}, 
    \qquad 
    \Lambda_\text{max} = \frac{\lambda_\text{max}^2}{(N_A - 1) \lambda_A}.
\end{equation}

To capture temporal neighborhood effects, we also implement MOSAIC-TN~B, where partner nodes are drawn non-uniformly: once $A_i$ is selected, previous partners $A_j$ are $w$ times more likely to be chosen. This modification reduces rejection overhead at large $w$ and remains valid for Pareto and Weibull IEDs (Supplementary Section~\ref{MOSAICbin}). By directly sampling more probable interactions, MOSAIC-TN~B avoids the sharp increase in rejected events that would otherwise occur (e.g., nearly tenfold for $w=9.4$ compared to the standard MOSAIC-TN).

\subsection{Modeling and evaluation of temporal social networks}  
\label{param_optimization}  

\noindent Face-to-face interactions were modeled using two kinetic channels parameterized by Pareto-distributed inter-event times. First, encounters between two agents $A_i$ and $A_j$ were represented as  
\begin{equation}
    A_i + A_j \xrightarrow{\mathcal{P}(\lambda_A, \alpha_A)} A_i + A_j + [A_i A_j],
\end{equation} 
where $[A_i A_j]$ denotes the formation of an interaction edge. Importantly, the agents themselves are not consumed by this process and remain available to initiate further interactions, so multiple simultaneous edges per node are possible. This process defines the \textit{node interduration}, corresponding to the time between successive interactions of a given agent. Parameters were estimated by fitting a Pareto distribution to the observed node interduration distribution in the IC2S2 conference data~\cite{genois2022combining}, yielding $\alpha_A = 0.76$ and $\lambda_A = 0.4$. 
Second, interaction termination was modeled as  
\begin{equation}
    [A_i A_j] \xrightarrow{\mathcal{P}(\lambda_\varnothing, \alpha_\varnothing)} \varnothing,
\end{equation}
corresponding to the removal of an edge. A Pareto fit to the empirical distribution of interaction durations gave $\alpha_\varnothing = 1.4$ and $\lambda_\varnothing = 0.61$.

Simulations were run with $N = 274$ nodes over $T = 7249$ time units, consistent with the IC2S2 dataset. To evaluate model fidelity, we compared simulated networks against the empirical data using five aggregated network metrics: clustering coefficient~\cite{watts1998collective}, modularity~\cite{newman2006modularity}, transitivity~\cite{newman2003structure}, network density~\cite{posfai2016network}, and degree assortativity~\cite{newman2002assortative}. After each simulation, pairwise interaction durations were summed, log-transformed, and normalized by their maximum to obtain an \textit{interaction score}. A graph was then constructed by connecting pairs of individuals with an interaction score greater than 0.4. Community detection was performed using the Louvain algorithm~\cite{blondel2008fast} with a resolution parameter of 1. Network metrics were computed in \texttt{networkx} (v2.8.8) using \texttt{nx.average\_clustering}, \texttt{nx.algorithms.community.quality.modularity}, \texttt{nx.transitivity}, \texttt{nx.density}, and \texttt{nx.degree\_pearson\_correlation\_coefficient}.

\subsection{AD modeling and Spanning Tree temporal network}
\noindent For AD modeling temporal networks, we utilized the V9 network from Le Bail \textit{et al.}~\cite{le2023modeling}, identified as the highest-performing model. Optimized simulation parameters for the conference dataset, along with the code for implementing AD modeling, were sourced from \url{https://github.com/DidierLeBail/Temporal-networks-PhD-code}. The conference dataset is labeled as \texttt{conf17} within this repository.

Spanning Tree temporal networks were generated by assigning node activity rates based on a Pareto distribution fitted to the conference data. The repository \url{https://github.com/anzhisheng/Temporal-networks-by-spanning-trees} was used for this purpose. Before generating the spanning trees, a Barabási-Albert graph~\cite{albert2002statistical} was constructed with $m = N/2$ neighbors per node, reflecting the observed conference data trend where nodes interact on average with half of the population at least once.

\subsection*{Data availability}
\noindent This study used previously published datasets on germinal center B-cell maturation~\cite{tas2016visualizing} and conference interactions~\cite{genois2022combining}. These data are available from the original publications and their associated repositories. All simulation data required to reproduce the findings are generated from the model definitions and parameters reported in the Methods.

\subsection*{Code availability}
\noindent The MOSAIC and DelaySSA Python implementations, together with the code and data needed to reproduce all figures, have been deposited on Zenodo at \url{https://doi.org/10.5281/zenodo.18346965}.

\printbibliography[title={References}]

\subsection*{Acknowledgement}
\noindent The authors thank Jonathan Karr, Farshid Jafarpour, Srividya Iyer-Biswas, and Peter Ashcroft for their valuable suggestions. 

\subsection*{Funding statement}
This research was supported by the COSMIC European Training Network, funded by the European Union’s Horizon 2020 research and innovation program under grant agreement No 765158 and from the Swiss National Science Foundation (Sinergia grant CRSII5 193832).  

\subsection*{Competing Interests Statement}
\noindent The authors declare no competing interest

\subsection*{Author contributions}
\noindent A.P. designed and implemented the MOSAIC and MOSAIC-TN frameworks, performed all simulations and benchmarks, and wrote the manuscript. N.B. and M.R.M. supervised the project and contributed to writing and editing the manuscript; M.R.M. additionally contributed to the design of MOSAIC. M.P. contributed to the implementation of MOSAIC and assisted with simulations.D.B. advised on dataset selection and the activity-driven modeling framework for social interactions, and clarified terminology.

\newpage

\onecolumn 

\appendix
\setcounter{figure}{0}
\renewcommand{\thefigure}{S\arabic{figure}}

\begin{refsection} 
\setcounter{page}{1}
\setcounter{figure}{0}
\setcounter{table}{0}
\renewcommand{\thefigure}{S\arabic{figure}}
\renewcommand{\thetable}{S\arabic{table}}
\renewcommand{\thepage}{S\arabic{page}}

{\centering \Huge \textbf{Supplementary Materials}}
\vspace{0.5cm}

\noindent {\huge Unifying Non-Markovian Dynamics and Agent Heterogeneity in Scalable Stochastic Networks}

\vspace{0.5cm}

\noindent
\textbf{Aurelien Pelissier}$^{1,2,3}$, 
\textbf{Miroslav Phan}$^{1,2}$, 
\textbf{Didier Le Bail}$^{4}$, 
\textbf{Niko Beerenwinkel}$^{2}$, 
\textbf{Mar\'ia
Rodr\'iguez Mart\'inez}$^{1,5}$

\vspace{0.5cm}

\noindent
\footnotesize
\setlength{\baselineskip}{0.95\baselineskip}
\hspace{-0.1cm}$^{1}$IBM Research Europe, 8803 Ruschlikon, Switzerland\\
$^{2}$Department of Biosystems Science and Engineering, ETH Zurich, 4058 Basel, Switzerland\\
$^{3}$Institute of Computational Life Sciences, Zurich University of Applied Sciences (ZHAW), 8820, Wadenswil, Switzerland\\
$^{4}$Centre de Physique Th\'eorique (CPT), Aix-Marseille University, CNRS, 13009 Marseille, France\\
$^{5}$Department of Biomedical Informatics \& Data Science, Yale School of Medicine, New Haven, CT, United States\\
$^\dagger$Corresponding author \href{maria.rodriguezmartinez@yale.edu}{maria.rodriguezmartinez@yale.edu}.

\normalsize

\section{Probability density functions in the MOSAIC framework}

\label{MOSAIC_distributions}

\subsection{Parametrization of inter-event time distributions}

\noindent For consistency with other works in the literature of stochastic simulations, we parameterize processes using their mean $\frac{1}{\lambda_0}$, where $\lambda_0$ would be the instantaneous rate if the distribution was exponential. Additionally, some distributions require a second parameter to describe their \textit{shape}. For the normal and log-normal distribution, we use $\gamma$, a scale-free standard deviation defined as the standard deviation over the mean. For the gamma and Weibull distributions we use $\alpha$, the shape parameter in their standard parametrization. 
Thus, all IEDs in this article are parametrized with two variables, ($\lambda_0, \alpha$) or ($\lambda_0, \gamma$).

\paragraph{Normal distribution.} The normal distribution is typically parametrized by the mean $\mu$  and  the standard deviation $\sigma$, as follows:
\begin{equation}
    normal(t; \mu,\sigma)  = \frac{1}{\sigma \sqrt{2 \pi}} \exp \left({-\frac{1}{2}\left(\frac{t-\mu}{\sigma}\right)^{2}}\right) . 
\end{equation}

\noindent The parameters $\mu$ and $\sigma$ can be chosen such as the mean inter-event time distribution $1/\lambda_0$ with:
$$\mu = \frac{1}{\lambda_0} \ \text{and} \ \sigma = \frac{\gamma}{\lambda_0}$$
Where $\gamma$ corresponds to the ratio between the standard deviation and the mean, representing a scale-free standard deviation. The alternative parametrization of the distribution for MOSAIC is then given by

\begin{equation}
\boxed{
    normal(t; \lambda_0,\gamma)  = \frac{\lambda_0}{\gamma \sqrt{2 \pi}} \exp \left({-\frac{1}{2}\left(\frac{\lambda_0 t-1}{\gamma}\right)^{2}}\right). 
}
\end{equation}

\paragraph{Log-normal distribution.} The log-normal distribution is typically parametrized by $\mu$ and $\sigma$, as follows:
\begin{equation}
    \textit{log-normal}(t; \mu,\sigma) = \frac{1}{t \sigma \sqrt{2 \pi}} \exp \left(-\frac{\left(\log (t)-\mu\right)^{2}}{2 \sigma^{2}}\right),
\end{equation}
The mean and variance are given by:
\begin{align*}
mean =  & \exp \left(\mu+\frac{\sigma^{2}}{2}\right) ,  \\
variance = & \left[\exp \left(\sigma^{2}\right)-1\right] \exp \left(2 \mu+\sigma^{2}\right) .
\end{align*}
\noindent If we want the mean inter-event time distribution to be $1/\lambda_0$, we chose $\mu$ and $\sigma$ as follows:
\begin{equation*}
\mu = \log \left(\frac{1}{ \lambda_0 \sqrt{1 + \gamma^2}}\right) \ \text{and} \ \sigma = \sqrt{\log \left(1+ \gamma^2\right)} \, . 
\end{equation*}
Where, as with the normal distribution, $\gamma$ corresponds to the ratio between the standard deviation and the mean, and thus, it is scale invariant. The alternative parametrization of the distribution for MOSAIC is then given by
\begin{equation}
\boxed{
    \textit{log-normal}(t; \lambda_0,\gamma) = \frac{1}{t \sqrt{2 \pi \log (1+ \gamma^2)}} \exp \left(-\frac{\left(\log (t)+\log \left( \lambda_0 \sqrt{1 + \gamma^2}\right)\right)^{2}}{2 \log (1+ \gamma^2)}\right),
}
\end{equation}
\label{SI-Gamma-params}
\paragraph{Gamma distribution.} The gamma distribution admits 2 constants, $\alpha$ and $\beta$:
\begin{equation}
    \label{Eq:gamma-param}
    gamma(t; \alpha,\beta) = \frac{\beta^{\alpha}}{\Gamma(\alpha)} t^{\alpha-1} e^{-\beta t} . 
\end{equation}
Mean and variance can be computed to be:
\begin{align*}
mean  = & \frac{\alpha}{\beta} \, , \\
variance  = & \frac{\alpha}{\beta^2} \, .
\end{align*}
The parameter $\beta$ can be chosen such as the mean inter-event time distribution is $1/\lambda_0$:
$$\beta= \alpha \lambda_0 \, . $$ 
Thus, we parametrize the gamma distribution as
\begin{equation}
\boxed{
    gamma(t; \lambda_0,\alpha) = \frac{(\alpha \lambda_0)^{\alpha}}{\Gamma(\alpha)} t^{\alpha-1} e^{-\alpha \lambda_0 t} . 
}
\end{equation}
We note that in this case the ratio between the standard deviation and the mean is related the the shape parameter $\alpha$ with:
\begin{equation*}
    \gamma = \frac{1}{\sqrt{\alpha}} \, , 
\end{equation*}

\paragraph{Weibull distribution.}

\label{weibull_appendix}
The Weibull distribution~\cite{jiang2011study} can be parametrized  with constants $\lambda$ and $\alpha$ as follows:
\begin{equation}
    Weibull(t; \lambda,\alpha) =  \frac{\alpha}{\lambda} \left( \frac{t}{\lambda} \right)^{\alpha-1} e^{-(t/\lambda)^\alpha} . 
\end{equation}
The mean and variance are:
\begin{align*}
mean = & \lambda \ \Gamma \left(1 + \frac{1}{\alpha} \right) \, , \\
variance = & \lambda^{2}\left[\Gamma\left(1+\frac{2}{\alpha}\right)-\Gamma^{2}\left(1+\frac{1}{\alpha}\right)\right] \, .
\end{align*}
If $t$ represents a "time-to-failure", the Weibull distribution gives a distribution for which the failure rate is proportional to a power of time. 
An alternative parametrisation often found in text books is  $\lambda = \left( \frac{\alpha}{\beta} \right) ^{\frac{1}{\alpha}}$, under which the PDF of the Weibull distribution becomes:

\begin{equation}
\label{Weibull_param}
    Weibull \left( t; \beta, \alpha \right) =  \beta t^{\alpha-1} \times \exp \left(- \frac{\beta t^{\alpha}}{\alpha}\right) . 
\end{equation}
To make the mean inter event time distribution equal to $1/\lambda_0$, $\beta$ has to be chosen as follows:
\begin{equation*}
\beta=\alpha\left[\lambda_{0} \Gamma\left(\frac{\alpha+1}{\alpha}\right)\right]^{\alpha}.
\end{equation*}
Thus, we parametrize the gamma distribution as
\begin{equation}
\boxed{
    Weibull \left( t; \lambda_0, \alpha \right) =  \alpha\left[\lambda_{0} \Gamma\left(\frac{\alpha+1}{\alpha}\right)\right]^{\alpha} t^{\alpha-1} \times \exp \left(- \frac{\alpha\left[\lambda_{0} \Gamma\left(\frac{\alpha+1}{\alpha}\right)\right]^{\alpha} t^{\alpha}}{\alpha}\right) . 
}
\end{equation}
We note that with this choice, the ratio between the standard deviation and the mean is related to $\alpha$ with:
\begin{equation*}
\gamma = \frac{\Gamma\left(\frac{\alpha+2}{\alpha}\right)}{\Gamma^{2}\left(\frac{\alpha+1}{\alpha}\right)}-1 \, .
\end{equation*}

\paragraph{Cauchy distribution.} The Cauchy distribution admits two parameters,  $\mu$ and $\sigma$:
\begin{equation}
    Cauchy(t; \mu,\sigma) = \frac{1}{\pi \sigma\left[1+\left(\frac{t-\mu}{\sigma}\right)^{2}\right]} \, . 
\end{equation}
The Cauchy distribution represents the distribution of the ratio of two independent and normally distributed random variables with mean zero. The mean and variance are 	undefined, as the integrals necessary to compute these values do not exist\footnote{ \small The integral associated with the mean, $\int_{-\infty}^\infty x f(x)\,dx $, 
does not exist. This can be proven, for instance, by noticing  that  	$\lim _{a\to \infty }\int _{-a}^{a} x f(x)\,dx $ and $\lim _{a\to \infty }\int _{-2a}^{a} x f(x)\,dx $ converge to different values. Similar arguments can be used to show that the variance does not exist either.}.
Intuitively, this happens because extremely large number can be drawn with non-zero probability. Nevertheless, we can choose the parameter $\mu$ as the inverse of the median inter event time $1/\lambda_0$:
$$\mu = \frac{1}{\lambda_0}\, . $$
Similarly, we can define $\gamma$ as the \textit{analogue} of ratio between the standard deviation and the mean, which is scale invariant:
$$\sigma = \frac{\gamma}{\lambda_0} \, .$$
Thus, we parametrize the Cauchy distribution as
\begin{equation}
\boxed{
    Cauchy(t; \lambda_0,\gamma) = \frac{1}{\pi \frac{\gamma}{\lambda_0}\left[1+\left(\frac{\lambda_0 t-1}{\gamma}\right)^{2}\right]} \, .
}
\end{equation}

\paragraph{Pareto distribution.}

The Pareto distribution admits two parameters,  $\mu$ and $\sigma$:
\begin{equation}
Pareto(t, \mu, \alpha ) = \begin{cases} \frac{\alpha \mu^\alpha}{t^{\alpha+1}} & t \geq \mu \\ 0 & t < \mu \end{cases}
\end{equation}
As with the Cauchy distribution, the Pareto distribution does not always have a finite mean, so we choose the parameter $\mu$ such as its median equals $1/\lambda_0$:
\begin{equation}
\label{Eq.Pareto_mu}
    \mu = \frac{1}{\lambda_0} 2^{-1/\alpha}\, . 
\end{equation}
Thus, we parametrize the Pareto distribution as
\begin{equation}
\boxed{
    Pareto(t, \lambda_0, \alpha ) = \begin{cases} \frac{\alpha}{t^{\alpha+1}} \cdot \frac{1}{2 \lambda_0^\alpha} & t \geq \frac{1}{\lambda_0} 2^{-1/\alpha} \\ 0 & t < \frac{1}{\lambda_0} 2^{-1/\alpha} \end{cases}
}
\label{pareto-equation}
\end{equation}
 
\subsection{Relationship between probability density functions and instantaneous rates}
\label{rate_proof}

\noindent We consider the survival distribution function (SDF) $\Psi(t) $ of a renewal process:
\begin{equation}
\Psi(t) =\int_{t}^{\infty} \psi\left(\tau\right) \mathrm{d} \tau 
\end{equation}
and its relationship with the probability distribution function (PDF) $\psi(t)= - \frac{d \Psi(t)}{d t} $. Using the definition of the instantaneous rate function:
\begin{equation}
    \lambda(t) = \frac{\psi(t)}{\Psi(t)} \, ,
\end{equation}
we can describe the time evolution of $\Psi(t)$ as a first order homogeneous differential equation:
\begin{equation}
    \lambda(t) \, \Psi(t) + \Psi'(t)   = 0  \, . 
\end{equation}
The general solution is easily written as~\cite{boyce2017elementary}:
\begin{equation}
    \label{Psi}
    \Psi\left(t\right)= K \exp \left({-\int^{t}_{0} \lambda(\tau) d\tau}\right) ,  
\end{equation}
where $K \in \mathbb{R}$ is an  integration constant. However, since by definition $\Psi(0) = 1$, we conclude that $K = 1$.
The PDF of that process can now by computed as follows: 
\begin{equation}
    \label{psi}
    \psi(t)
    = - \frac{d \Psi}{d t} (t)
    = \lambda(t) \, \exp \left({-\int^{t}_{0} \lambda(\tau) d\tau}\right)
\end{equation}


Note that $\Psi(t)$ verifies the additional normalization condition $\Psi(\infty) = 0$. This implies:
\begin{equation}
\begin{aligned}
    \Psi(\infty)  =  & \ \exp \left({-\int^{\infty}_{0} \lambda(\tau) d\tau}\right) = 0\\
    \Rightarrow & \ \int^{\infty}_{0} \lambda(\tau) d\tau = \infty\\  \, .
\end{aligned}
\end{equation}
This means that $\lambda(t)$ has to be chosen such as its definite integral from 0 to $\infty$ is infinite, otherwise $\psi(t)$ and $\Psi(t)$ do not represent a renewal process.

\subsection*{Examples} 
\noindent In general, the instantaneous rate for any distribution can be computed as $\lambda(t) = \frac{PDF(t)}{SDF(t)}$, where the survival distribution function (SDF) is related to the cumulative distribution function (CDF) according to $SDF = 1 - CDF$. 

We provide here a few examples of instantaneous rate functions and their associated PDFs:
\begin{itemize}

    \item $\lambda(t) = a_0$ leads to $PDF =a_0 \times \exp \left( - a_0 t \right)$ and $SDF = \exp \left( - a_0 t \right)$, which represent an exponential distribution.

    \item $\lambda(t) = \beta t^{\alpha-1}$ leads to $PDF =\beta t^{\alpha-1} \times \exp \left( - \frac{ \beta t^{\alpha}}{\alpha} \right)$ and $SDF =  \exp \left( - \frac{ \beta t^{\alpha}}{\alpha} \right)$, associated with the Weibull distribution~\cite{jiang2011study}.
    
    \item $\lambda(t) = \dfrac{c^2 t}{1+ct}$ leads to $ PDF = c^2 t \times \exp \Bigl(-ct\Bigr)$ and $SDF = (1+ct) \times \exp \Bigl(-ct\Bigr)$.
    
    \item Many important distributions do not have a simple analytic form for the instantaneous rate. For instance, the normal distribution, $PDF = \frac{1}{\sigma \sqrt{2 \pi}} e^{-\frac{1}{2}\left(\frac{t-\mu}{\sigma}\right)^{2}}$ and $SDF = \frac{1}{2}\left[1-\operatorname{erf}\left(\frac{t-\mu}{\sigma \sqrt{2}}\right)\right]$, with \textit{erf} being the error function \cite{gautschi1972error}, results in an instantaneous rate that cannot be expressed in terms of basic functions. An approximation is however possible at large times, where the instantaneous rate asymptotically approximates a linear function $\lambda(t) \approx \frac{t-\mu}{\sigma^{2}}$.

\end{itemize}

\begin{figure*}[h!t]
    \centering
    \includegraphics[width=0.85\linewidth]{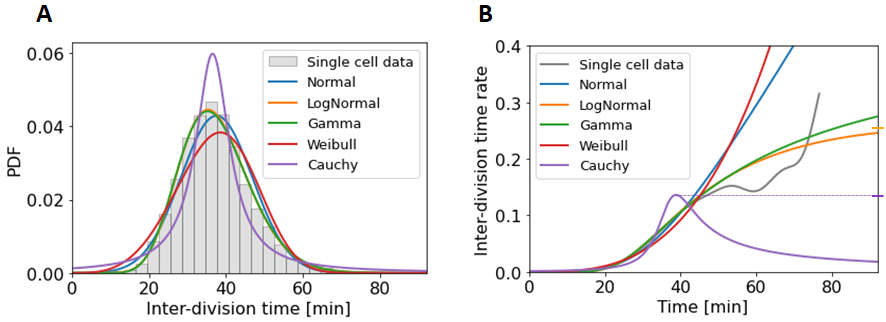}
    \caption{\small (A) PDF of several  distributions typically used to represent biochemical waiting times, where the parameters have been chosen to fit the measured inter-division time of \textit{Bacillus subtilis}~\cite{sauls2019control} at constant temperature. (B) Although the PDFs are relatively similar, the instantaneous rates show markedly different behaviors as a function of time. The single-cell instantaneous rate was estimated with a Gaussian kernel density estimator. When finite, the maximum instantaneous rates are displayed on the right of the figure.}
    \label{fig:distributions}
\end{figure*}

In Figure~\ref{fig:distributions}B, we show that the normal, Weibull ($\alpha \geq 1$) and gamma ($\alpha \geq 1$) distributions have monotonically increasing rates, while the Cauchy and log-normal distributions exhibit a maximum.

\newpage

\section{Modeling of Stochastic Agents with Individual Complexity}

\label{MOSAIC_proof}

\noindent We consider $N$ Poisson processes running in parallel, each with their respective reaction rate $\lambda_j$ $(1 \leq j \leq N)$, and denote by $a_0 = \sum \lambda_j$ the sum of the individual rates. The standard Gillespie (SG) algorithm is a popular stochastic simulation framework that can generate statistically correct trajectories of a stochastic equation system. The algorithm assumes that the reaction rates are known and constant, under which assumption, a trajectory can be obtained using the following iterative rules (see~\cite{masuda2018gillespie} for the full derivation):

\renewcommand{\labelenumi}{\theenumi}
\renewcommand{\theenumi}{(\roman{enumi})}

\begin{enumerate}
    \item Draw $u \in \mathcal{U}^{[0,1]}$ a random variate from the uniform density on the interval $[0, 1]$, and compute the time increment to the next event with
    \begin{equation}
        \Delta t = \frac{\ln (1/u)}{a_0},
        \label{eq:deltat2}
    \end{equation}
    
    \item Draw the process $j$ that has produced the event with probability
    \begin{equation}
        P_j = \frac{\lambda_j}{a_0},
        \label{eq:deltat3}
    \end{equation}
\end{enumerate}

In the main text, we introduced MOSAIC with a Rejection Gillespie approach, that is statistically exact in the limit of $\Delta t \rightarrow 0$. We provide the algorithm here again for reference. 

\paragraph{MOSAIC Algorithmic procedure.}  
Let $t_{j}$ denote the time elapsed since the last event of the $j$th process $(1 \leq j \leq N)$, and $\lambda_{j}(t_j)$ its time-dependent rate. Each iteration of MOSAIC proceeds in four steps:

\begin{enumerate}
    \item \textit{Set the global maximum rate.}  
    Choose $\lambda_\text{max}$ such that
    \begin{equation}
        \lambda_\text{max} \geq \max_{1 \leq j \leq N} \lambda_j(t_j).
    \end{equation}

    \item \textit{Advance the time.}  
    Draw $u \sim \mathcal{U}[0,1]$ and set
    \begin{equation}
        \Delta t = \frac{\ln (1/u)}{N \cdot \lambda_\text{max}}.
    \end{equation}

    \item \textit{Select a candidate process.}  
    Choose $j$ uniformly at random:
    \begin{equation}
        p_j = \frac{1}{N}.  
    \end{equation}

    \item \textit{Accept or reject the event.}  
    Accept with probability
    \begin{equation}
        p_\text{accept} = \frac{\lambda_j(t_j)}{\lambda_\text{max}}, 
    \end{equation}
    in which case the reactants are updated. If rejected, the event is empty (no population change), but time still advances by $\Delta t$.
\end{enumerate}

Below, we provide the proof, which we decompose into two independent subparts. First, we show that the introduction of rejected steps in the Gillespie model is mathematically equivalent to the standard Gillespie (SG) algorithm. Second, we show that locally considering each process as Poissonian during the time step $\Delta t$ at each Gillespie iteration yields statistically exact results for non-Markovian simulations when $\Delta t \rightarrow 0$. We note that both of these aspects were already proven separately in reference \cite{thanh2014efficient} and \cite{boguna2014simulating}, respectively. Here we put them together for the reader's convenience. As the rejection framework allows for the arbitrary reduction of $\Delta t$ by increasing $\lambda_\text{max}$, MOSAIC is exact in the limit  $\lambda_\text{max} \rightarrow \infty$ or $N \rightarrow \infty$.

\subsection{Proof for the Rejection-based Gillespie Algorithm}
\label{rejection_proof}
\noindent The rejection-based Gillespie algorithm  follows the same distribution as the Standard Gillespie (SG) if and only if (iff): 

\begin{itemize}[label={}]
    \item (i) The reaction $R_j$ occurs with probability $P = {\lambda_j}/{a_0}$.
    \item (ii) The $\Delta t$ time increment follows the same exponential distribution as in SG, i.e. $f_{\Delta t}(x) = a_0 \cdot \exp({-a_0 x})$.
\end{itemize}

\subsection*{(i) Reaction $R_j$ accepted with the same probability as in SG} 
\noindent We define $p_\text{accept}(R_j)$ as the joint probability of $R_j$ being first selected and then accepted, $\lambda_{\text{max}}$ as the upper propensity bound for all reactions, and $a_{0, \text{max}} = N \lambda_{\text{max}}$. We can write:
$$
p_\text{accept}(R_j) = \frac{\lambda_{\text{max}}}{a_{0, \text{max}}} \times \frac{\lambda_j}{\lambda_\text{max}} = \frac{\lambda_j}{a_{0, \text{max}}}
$$
We then denote by $p_\text{accept}(R)$ the probability of any reaction being accepted:
$$p_\text{accept}(R) = \frac{a_0}{a_{0, \text{max}}}.
$$
Finally, we write the conditional probability of reaction $R_j$ being accepted given that some reaction had been accepted as:
$$
p_{\text{accept}}(R_j \mid R) = \frac{p_{\text{accept}}(R_j)}{p_{\text{accept}}(R)} = \left(\frac{\lambda_j}{a_{0, \text{max}}}\right) / \left(\frac{a_0}{a_{0, \text{max}}}\right) = \frac{\lambda_j}{a_0} \, , 
$$ 
\noindent in agreement with SG framework. 

\subsection*{(ii) The time increment $\Delta t$ follows an exponential PDF}
\noindent We denote by $k$ the number of trials until the reaction is accepted (thus, there are $k-1$ rejections until success), with time being advanced by an increment of $\Delta t = - \ln{(u)} / a_{0,\text{max}}$ after each attempt. It follows that after $k$ attempts, the time increment is:
\begin{equation}
  \label{eqSI:t_erlang}
  \Delta t = - \frac{1}{a_{0, \text{max}}}\ln \left(\prod_{i=1}^{k}{u_i}\right)
\end{equation}
which corresponds to an Erlang distribution with parameters $k$ and $a_{0, \text{max}}$. This distribution represents the time elapsed until the $k$th event of a Poisson process with rate  $a_{0, \text{max}}$. In addition, $k$ is geometrically distributed with probability $p_\text{accept}(R)$, i.e. $$P(X=k) = (1 - p_\text{accept}(R))^{k-1} \times p_\text{accept}(R).$$

\noindent As the PDF for $\Delta t$ can be expressed as the derivative of its CDF, we can write
\begin{align*}
   f_{\Delta t}(x) &= \frac{d}{dx}F_{\Delta t}(x)\\
                   &= \frac{d}{dx}P(\Delta t \leq x),\\
   \intertext{ where $P(\Delta t \leq x)$ can be partitioned for values of k:}
                   &= \frac{d}{dx}\sum_{k=1}^{\infty}P(\Delta t \leq x \mid X=k)P(X=k)\\
                   &= \frac{d}{dx}\sum_{k=1}^{\infty}P(\Delta t \leq x \mid X=k)\left( 1 - \frac{a_0}{a_{0, \text{max}}} \right)^{k-1}\frac{a_0}{a_{0, \text{max}}}\\
   \intertext{ and as shown in Eq.~\ref{eqSI:t_erlang}, the distribution of $\Delta t$ parametrized by $k$ follows an Erlang distribution:}
                   &= \sum_{k=1}^{\infty}\frac{d}{dx}F_{\text{Erlang}(k, \lambda_{0,\text{max}})}\left( 1 - \frac{a_0}{a_{0, \text{max}}} \right)^{k-1}\frac{a_0}{a_{0, \text{max}}}\\
                   &= \sum_{k=1}^{\infty}f_{\text{Erlang}(k, \lambda_{0,\text{max}})}\left( 1 - \frac{a_0}{a_{0, \text{max}}} \right)^{k-1}\frac{a_0}{a_{0, \text{max}}}\\
                   &= \sum_{k=1}^{\infty}\frac{a_{0,\text{max}}^k \cdot x^{k-1} \cdot \exp({-a_{0,\text{max}} x}) }{(k-1)!} \cdot \left( \frac{a_{0, \text{max}} - a_0}{a_{0, \text{max}}} \right)^{k-1} \cdot \frac{a_0}{a_{0, \text{max}}}\\
                   &= a_0 \exp({-a_{0, \text{max}}x}) \sum_{k=1}^{\infty}\frac{({a_{0, \text{max}} - a_0})^{k-1} \cdot x^{k - 1}}{(k-1)!}\\
                   &= a_0 \exp({-a_{0, \text{max}}x}) \cdot \exp(x \cdot (a_{0, \text{max}} - a_0))\\
                   &= a_0 \cdot \exp({-a_0x}).
\end{align*}
Hence, in Rejection Gillespie, $\Delta t$ follows the same exponential distribution as in the SG.

\subsection{Proof for the Non-Markovian Gillespie Algorithm (nMGA)}
\label{nMGA}
\noindent In this section, we summarize the proof given by Boguna~\cite{boguna2014simulating}, and consider a second-order approximation of their algorithm. 
We consider $N$ renewal processes running in parallel, and denote by $t_{j}$ the time elapsed since the last event of the $j$th process $(1 \leq j \leq N)$. We denote by $\psi_{j}(t_j)$ the probability density function of inter-event times for the $j$th process, and by
\begin{equation}
    \Psi_{j}\left(t_{j}\right)=\int_{t_{j}}^{\infty} \psi_{j}\left(\tau\right) \mathrm{d} \tau  \, , 
\end{equation}
the survival function of the $j$th process, i.e., the probability that the inter-event time is larger than $t_{j}$. We also set
\begin{equation}
    \label{eq:MOSAIC_prob_next_event}
    \Phi\left(\Delta t \mid\left\{t_{j}\right\}\right)=\prod_{j=1}^{N} \frac{\Psi_{j}\left(t_{j}+\Delta t\right)}{\Psi_{j}\left(t_{j}\right)}  \, 
\end{equation}
which is the probability that no process generates an event for time $\Delta t$~\cite{masuda2018gillespie}. Then in the non-Markovian Gillespie algorithm (nMGA), the time until the next event, $\Delta t$, is computed by solving $\Phi\left(\Delta t \mid\left\{t_{j}\right\}\right) = u$, where $u \in \mathcal{U}^{[0,1]}$ is a random variate drawn from the uniform density on the interval $[0, 1]$. This can be time-consuming for some distributions~\cite{boguna2014simulating}. In the limit of a large number of processes $N\to\infty$, we can simplify the numerical computation of the time $\Delta t$ needed in the algorithm. We start by rewriting the function $\Phi(\Delta t |{tj})$ as:
\begin{equation}
\label{eqSI:nMGA_approx}
    \Phi\left(\Delta t \mid\left\{t_{j}\right\}\right)=\exp \left[-\sum_{j=1}^{N} \ln \left(\frac{\Psi_{j}\left(t_{j}\right)}{\Psi_{j}\left(t_{j}+\Delta t\right)}\right)\right]
\end{equation}
The sum within the exponential function is a sum of $N$ monotonously increasing functions of $\Delta t$. Therefore, when $N\to\infty$, the survival probability $\Phi\left(\Delta t \mid\left\{t_{j}\right\}\right)$ is close to zero everywhere except when $\Delta t \sim 0$. Hence, we only need to consider $\Phi\left(\Delta t \mid\left\{t_{j}\right\}\right)$ around $\Delta t=0$. In this neighborhood, we can perform a Taylor expansion for small $\Delta t$, namely,  $\Psi_{j}\left(t_{j}+\Delta t\right) \approx \Psi_{j}\left(t_{j}\right)-\psi_{j}\left(t_{j}\right) \Delta t+O\left(\Delta t^{2}\right)$. Plugging this expression into Eq. \ref{eqSI:nMGA_approx}, using the approximation $\ln(1+x) \approx x + O\left(x^{2}\right)$ and $1/(1-x) \approx 1 + x + O\left(x^{2}\right)$ when $x \rightarrow 0$, we can write:
\begin{equation}
    \begin{aligned}
    \Phi\left(\Delta t \mid\left\{t_{j}\right\}\right) &=\exp \left[-\sum_{j=1}^{N} \ln \frac{\Psi_{j}\left(t_{j}\right)}{\Psi_{j}\left(t_{j}+\Delta t\right)}\right] \\
    & \approx \exp \left[-\sum_{j=1}^{N} \ln \frac{\Psi_{j}\left(t_{j}\right)}{\Psi_{j}\left(t_{j}\right)-\psi_{j}\left(t_{j}\right) \Delta t+O\left(\Delta t^{2}\right)}\right]\\
    &\approx \exp \left[-\Delta t\left(\sum_{j=1}^{N} \lambda_j(t_{j})\right)  +O\left(\Delta t^{2}\right) \right] \, ,
    \end{aligned}
    \label{eqSI:Taylor_appr}
\end{equation}
where the instantaneous rate $\lambda_j$ is defined as:
$$\lambda_j(t_j) = \frac{\psi_{j}(t_j)}{\Psi_{j}(t_j)}.$$
With this approximation, the time until the next event is determined by 

\begin{equation}
    \Phi\left(\Delta t \mid\left\{t_{j}\right\}\right) \approx \exp \left[-\Delta t\left(\sum_{j=1}^{N} \lambda_j\left(t_{j}\right)\right)\right] . 
    \label{eq:38}
\end{equation}
Denoting by $ u $  a uniform random variable sampled from $[0,1]$
to  represent $\Phi\left(\Delta t \mid\left\{t_{j}\right\}\right)$, we can   approximate $\Delta t$ as follows:
%
%
\begin{equation}
    \Delta t \approx \frac{\ln (1/u)}{\sum_{j=1}^{N} \lambda_j(t_{j})}.
    \label{eqSI:nMGA}
\end{equation}
We note that when  $\lambda_j(t_{j})$ depends on time,  the uniform random variable $u$ may not directly correspond to the distribution $\Phi\left(\Delta t \mid\left\{t_{j}\right\}\right)$. However, the approximation remains valid within a small neighborhood around $\Delta t = 0$ and when $\lambda_j(t_{j})$ changes slowly with time.
By eliminating the time dependency, i.e. setting $\lambda_{j}\left(t_{j}\right)=\lambda_{j}$, we recover the SG algorithm, where  $\Delta t$ is exponentially distributed. In the limit where $\lambda_j(t_{j})$ changes slowly with time, $\Delta t$ is still approximately exponentially distributed. nMGA exploits this approximation and locally treats each process as Poissonian during the time step $\Delta t$.

\subsection*{Second-order approximation}
\noindent We can also consider a quadratic expansion of Eq.~\ref{eqSI:nMGA_approx}, which results in a second-order approximation of  the time interval $\Delta t$. To do so, we first expand up to second order $\Psi_{j}\left(t_{j}+\Delta t\right)$:
\begin{equation}
    \Psi_{j}\left(t_{j}+\Delta t\right)
    \approx \Psi_{j}\left(t_{j}\right)-\psi_{j}\left(t_{j}\right) \Delta t - \psi_{j}'\left(t_{j}\right) \Delta t^{2}/2 +O\left(\Delta t^{3}\right), 
\end{equation}
\noindent where we remind the reader that $\Psi_j' = - \psi_j$. Additionally, the instantaneous rate $\lambda_j = \psi_j / \Psi_j$ is related to its derivative $\lambda'_j$ through the relation:

\begin{equation}
    \lambda'_j = \left(\frac{\psi_j }{\Psi_j}\right)' = \frac{\psi_j' \Psi_j - \psi_j \Psi_j'}{\Psi_j^2} = \frac{\psi'_j }{\Psi_j} + \left(\frac{\psi_j }{\Psi_j}\right)^2 = \frac{\psi'_j }{\Psi_j} + \lambda_j^2.
\end{equation}

\noindent Plugging this expression into Eq.~\ref{eqSI:nMGA_approx}, using the approximation $1/(1-x) = 1 + x + x^2 +  O\left(x^{3}\right)$ and $\ln(1+x) = x - x^{2}/2 + O\left(x^{3}\right)$ for $x \rightarrow 0$, we obtain:
\begin{equation}
    \label{Eq:MOSAIC_second_order}
    \begin{aligned}
    \Phi\left(\Delta t \mid\left\{t_{j}\right\}\right) &
    \approx \exp \left[-\sum_{j=1}^{N} \ln \frac{1}{1-\lambda_{j}\left(t_{j}\right) \Delta t  - \left(\lambda_j'(t_{j}) - \lambda_j^2(t_{j}) \right) \dfrac{\Delta t ^2}{2} + O\left(\Delta t^{3}\right)}\right]\\
    & \approx \exp \left[-\sum_{j=1}^{N} \ln \left( 1+\lambda_{j}\left(t_{j}\right) \Delta t  + \left(\lambda_j'(t_{j}) + \lambda_j^2(t_{j})\right) \frac{\Delta t ^2}{2} + O\left(\Delta t^{3}\right) \right)\right]\\
    & \approx \exp \left[-\sum_{j=1}^{N} \left(  \lambda_{j}\left(t_{j}\right) \Delta t  + \lambda_j'(t_{j})\frac{\Delta t ^2}{2} + O\left(\Delta t^{3}\right) \right) \right]\\
    \end{aligned}
\end{equation}


\noindent Solving $\Phi\left(\Delta t \mid\left\{t_{j}\right\}\right) = u$ to determine the next time increment $\Delta t$, we get a quadratic equation for which we take the positive solution:
\begin{equation}
\label{eqSI:dt_quadratic}
    \Delta t \approx \frac{- \sum_{j=1}^{N} \lambda_j (t_j) + \sqrt{\left(\sum_{j=1}^{N} \lambda_j(t_j)\right)^2 + 2 \left(\sum_{j=1}^{N} \lambda_j^{'}(t_j) \right) \cdot \ln(1/u)}}{\left(\sum_{j=1}^{N} \lambda_j^{'}(t_j) \right)} \, . 
\end{equation}
Once again,  Eq.~\ref{eqSI:dt_quadratic} makes explicit that $\Delta t$ is no longer exponentially distributed when $\lambda_j^{'}(t_j) \neq 0$, and hence, the rejection framework described in the first part cannot be applied with the second-order approximation of nMGA.

\subsection{Quantifying the errors of nMGA and MOSAIC }

\label{quant_error_MOSAIC}

\noindent Let us derive the error for both nMGA and MOSAIC. As a reminder, at each time step, both nMGA and MOSAIC utilize the first-order approximation of the survival distribution function to generate the next event with
\begin{equation}
    \Psi_{j}\left(t_{j}+\Delta t\right)=\Psi_{j}\left(t_{j}\right)-\psi_{j}\left(t_{j}\right) \Delta t+O\left(\psi_j'(t_j) \Delta t^{2}\right),
    \label{eq:Taylor_approx}
\end{equation}
from which the probability that no process generates an event for time $\Delta t$ with (Eq.~\ref{eq:MOSAIC_prob_next_event}) is computed from
\begin{equation}
    \Phi\left(\Delta t \mid\left\{t_{j}\right\}\right)=\prod_{j=1}^{N} \frac{\Psi_{j}\left(t_{j}+\Delta t\right)}{\Psi_{j}\left(t_{j}\right)},
\end{equation}
where $N$ is the total number of processes being considered. The error in this approximation is given by the difference between the actual value $\Phi\left(\Delta t \mid\left\{t_{j}\right\}\right)$ and the approximated value derived from the first term of the Taylor expansion. According to Taylor's theorem, and looking at the first and second order expression of Eq.\ref{Eq:MOSAIC_second_order}, we write the error term for the first-order approximation of $\ln \left[ \Phi\left(\Delta t \mid\left\{t_{j}\right\}\right) \right]$ as the second-order remainder~\cite{taylor1717methodus}:
\begin{equation}
    R_2^{\ln \Phi} = \sum_{j=1}^{N} \lambda_j'(t_j) \frac{\Delta t^2}{2} + O(\Delta t^3),
\end{equation}
where $\lambda_j'(t_j)$ is the derivative of the instantaneous rate function at time $t_j$.

\paragraph{Error of nMNGA.} Plugging in the value of $\Delta t$ of nMGA (Eq.\ref{eqSI:nMGA}), we estimate the error $R^{\ln \Phi}_2$ in nMGA per time step as
\begin{equation}
   \left( R^{\ln \Phi}_2 \right)_{ \mbox{\tiny nMGA}} \sim \sum_{j=1}^{N} \frac{\lambda_j'(t_j)}{2} \left( \frac{\ln (1/u)}{\sum_{i=1}^{N} \lambda_i(t_i)} \right)^2 \sim \frac{\ln^2(1/u)}{2 N} \cdot \frac{\langle \lambda' \rangle}{\langle \lambda \rangle^2},
    \label{eq:error_nmga}
\end{equation}
where we have introduced $\langle \lambda \rangle$ and $\langle \lambda' \rangle$, the average rate and the average derivative of the rates respectively, i.e.:
\begin{equation}
  \langle \lambda \rangle =  \frac{1}{N} \sum_{j=1}^{N}   \lambda_j(t_j) \ \ \text{and} \ \ \langle \lambda' \rangle =  \frac{1}{N} \sum_{j=1}^{N}   \lambda_j'(t_j) \, .
\end{equation}

\paragraph{Error of MOSAIC.} Using instead MOSAIC's time increment (Eq.\ref{eq:deltat}),  
\begin{equation}
    \Delta t = \frac{\ln(1/u)}{N \lambda_\text{max}} \, ,
\end{equation}
we obtain MOSAIC's error estimate: 
\begin{equation}
    \left( R^{\ln \Phi}_2 \right)_{ \mbox{\tiny MOSAIC}}
    \sim \frac{\ln^2(1/u)}{2 N} \cdot \frac{\langle \lambda' \rangle}{\lambda_\text{max}^2} \, .\\
    \label{eq:error_MOSAIC}
\end{equation}
We note that Eqs.~\ref{eq:error_nmga} and \ref{eq:error_MOSAIC} depend on $\ln^2(1/u)$, which does not have an upper bound so the error in a particular iteration can become arbitrarily large. However, we can write the expected error by substituting $\ln^2(1/u)$ with $\mathbb{E}[\ln^2(1/u)] = \int_0^{\infty} t^2 e^{-t} \, dt = 2$, where
\begin{equation}
    \label{eq:phi_error}
    \mathbb{E} \left[ \left( R^{\ln \Phi}_2 \right)_{ \mbox{\tiny nMGA} } \right] 
    \sim \frac{\langle \lambda' \rangle}{N \cdot \langle \lambda \rangle^2} \ \ \ \ \text{and} \ \ \ \     \mathbb{E} \left[ \left( R^{\ln \Phi}_2 \right)_{ \mbox{\tiny MOSAIC} }\right] 
    \sim \frac{\langle \lambda' \rangle}{N \cdot \lambda_\text{max}^2} \, .
\end{equation}
Here, we make several noteworthy observations. First, in both nMGA and MOSAIC, the error is directly proportional to the first derivative of the rate function, so both methods yield exact results for the exponential distributions where $\lambda_j'(t_j) = 0$.  In contrast, distributions with rapidly changing rates, such as the Weibull distribution with a high shape parameter, will exhibit significantly higher errors compared to distributions with more gradually varying rates, such as long-tailed distributions like Cauchy or Pareto.

Then, the primary difference between the two methods lies in the denominators of the expected error, which dictate how errors scale. Specifically, the error in nMGA scales with $1/\langle \lambda \rangle^2$, while MOSAIC's scales with $1/\lambda_\text{max}^2$. Consequently, nMGA is particularly prone to large errors when agents are initialized with low instantaneous rates, as $\langle \lambda \rangle$ can be very small at the start of the simulation. This issue is not encountered in MOSAIC, as $\lambda_\text{max}$ can be chosen independently, providing greater flexibility and accuracy.


\paragraph{Error in the time increment $\mathbf{\Delta t}$.} The errors derived above reflect the uncertainty in estimating $\Phi(\Delta t)$ at a specific time step, the probability that no events occur within the interval $\Delta t$. To quantify the error in the time step increment itself, we consider the sensitivity of $\Delta t$ to variations in $\ln \Phi(\Delta t)$. This error propagation can be determined by applying the chain rule~\cite{ku1966notes}:
\begin{equation}
    \text{Error}[\Delta t] \sim \left| \frac{\partial \Delta t}{\partial \ln \Phi(\Delta t)} \right| \cdot \text{Error}[\ln \Phi(\Delta t)],
\end{equation}
where the term  $\frac{\partial \Delta t}{\partial \ln \Phi(\Delta t)}$ represents how sensitive the time increment is to changes in $\ln \Phi(\Delta t)$. Here, we have ignored the sign, focusing only on the magnitude of the error. To analytically derive the error of $\Delta t$ in MOSAIC, we differentiate
\begin{equation}
    \Phi_{\mbox{\tiny MOSAIC}}(\Delta t) \sim \exp \left[-\Delta t N \lambda_\text{max} \right] 
\end{equation}
and find that
\begin{equation}
    \left| \frac{\partial \ln \Phi_{\mbox{\tiny MOSAIC}}(\Delta t)}{\partial \Delta t} \right| \sim N \lambda_\text{max}.
\end{equation}
The error in $\Delta t$ can thus be propagated as
\begin{equation}
    \text{Error}[\Delta t_\text{MOSAIC}] \sim \frac{1}{N \lambda_\text{max}} \cdot \text{Error}[\ln \Phi_{\mbox{\tiny MOSAIC}}(\Delta t)],
\end{equation}
which we also write as
\begin{equation}
    \text{Error}[\Delta t_\text{MOSAIC}] \sim \frac{1}{N \lambda_\text{max}} \cdot \left( R^{\ln \Phi}_2 \right)_{ \mbox{\tiny MOSAIC}}.
\end{equation}
Applying the same approach for nMGA, we write
\begin{equation}
    \Phi_{ \mbox{\tiny nMGA}}\left(\Delta t\right) \sim \exp \left[-\Delta t\left(\sum_{j=1}^{N} \lambda_j\left(t_{j}\right)\right)\right] \sim \exp \left[-\Delta t N \langle \lambda \rangle \right] 
\end{equation}
and obtain
\begin{equation}
    \text{Error}[\Delta t_\text{nMGA}] \sim \frac{1}{N \langle \lambda \rangle} \cdot \left( R^{\ln \Phi}_2 \right)_{\mbox{\tiny nMGA}}.
\end{equation}

\paragraph{Accumulated error in the IED.} In a stochastic simulation, errors accumulate across multiple time steps, denoted here as $n$. The total error in the simulated IED can be expressed as the sum of errors over all time steps:
\begin{equation}
    \text{Error}[\text{IED}] = \sum_{i=0}^n \text{Error}[\Delta t_i] \sim n \cdot \mathbb{E}\Bigl[\text{Error}[\Delta t]\Bigr].
\end{equation}
In the case of the standard Gillespie algorithm and nMGA, the number of time steps required to simulate a system for a fixed time $T_\text{end}$ scales with the number of processes, $N$, and the mean reaction rate, $\langle \lambda \rangle$, such that $ n \sim N \langle \lambda \rangle $. Consequently, the accumulated error for nMGA can be approximated as:
\begin{equation}
    \text{Error}[\text{IED}_\text{nMGA}] \sim N \langle \lambda \rangle \cdot \text{Error}[\Delta t_\text{nMGA}] \sim \mathbb{E}\Bigl[\bigl(R^{\ln \Phi}_2\bigr)_{\mbox{\tiny nMGA}}\Bigr].
\end{equation}
In the case of MOSAIC, where some reactions are rejected, the number of time steps increases compared to nMGA, scaling as $\sim N \lambda_\text{max}$. Substituting this relation, the accumulated error for MOSAIC can be written as:
\begin{equation}
    \text{Error}[\text{IED}_\text{MOSAIC}] \sim N  \lambda_\text{max} \cdot \text{Error}[\Delta t_\text{MOSAIC}] \sim \mathbb{E}\Bigl[\bigl(R^{\ln \Phi}_2\bigr)_{\mbox{\tiny MOSAIC}}\Bigr].
\end{equation}
Here we observe that, while MOSAIC involves a greater number of time steps due to the rejection of some reactions, the total accumulated error over the course of the simulation remains comparable with nMGA, since they both equal $R_2^{\ln\Phi}$. This equivalence can be intuitively understood as a balance: the increase in the number of time steps in MOSAIC is effectively counteracted by the proportional reduction in error magnitude per time step. Substituting the expression for $R_2^\Phi$ derived in Eq.~\ref{eq:phi_error}, the final expressions for the errors are:
\begin{equation}
    \label{eq:MOSAIC_IED_error}
    \text{Error}[\text{IED}_\text{nMGA}] 
    \sim \frac{\langle \lambda' \rangle}{N \cdot \langle \lambda \rangle^2} \ \ \ \ \text{and} \ \ \ \ 
    \text{Error}[\text{IED}_\text{MOSAIC}]
    \sim \frac{\langle \lambda' \rangle}{N \cdot \lambda_\text{max}^2}.
\end{equation}\\
We point out that, for illustrative purposes, we assumed here that $\langle \lambda' \rangle$, $\langle \lambda \rangle$, and $\lambda_\text{max}$ remain approximately constant throughout the simulation, a reasonable assumption for systems that quickly reach a steady state, where these parameters do not fluctuate significantly. However, this framework can also be generalized to systems with time-varying parameters. In such cases, the IED error at any given time step can be interpreted as the cumulative error that would result if the conditions (i.e., the values of $\langle \lambda' \rangle$, $\langle \lambda \rangle$, and $\lambda_\text{max}$) at that specific time step were held constant and applied across the entire simulation.

\paragraph{Accumulated error in the population dynamics.} While we successfully derived the expected error for the simulated IED, most non-Markovian studies report observables in terms of population dynamics rather than the IED, as population dynamics are generally more accessible experimentally~\cite{stumpf2017stem, england2010global, zeisel2011coupled} due to the difficulty of directly measuring IED in experimental systems. Here, we describe how errors in the time step $\Delta t$ propagates into population count dynamics over the course of the simulation. We consider $N$ processes and define the total rate as $\lambda_{\text{total}} = N \lambda_\text{max}$. For a given iteration, we estimate that the error in the time step propagates to the population as 
\begin{equation}
    \text{Error}[P] \sim \text{Error}[\Delta t] \cdot \frac{\lambda_{\text{total}}}{N} \sim \text{Error}[\Delta t] \cdot \lambda_\text{max}.
\end{equation}
Here, we normalize the population count error by the number of processes because we are interested in the relative population rather than the absolute errors in the counts. Due to the accumulated error during the simulation, we write the deviation of the population at time $t$ from the ground truth as
\begin{equation}
    \Delta P (t) = \sqrt{t} \cdot \text{Error}[P],
\end{equation}
where we have assumed that the error propagates through time like a diffusing behavior. Then, we can write the total error of the simulation as a sum through the number of total events $T$ 
\begin{align}
    \text{Error}[\text{POP}_\text{MOSAIC}] &= \sum_{k=0}^T  \Delta P (t_k)\\
    &= \text{Error}[P] \left(\sum_{k=0}^T \sqrt{t_k} \right)\\
    &\sim \lambda_\text{max} T^{3/2} \cdot \text{Error}[\Delta t_\text{MOSAIC}] \cdot
\end{align}\\
where the series scales approximately as $T^{3/2}$ due to the summation of $\sqrt{t}$ term. Substituting $T \sim \lambda_\text{max} N$ and the time step error derived under the MOSAIC framework
\begin{equation}
    \text{Error}[\Delta t_\text{MOSAIC}] \sim \frac{\langle \lambda' \rangle}{N^2 \cdot \lambda_\text{max}^3},
\end{equation}
we obtain:
\begin{equation}
    \text{Error}[\text{POP}_\text{MOSAIC}] \sim \sqrt{\frac{1}{N \lambda_\text{max}}} \langle \lambda' \rangle.
\end{equation}

\newpage

\subsection{Rejection sampling allows for arbitrary approximation accuracy}

\label{arbitrary_accuracy}

\begin{figure*}[h!t]
    \centering
    \captionsetup{width=1\linewidth}
    \includegraphics[width=1\linewidth]{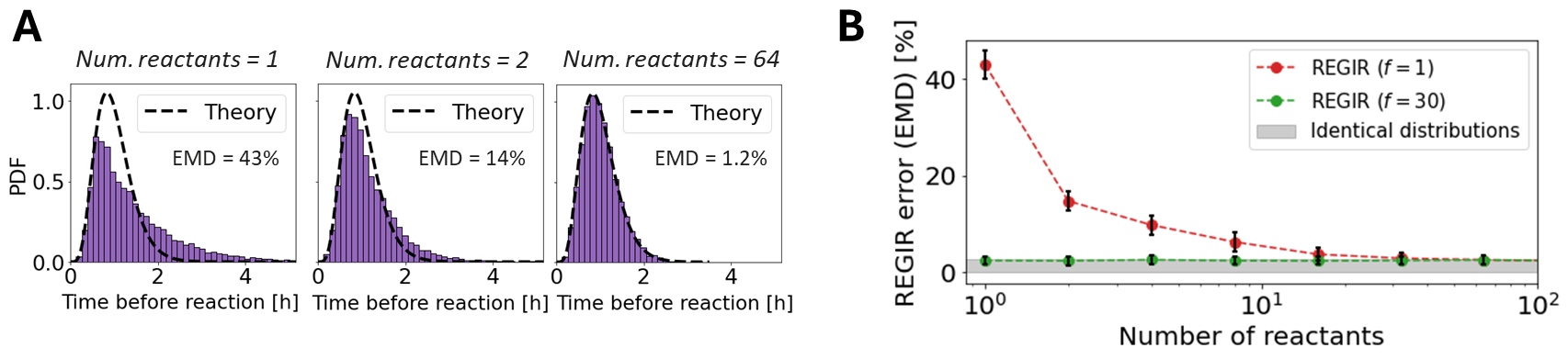}
    \caption{(A) MOSAIC ($f = 1$) approximation accuracy on a toy reaction $A \rightarrow \varnothing$ with a gamma inter-event distribution of shape parameter 6, visualized for different population sizes. The accuracy is computed using the earth mover distance (EMD) between the theoretical and simulated distributions, given in units of the distribution mean $1/\lambda_0$. In (B), we show how the EMD scales with the population size for two variations of MOSAIC, where the difference between the two lies in the additional parameter $f$ used to define $\lambda_\text{max}$.}
    \label{fig:MOSAIC_accuracy}
\end{figure*}

\noindent In this section, we discuss the differences between nMGA~\cite{boguna2014simulating} and MOSAIC in terms of simulation accuracy. We recall that both MOSAIC and nMGA are exact only when $\Delta t \rightarrow 0$, as they use a first-order Taylor approximation of the survival distribution function. Indeed, the survival distribution function can be Taylor approximated as follows:
\begin{equation}
    \Psi_{j}\left(t_{j}+\Delta t\right)=\Psi_{j}\left(t_{j}\right)-\psi_{j}\left(t_{j}\right) \Delta t+O\left(\psi_j'(t_j) \Delta t^{2}\right).
    \label{eq:Taylor_approx}
\end{equation}
The approximation breaks down when the inequality $\psi_j(t_j) \gg \psi_j'(t_j)\Delta t$ is no longer verified. The main difference between MOSAIC and nMGA lies in the introduction of the rejection step. In nMGA, the time increment until the next event is computed as:
\begin{equation}
    \Delta t\ \text{(nMGA)}= \frac{\ln (1/u)}{\sum_{j=1}^{N} \lambda_j(t_{j})} \,,
    \label{eq:nMGA_increment}
\end{equation}
while in MOSAIC, the expression becomes:
\begin{equation}
     \Delta t \ \text{(MOSAIC)} = \frac{\ln (1/u)}{N \cdot \lambda_\text{max}} \, .
     \label{eq:MOSAIC_increment}
\end{equation}
From Eq.\ref{eq:nMGA_increment}, it is clear that low rates are associated with large time increments $\Delta t$, and that in such a regime, the linear approximation nMGA used to compute $\Delta t$ might fail (Eq.\ref{eq:Taylor_approx}). Indeed, nMGA is only exact in the limit of an infinite number of processes ($N \rightarrow \infty$) where it can be assumed that $\sum_{j=1}^{N} \lambda_j(t_{j}) \rightarrow \infty$. For processes characterized by low rates at some time points (such as gamma distribution at $t=0$ for $\alpha > 1$), this approximation can be poor even in the limit of a large number of processes. MOSAIC circumvents this problem by setting $\lambda_\text{max}$ to an arbitrary large value (for example $\lambda_\text{max} \geq \lambda_0$), such that the time increment $\Delta t$ remains \textit{small enough} for the first order Taylor approximation (Eq.~\ref{eq:Taylor_approx}) to hold during the entire simulation. Still, this condition may not be sufficient when the number of processes $N$ is too low (Supplementary Figure~\ref{fig:MOSAIC_accuracy}A). To handle these cases, we can set $\lambda_\text{max}$ at each iteration such that:
\begin{equation}
    \label{eq:f_formula}
    \lambda_\text{max} \geq \lambda_0 \cdot \min \left\{ \frac{f}{N}, \ 1 \right\},
\end{equation}
where $\lambda_0$ is the inverse of the mean inter-event
time distribution, and $f$ a factor defined by the user to guarantee a desired upper bound for $\Delta t$, e.g. such as 
$\psi_j(t_j) \gg \psi_j'(t_j)\Delta t$ always holds:
\begin{equation}
    \Delta t \ \text{(MOSAIC)} \leq \frac{\ln (1/u)}{f \cdot \lambda_0}.
    \label{eq:MOSAIC_adjusted}
\end{equation}
An important consideration is that this definition will affect the system only when the systems contains a low number of processes. For systems with a large number of processes ($N \gg f$), the relationship
\begin{equation}
    \max_{\{j \in [1,N]\}} \lambda_j (t_j) \geq f \cdot \frac{\lambda_0}{N}
\end{equation}
is always verified with high probability, so that 
\begin{equation}
\lambda_\text{max} \geq \max_{\{j \in [1,N]\}} \lambda_j (t_j)
\end{equation}
regardless of the value of $f$. On the other hand, for systems with a low number of processes ($N < f$), increasing $f$ results in increased accuracy, but at the cost of additional computational time. More precisely, the computational cost of running MOSAIC will increase from $O(r N)$ to $O(r f)$, where $r$ is the attempted-over-accepted ratio of the system for $f=1$. Using the general equation for the MOSAIC error (Eq.~\ref{eq:MOSAIC_IED_error}), and substituting $\lambda_\text{max}$ from Eq.~\ref{eq:f_formula} we get that for $f \geq N$,
\begin{equation}
    \text{Error}[\text{IED}_\text{MOSAIC}]
    \leq \frac{N \cdot \langle \lambda' \rangle}{f^2 \cdot \lambda_0^2}.
\end{equation}\\
and otherwise ($f < N$)
\begin{equation}
    \text{Error}[\text{IED}_\text{MOSAIC}]
    \leq \frac{\langle \lambda' \rangle}{N \cdot \lambda_\text{0}^2}.
\end{equation}\\

\begin{figure}[h!t]
    \centering
    \includegraphics[width=0.8\linewidth]{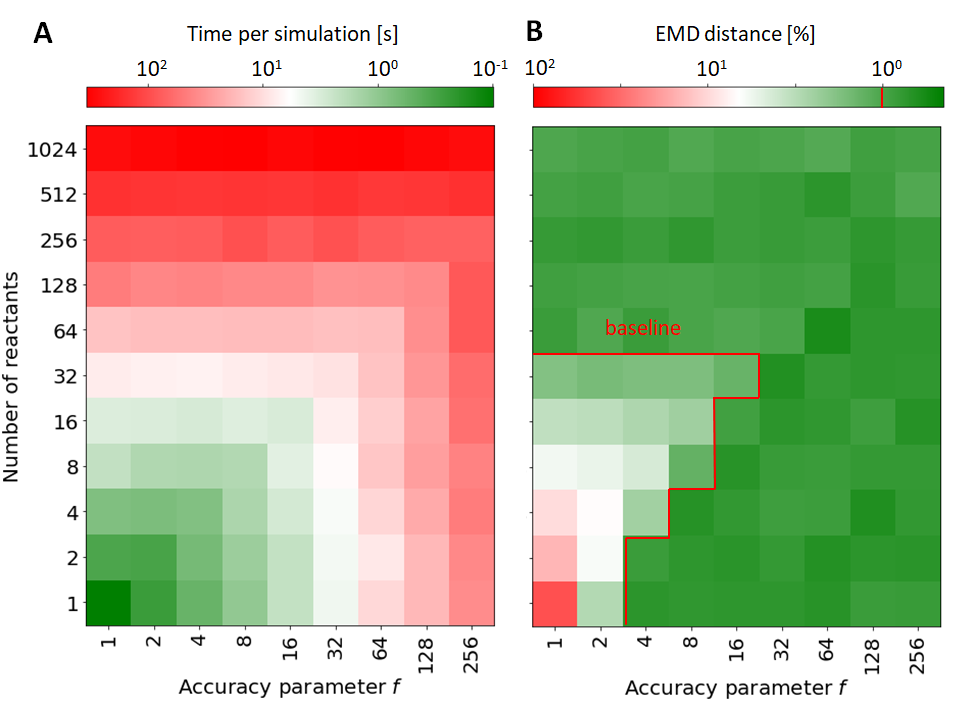}
    \caption{\small Trade-off between computational cost and accuracy. The toy reaction $A \rightarrow \varnothing$ is simulated with a gamma inter-event time distribution with parameters ($\lambda_0 = 1$, $\alpha = 6$). We display the heatmap of (A) the computational time for one simulation and (B) the EMD distance from the theoretical distribution. All values are averaged over 100 simulations, and EMD distances are given with a $\pm 10\%$ confidence interval. The red lines mark the border where the EMD becomes below the baseline EMD (defined as the EMD between two equal distributions with 10k sampling, equal here to 1\%).}
    \label{fig:EMD_tradeoff}
\end{figure}

Interestingly, our empirical investigation shows that such consideration only significantly affect the simulation accuracy when $N < 30$ (Supplementary Figure~\ref{fig:MOSAIC_accuracy}B \& Figure~\ref{fig:EMD_tradeoff}), which falls well below the number of reactants typically involved in most practical scenarios. In particular, the choice $f$ did not impact computational cost and accuracy the systems we discussed in this main article.

\newpage

\subsection{Rejection sampling reduces the computational complexity of MOSAIC to $\mathbf{\textit{O(N)}}$}
\label{proof_complexity}
\noindent We consider a simple reaction channel with $N$ reactants and an arbitrary IED over a duration $T$. Let us examine two different aspects of this simulation: (i) the computational complexity of a single time step and its scaling properties, and (ii) the number of  steps required to  simulate a fixed interval of length $T$. Let us first focus on the scaling properties of a single-time step. During a large simulation, two factors contribute to its computational cost, the calculation of the time step to the next reaction, and the update of the reactant populations~\cite{sanft2015constant}. For SG with one reaction channel, both of these steps are $O(1)$. In the case of nMGA however, these updates carry a computational cost of $O(N)$, as each reactant has its own reaction rate, which is equivalent to having its own reaction channel. In particular, the update of the instantaneous rates can be very expensive, as it requires recalculating them for each process in the entire population after every simulation iteration. On the other hand, MOSAIC reduces the computational complexity to $O(1)$ using a rejection base approach, where only the rate of the drawn reactant is computed at each step. The maximum rate $\lambda_\text{max}$ is either kept constant throughout the simulation or is updated using an ordered data structure for storing the $t_j$ values, thus also $O(1)$ (See Methods~\ref{Methods_MOSAIC}).

Regarding the number of steps required to simulate a fixed interval, SG, nMGA, Laplace, and DelaySSA scale linearly with the number of reactants $N$, because the time step becomes increasingly smaller as more channels are added to the simulation (Eq.\ref{eq:deltat}). However, the rejection approach of MOSAIC results in a larger number of time steps required to simulate a $ T_\text{end}$ interval, as a fraction of the time steps are rejected by the algorithm. The complexity then becomes $O(N) + O(R)$, where $R$ refers to the number of rejected steps. From Eq.\ref{eq:choice} and Eq.\ref{eq:rejection}, we can compute the probability of rejection for a given iteration as:
\begin{equation}
\begin{aligned}
  p_{reject} &  = 1 - p_{accept} \\
  & =  1 - \frac{\sum_{j=1}^{N} p_j \lambda_j(t_j) }{\lambda_\text{max} } \\
  & =  1 - \frac{1}{N \lambda_\text{max}}\sum_{j=1}^{N}  \lambda_j(t_j) \\
  & =  1 - \frac{\lambda_0}{ \lambda_\text{max}}, 
\end{aligned}
\end{equation}

\noindent where $\lambda_0$ is the average observed rate. Then, the expected number of rejections before the first accepted reaction ($A$) is given by the mean of the geometric distribution with success probability $p_\text{accept}$:

\begin{equation}
 \frac{R}{A}  = \frac{1}{p_{accept}} - 1 = \frac{(\lambda_\text{max} - \lambda_0)}{ \lambda_0}\\
\end{equation}

\noindent Thus we can conclude that the scaling of SG and MOSAIC running times, $T_{\text{R, SG} }$ and $T_{\text{R, MOSAIC}}$, are proportional according to the relation:

\begin{equation}
    \frac{T_{\text{R, MOSAIC}}}{T_{\text{R, SG}}} = \frac{A+R}{A} = \frac{\lambda_\text{max}}{\lambda_0} 
\end{equation}

\begin{figure*}[h!t]
    \centering
    \captionsetup{width=0.8\linewidth}
    \includegraphics[width=0.5\linewidth]{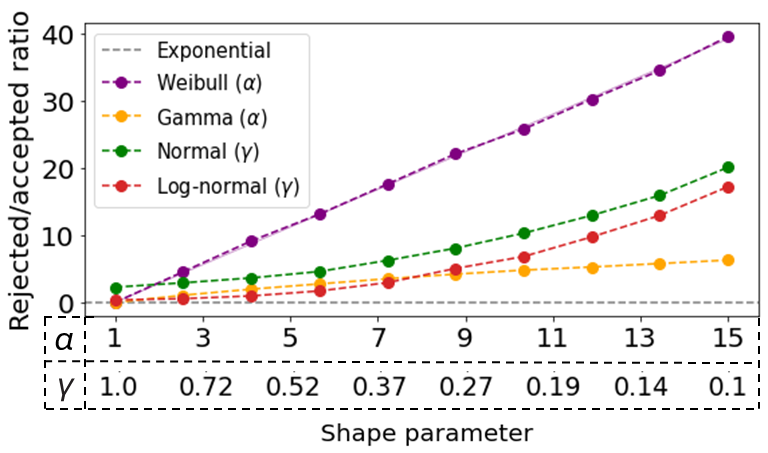}
    \caption{$R/A$ ratio for different shape parameters of different distributions, averaged over 100 simulations of on a toy reaction $A \rightarrow \varnothing$. A standard deviation of $\sim \pm 12\%$ was observed for all measurements. Note that a direct comparison of the ratio across distributions is not meaningful as they have different standard deviations. We also note that $R/A$ is scale-invariant and thus is independent of the distribution mean $1/\lambda_0$.}
    \label{fig:MOSAIC_ratio}
\end{figure*}

\noindent Supplementary Figure~\ref{fig:MOSAIC_ratio} shows the rejected over accepted reactions ratio ($\nicefrac{R}{A}$) for various distribution and shape parameters. We note that the ratio significantly varies with the choice of distribution and the parameters. For instance, an exponential distribution has a ratio of 0 (since $\hat\lambda =  \lambda_\text{max} = \lambda_i \; \forall i$, no reaction is rejected). On the other side of the spectrum, the Weibull instantaneous rate increases polynomially with time, so the maximum rate $\lambda_\text{max}$ increases quickly with $\alpha$, thus increasing the number of rejections. Other longer-tailed distributions with reaction rates increasing sub-linearly with time, e.g. the gamma distribution, will be less affected by changes in their respective shape parameter. In general, simulating a distribution with smaller variance will increase the maximum rate and as a result also increase the computational cost. This is intuitively clear from Supplementary Figure~\ref{fig:MOSAIC_ratio}, where the rejected over accepted reaction ratio ($R/A$) monotonically increases with the shape parameter of the gamma and Weibull distributions, which inversely correlate to the variance of their respective distribution. On the other hand, as both $R$ and $A$ are proportional to the rate $\lambda_0$, the ratio $\nicefrac{R}{A}$ is independent of the mean IED $\nicefrac{1}{\lambda_0}$ and thus scale invariant.

\newpage

\section{Relationship between the node and the pairwise interactions IEDs}

\label{distribution_relationship}

\noindent We consider two populations two type of reactants A and B, and a reaction channel \textbf{A + B} $\to$ \textbf{X}. When a reactant $A_i$ interacts at a given time $t$, the probability of pairing with a specific reactant $B_j$ is weighted by $w_{B_j}$, which represents $B_j$'s internal state and history of interactions. Specifically, the probability of $A_i$ to pair with $B_j$ at time $t$ is given by the ratio:
\begin{equation}
    \frac{w_{B_j}(t)}{\sum_{m} w_{B_m}(t)},
\end{equation}
where the denominator ensures normalization across all potential partners $B_m$. 

To compute the inter-event time distribution for the pair $(A_i, B_j)$, we must account for all possible scenarios in which $A_i$ does not pair with $B_j$ for $k$ consecutive interactions before finally pairing with $B_j$ on the $(k+1)$-th interaction, which we write as:
\begin{equation}
    P(\text{pairing with } B_j \text{ after } k \text{ steps}) =  \prod_{i=1}^k \left(1 - \frac{w_{B_j}(t_i)}{\sum_{m} w_{B_m}(t_i)}\right) \cdot \frac{w_{B_j}(t_{k+1})}{\sum_{m} w_{B_m}(t_{k+1})}.
\end{equation}
Here, the first term (the product) represents the probability that $A_i$ did not pair with $B_j$ during the first $k$ interactions, while the second term represents the probability of pairing with $B_j$ at the $(k+1)$-th interaction.

The inter-event time distribution for the pair $(A_i, B_j)$, denoted $\psi_{AB}(t)$, is influenced by the activity pattern of $A_i$, represented by $\psi_A(t)$, and the probabilities of pairing $(A_i, B_j)$ after $k$ prior interactions. Specifically, it is expressed as a weighted sum of convolutions of $\psi_A(t)$, where the weights are determined by the pairing probabilities:
\begin{equation}
    \psi_{AB}(t) = \sum_{k=0}^\infty \left[\prod_{i=1}^k \left(1 - \frac{w_{B_j}(t_i)}{\sum_{m} w_{B_m}(t_i)}\right) \cdot \frac{w_{B_j}(t_{k+1})}{\sum_{m} w_{B_m}(t_{k+1})}\right] \cdot \psi_A^{*(k+1)}(t),
\end{equation}
where $\psi_A^{*(k+1)}(t)$ denotes the $(k+1)$-fold convolution of $\psi_A(t)$. This formulation highlights how the inter-event time distribution of $(A_i, B_j)$ depends on both $A_i$'s activity and the probabilistic nature of its pairings with $B_j$.

Assuming $N_B$ a constant number of nodes over time, the term $\frac{w_{B_j}(t_i)}{\sum_m w_{B_m}(t_i)}$ can be approximated as $\frac{1}{N_B}$ on average. Under this assumption, the pairing probability simplifies, and the likelihood of $A_i$ pairing with $B_j$ after $k$ prior interactions follows a geometric distribution:
\begin{equation}
    P(\text{pairing with } B_j \text{ after } k \text{ steps}) = \frac{1}{N_B} \left(1 - \frac{1}{N_B}\right)^k.
\end{equation}
As a result, the inter-event time distribution for $(A_i, B_j)$ simplifies to:
\begin{equation}
    \psi_{AB}(t) = \sum_{k=0}^\infty \frac{1}{N_B} \left(1 - \frac{1}{N_B}\right)^k \cdot \psi_A^{*(k+1)}(t).
\end{equation}
As the mean of $k$ convoluted distribution is the sum of their respective mean, i.e. $\mathbb{E}\left[\psi_A^{*(k)}\right] = k \cdot \mathbb{E}\left[\psi_A\right]$, and recognizing the derivative of a geometric series, we conclude using the linearity of the expected value that the means of the distributions $\psi_A(t)$ and $\psi_{AB}(t)$ are related by:
\begin{equation}
    \mathbb{E}[\psi_{AB}] = N_B \cdot \mathbb{E}[\psi_A].
\end{equation}
For systems where the nodes' activity follows a Poisson process, i.e. $\psi_A(t) = \lambda_A \exp \left( - \lambda_A t  \right)$, we can also write
\begin{equation}
    \psi_{AB}(t) = \frac{\lambda_A}{N_B} \exp \left( - \frac{\lambda_A}{N_B} \cdot t \right).
\end{equation}

For non-Poissonian activity patterns, however, the combination of convolutions and geometric probabilities makes $\psi_{AB}(t)$ challenging to compute explicitly, often requiring numerical methods or approximations even if we assume a uniform probability of selected $B_j$. Nevertheless, we can still make general statements about $\psi_{AB}(t)$, particularly when $\psi_A(t)$ exhibits long-tailed behavior. If $\psi_A(t)$ follows a heavy-tailed distribution, such as a power law, the convolution $\psi_A^{*(k+1)}(t)$ retains this heavy-tailed nature, with a progressively slower decay as $k$ increases. This means that even without explicitly computing $\psi_{AB}(t)$, we know it will inherit the heavy-tailed characteristics of $\psi_A(t)$. Still, the link between the two distributions is not straightforward, and controlling one does not necessarily translate to control over the other. Given the complexity of the relationship between the inter-event time distribution for individual pairs and that of individual nodes in the network, it is crucial to clearly define and prioritize the specific aspect to focus on in the modeling process.

\section{MOSAIC for node-driven temporal networks (MOSAIC-TN)}

\label{MOSAIC-TN}

\subsection{Algorithm}

\label{MOSAIC-TN-algorithm}

\noindent Let us consider a reaction channel \textbf{A + A} $\to$ \textbf{X}. We denote by $N_A$ the number of nodes. We first consider a fully connected topology, so we have $N = N_A (N_A-1)/2$ processes. For each individual node $\text{A}_i$  $(1 \leq i \leq N_A)$ and $\text{A}_j$  $(1 \leq j \leq N_A)$, we consider $t_i$ and $t_j$ the time elapsed since the last interaction involving node $A_i$ and $A_j$, respectively. At each iteration, two nodes $A_i$ and $A_j$ are drawn, and their interaction propensity is determined by a time-dependent pairwise interaction rate $\Lambda_{ij}(t_i, t_j) > 0$, which accounts for the individual times of both nodes. Importantly, $\Lambda_{ij}$ can also incorporate pair-specific properties or interaction history of the node pair $(i, j)$, enabling the modeling of \textit{temporal neighborhood effects} ,  that is, interaction dynamics shaped by the past interactions or unique relationship between the two nodes. The MOSAIC algorithm for node-driven temporal networks (MOSAIC-TN) involves four steps:
\begin{enumerate}

    \item Set $\Lambda_\text{max}$, the maximum reaction rate over all processes, such that:
    
    \begin{equation}
        \Lambda_\text{max} \geq \ \max_{\{(i,j) \in [1,N_A], \ i \neq j\}} \Lambda_{ij}(t_i, t_j).
    \end{equation}

    \item Compute the time increment to the next event using $\Lambda_\text{max}$. Namely, a random variable  is uniformly drawn from the interval $[0, 1]$, i.e. $u \in \mathcal{U}^{[0,1]}$, and the time increment is computed as:  
    
    \begin{equation}
        \Delta t = \frac{2 \ln (1/u)}{N_A (N_A-1) \cdot \Lambda_\text{max}}.
        \label{deltat_TN_sup}
    \end{equation}

    \item Select the two reactants $A_i$ and $B_j$ for the next event. All reactants have an equal probability of being drawn, and therefore, the probability of selecting $A_i$ and $A_j$ are:
    
    \begin{equation}
        p_i = \frac{1}{N_A} \ \text{and} \ p_j = \frac{1}{N_A - 1},  \ \ \text{respectively}
    \end{equation}

    \item Accept the process with probability $p_\text{accept}$, given by:
    \begin{equation}
        \label{Eq:MOSAIC_compute_pairwise_rate}
        p_\text{accept} = \frac{\Lambda_{ij} (t_i, t_j)}{\Lambda_\text{max}}, 
    \end{equation}
    and update the reactants' population accordingly. If the process is rejected, the next event is set to an empty event, i.e. the reactant populations remain unchanged.

\end{enumerate}
Here, the definition of the pairwise rate $\Lambda_{ij}$ is crucial, as it directly governs both the interaction rate between node pairs and the resulting inter-event time distribution for individual nodes. We assume that, each node possesses an \textit{intrinsic} instantaneous rate, denoted as $\lambda_{i}(t_i)$ and $\lambda_{j}(t_j)$ for nodes $A_i$ and $A_j$, respectively. These rates, akin to activity-driven models, quantify the node's propensity to interact and can be conceptualized as a broadcast signal indicating its willingness to engage in interactions. We also introduce a scale free rate  $\tilde{\lambda}(t) = \lambda(t) / \lambda_0$ and an arbitrary scaling constant $\Lambda_0$ for clarity.

For a given pair of nodes, we can consider different interaction scenarios. In one scenario, interactions are dominated by the most active node, reflecting an independent additive process. Here, the pairwise rate can be expressed as $\Lambda_{ij} \propto \Lambda_0 \left[\tilde{\lambda}_i(t_i) + \tilde{\lambda}_j(t_j)\right]$, making it well-suited for modeling directed communication networks, such as epidemic spreading, email exchanges or message transmissions. Alternatively, in a scenario where interactions require synchronous availability between the two nodes, a multiplication emerges as a natural choice, $\Lambda_{ij} \propto \Lambda_0 \left[\tilde{\lambda}_i(t_i) \cdot \tilde{\lambda}_j(t_j)\right]$, accounting for the fact that the interaction rate is limited by the node with the lower availability, as neither can sustain the interaction independently. This type of temporal system, requiring synchronization or simultaneous availability of interacting agents, is prevalent across various fields. In neuroscience, the precise timing of signals between neurons is critical for effective communication, as synaptic transmission often relies on the synchronized release of neurotransmitters and receptor activation~\cite{kelly2012framework}. Similarly, in social interactions, effective engagement typically requires both participants to be active simultaneously~\cite{sheng2023constructing}. In molecular biology, ligand-receptor binding events also depend on synchronization, as successful interactions occur only when both molecules are available at the same time~\cite{guevara2014use}.

In Supplementary Section~\ref{MOSAIC_paired_rate}, we explore different formulations of the pairwise interaction rate $\Lambda_{ij}$ and their influence on the inter-event time distribution of individual nodes. With the additive rate, nodes can still engage in interactions even when their intrinsic rate is zero, as they may be randomly paired with a highly active node. As a result, there remains a nonzero probability of events occurring at $t_i \approx 0$, limiting the flexibility in shaping the inter-event time distribution. This constraint is particularly relevant for distributions such as Gamma or Weibull ($\alpha > 1$), which inherently assign zero probability to events occurring at $t=0$. On the other hand, we prove in Supplementary Section~\ref{MOSAIC_paired_rate_AA} that defining a multiplicative pairwise rate $\Lambda_{ij}$ with
\begin{equation} 
    \label{Eq:paired_rate}
    \Lambda_{ij} (t_i, t_j) = \frac{N_A}{N_A-1} \cdot \frac{\lambda_i(t_i) \cdot \lambda_j(t_j)}{\sum_{k=1}^{N_A} \lambda_k(t_k)}
\end{equation}
allows for the inter-event time distribution of node $A_j$ to converge to the PDF of its intrinsic rate as $N_A \to \infty$, through the relation:
\begin{equation}
    \psi_A{_j}(t) \underset{N_A \to \infty}{=} \lambda_j(t) \times \exp \left({-\int^{t}_{0} \lambda_j(\tau) \ d\tau}\right).
    \label{eq_PDF_TN}
\end{equation}
The formulation of the multiplicative rate $\Lambda_{ij}$, emerges naturally from considering the minimum event times of $N_A$ independent processes, governed by a total rate of $\sum_{k=1}^{N_A} \lambda_k(t_k)$ (Supplementary Section~\ref{pairwise_rate}). It reflects the combined likelihood of both nodes being available, weighted by their respective contributions to the total interaction rate. The additional term $N_A/(N_A-1)$ is a correction factor to account for the fact that a node cannot interact with itself, as derived in Supplementary Section~\ref{MOSAIC_paired_rate_AA}.

\subsection{Approximation of the mean observed rate}

\label{MOSAIC_total_approximation}

\noindent While computing the term $N_A \langle \lambda \rangle = \sum_{k=1}^{N_A} \lambda_k(t_k)$ may increase computational cost due to the need to iterate over all nodes at each step, it can be approximated by introducing the mean observed instantaneous rate, $\lambda_0 = \frac{1}{N_A} \left\langle \sum_{k=1}^{N_A} \lambda_k(t_k) \right\rangle$. Using this approximation, $\Lambda_{ij}$ can be expressed as:
\begin{equation} 
    \Lambda_{ij} (t_i, t_j) = \frac{N_A}{N_A-1} \cdot \frac{\lambda_i(t_i) \cdot \lambda_j(t_j)}{\left\langle \sum_{k=1}^{N_A} \lambda_k(t_k) \right\rangle} \approx \frac{\lambda_i(t_i) \cdot \lambda_j(t_j)}{(N_A-1) \cdot \lambda_0}.
\end{equation}
In Supplementary Section~\ref{mean_observed_rate}, we show that, in the steady state, the mean observed rate is given by $\lambda_0 = \frac{1}{N_A} \sum_{i=1}^{N_A} \lambda_{0_i}$, where $\lambda_{0_i} = 1/\mathbb{E}[\tau_i]$, and $\mathbb{E}[\tau_i]$ denotes the mean inter-event time of the intrinsic PDF for each individual node,
\begin{equation} 
    \mathbb{E}[\tau_i] = \int_{0}^{\infty} t \cdot \psi_A{_j}(t) \, dt.\\
\end{equation}

\subsection{Choice of the pairwise interaction rate function $\Lambda_{ij}$}

\label{MOSAIC_paired_rate}

\noindent We provide here the mathematical guarantees underlying MOSAIC-TN, as well as the rationale behind our choice of the pairwise rate formula. Here we consider a system in which each node, belonging to class $A$ or $B$, follows its own intrinsic rate function, $\lambda_A(\tau_A)$ and $\lambda_B(\tau_B)$, respectively, and that their pairwise interaction rate is given by $\Lambda(\tau_A, \tau_B)$. Here, $\tau_A$ and $\tau_B$ represent the internal times of reactants $A$ and $B$, respectively.\\

\noindent \textbf{Effective inter-event time distributions:} Let $\psi_A^\text{eff}$ and $\psi_B^\text{eff}$ denote the \textit{effective} PDF of inter-event time distribution of individual reactants $A$ and $B$, respectively. These PDFs are referred to as effective because they cannot be directly expressed as functions of the intrinsic rates $\lambda_A(t)$ and $\lambda_B(t)$, due to the coupling introduced by the interaction term $\Lambda$. These effective PDFs differ from the PDFs the reactants $A$ and $B$ in isolation $\psi_A (t) = \lambda_A (t) \cdot \exp(-\int_0^\infty \lambda_A (t))$ and $\psi_B = \lambda_B(t) \cdot \exp(-\int_0^\infty \lambda_B(t))$, which we refer to as the \textit{intrinsic} PDFs.

In the general case, the effective inter-event time distribution for reactant $A$ is determined by the minimum inter-event time over all possible interactions with $B$, which corresponds to the combined instantaneous rates of all such interactions, represented by the sum $\sum_{k=1}^{N_B} \Lambda(\tau_A, \tau_{Bk})$. Note that, while we focus here on the effective dynamics of reactant $A$, the same formulation can be applied to $B$ symmetrically without loss of generality. The effective SDF for $A$, denoted as $\Psi_A^\text{eff}(t)$, can thus be expressed as:
\begin{equation}
    \Psi_A^\text{eff}(t) = \exp\left(-\int_0^t \sum_{k=1}^{N_B} \Lambda(\tau_A, \tau_{Bk}) \, d\tau_A \right), \ \ \ \text{where} \ \ \ \tau_{Bk} \sim \psi_B^\text{obs} (\tau_B)
\end{equation}
represents the internal times of the $k$-th reactant $B$, sampled according to the \textit{observed} PDF $\psi_B^\text{obs}(\tau_B)$. Formally, $\psi_B^\text{obs} (\tau_B)$ is the PDF of the backward recurrence time associated with $\psi_B^\text{eff} (\tau_B)$, as we discuss in Supplementary Section~\ref{mean_observed_rate}.\\


\noindent \textbf{Poisson processes.} The simplest case arises when the rates of both reactants, $\lambda_A$ and $\lambda_B$, are constant. In this scenario, for any pairwise rate function $\Lambda$, the effective survival function of $A$ is given by

    \begin{equation}
        \Psi_A^\text{eff} (t) = \exp \left({- N_B \mathlarger{\int}_{0}^{t} \Lambda \biggl(\lambda_A,  \lambda_B \biggr) \ d\tau}\right) = e ^ { - N_B \cdot \Lambda \bigl(\lambda_A,  \lambda_B \bigr) \cdot t}.
    \end{equation}
    If we account for individual properties within the population of $B$, where each individual reactant $B_k$ has its own rate $\lambda_{Bk}$, the observed inter-event time distribution for $A$ is influenced by the heterogeneity in $B$'s rates. In this case, the effective survival function for $A$ reflects the cumulative contributions of all interactions with $B_k$, and is given by
    \begin{equation}
        \Psi_A^\text{eff} (t) = e ^ { - \sum_{k=1}^{N_B} \Lambda \bigl(\lambda_A,  \lambda_{Bk} \bigr) \cdot t}.
    \end{equation}.

\noindent \textbf{Multiplicative rates:} For multiplicative rate, $\Lambda(\tau_A, \tau_B) = c \cdot \lambda(\tau_A) \cdot \lambda(\tau_B)$, we can write

    \begin{align}
        \Psi_A^\text{eff} (t) &= \exp \left({- c \mathlarger{\int}_{0}^{t} \sum_{k=1}^{N_B} \lambda(\tau_{Bk}) \cdot \lambda(\tau_A) \ d\tau_A}\right),\\
        \Psi_A^\text{eff} (t) &= \exp \left({- c \sum_{k=1}^{N_B} \lambda(\tau_{Bk}) \mathlarger{\int}_{0}^{t} \lambda(\tau_A) \ d\tau_A}\right),\\
    \end{align}
    Defining the mean observed rate $\langle \lambda_B \rangle = \frac{1}{N_B} \sum_{k=1}^{N_B} \lambda(\tau_{Bk})$, we can write
    \begin{align}
        \Psi_A^\text{eff} (t) &= \exp \left({- \mathlarger{\int}_{0}^{t} \lambda(\tau_A) \ d\tau_A}\right) ^ {c N_B  \cdot \langle \lambda_B \rangle }\\
       \Psi_A^\text{eff} (t) &= \biggl[\Psi_A (t)\biggr] ^ {c N_B  \cdot \langle \lambda_B \rangle }
    \end{align}
    Finally, we can conclude that, by setting $c = 1 / (N_B \cdot \langle \lambda_B \rangle)$, we always have $\Psi_A^\text{eff} (t) = \Psi_A (t)$.\\

\noindent \textbf{Additive rates:} For additive rate, $\Lambda(\tau_A, \tau_B) = c \cdot [\lambda(\tau_A) + \lambda(\tau_B)]$, we can write 

    \begin{align}
        \Psi_A^\text{eff} (t) &= \exp \left({- c \mathlarger{\mathlarger{\int}}_{0}^{t} \sum_{k=1}^{N_B} \lambda(\tau_A) + \lambda(\tau_{Bk}) \ d\tau_A}\right),\\
        \Psi_A^\text{eff} (t) &= \exp \left({- c N_B \mathlarger{\int}_{0}^{t} \lambda(\tau_A) \ d\tau_A}\right) \cdot \exp \left({- c N_B \mathlarger{\int}_{0}^{t} \langle \lambda_B \rangle \ d\tau_A}\right)\\
         \Psi_A^\text{eff} (t)  &\approx \biggl[ \Psi_A (t) \biggr]^{c N_B} \cdot \exp \biggl({- c N_B \langle \lambda_B \rangle \cdot t}\biggr)
    \end{align}
    Setting the constant $c = \nicefrac{1}{N_B}$, we obtain
    \begin{align}
         \Psi_A^\text{eff} (t)  &\approx \biggl[ \Psi_A (t) \biggr] \cdot e^ {- \langle \lambda_B \rangle t},
    \end{align}
    from which we derive the effective PDF as 
    \begin{equation}
        \psi_A^\text{eff} (t) \approx \big[ \Psi_A (t) \big] \cdot \langle \lambda_B \rangle e^{- \langle \lambda_B \rangle t}  +  \psi_A(t) \cdot e^{- \langle \lambda_B \rangle t}.
    \end{equation}
    Interestingly, this is the mixture of two PDFs. The first behave like an exponential distribution for low time, since $\Psi_A (0) = 1$, while the second is the intrinsic PDF but scaled with an additional exponential tail.

\subsection{The reaction A$_i$ + A$_j$ $\rightarrow$ X}
\label{MOSAIC_paired_rate_AA}

\noindent In the particular case of two reactants of the same type reacting with each others, the MOSAIC algorithm is slightly modified. The reaction $\text{A} + \text{A} \rightarrow X$ involve $\frac{1}{2} N_A (N_A - 1)$ processes, and the probabilities of choosing reactants $A_i$ and $A_j$ at each iteration are given by $\frac{1}{N_A}$ and $\frac{1}{N_A-1}$, respectively. In this case, the observed SDF is directly related to the observed PDF with
    \begin{equation}
        \Psi_A^\text{eff}(t) = \exp\left(-\mathlarger{\mathlarger{\int}}_0^t \sum_{k=1}^{N_A - 1} \Lambda(\tau_A, \tau_{Ak}) \, d\tau_A \right), \ \ \ \text{where} \ \ \ \tau_{Ak} \sim \psi_A^\text{obs} (\tau_A)
    \end{equation}
    That summation includes the reaction rates of $N_A - 1$ nodes, excluding the node of interest $A$. In the case of the multiplicative rate, $\Lambda(\tau_{A_i}, \tau_{A_j}) = c \cdot \lambda(\tau_{A_i}) \cdot \lambda(\tau_{A_j})$, we can write
    \begin{equation}
        \Psi_A^\text{eff} (t) = \exp \left({- c \sum_{k=1}^{N_A-1} \lambda(\tau_{Ak}) \mathlarger{\int}_{0}^{t} \lambda(\tau_A) \ d\tau_A}\right),
    \end{equation}
    Defining the mean observed rate $\langle \lambda_A \rangle = \frac{1}{N_A} \sum_{k=1}^{N_A}  \lambda(\tau_{Ak})$, we have
    \begin{equation}
        \sum_{k=1}^{N_A-1} \lambda(\tau_{Ak}) = \sum_{k=1}^{N_A} \lambda(\tau_{Ak}) - \lambda(\tau_{A}) = N_A \langle \lambda_A \rangle - \lambda(\tau_{Ak})
    \end{equation}
    and we can thus rewrite
    \begin{align}
        \Psi_A^\text{eff} (t) &= \exp \left({- \mathlarger{\int}_{0}^{t} \lambda(\tau_A) \ d\tau_A}\right) ^ {c \cdot [N_A \langle \lambda_A \rangle - \lambda(\tau_{Ak})]}.
    \end{align}
    Defining $c$ as
    \begin{align}
        c &= \frac{N_A}{N_A-1} \cdot \frac{1}{N_A \cdot \langle \lambda_A \rangle} \ ,
    \end{align}
    we get an effective SDF
    \begin{align}
        \Psi_A^\text{eff} (t) &= \left[\exp \left({- \mathlarger{\mathlarger{\int}}_{0}^{t} \lambda(\tau_A) \ d\tau_A}\right)\right] \text{{\LARGE \^{}}} \left(\frac{N_A - \frac{\lambda(\tau_A)}{\langle \lambda_A \rangle}}{N_A - 1}\right) \underset{N_A \to \infty}{=} \Psi_A (t),
    \end{align}
    which converge to the intrinsic density distribution for large $N_A$, as the factor involving self reaction becomes diluted by the number of processes and becomes negligible.

\newpage

\subsection{MOSAIC-TN on constrained topologies}
\label{MOSAIC_constrained_topologies}

\noindent In many applications, interactions are constrained by an underlying static network (e.g., a contact graph), so that only a subset of all node pairs can interact. Let $N_A$ denote the number of nodes and let $\mathcal{E}$ be the set of admissible unordered node pairs $(i,j)$ consistent with the prescribed topology. We write $N = |\mathcal{E}|$ for the number of such pairs; in the fully connected case, $\mathcal{E}$ contains all node pairs and $N = N_A(N_A - 1)/2$.

Algorithmically, MOSAIC-TN extends straightforwardly to this setting. We first choose an upper bound
\begin{equation}
    \Lambda_\text{max} \;\geq\; \max_{(i,j)\in\mathcal{E}} \Lambda_{ij}(t_i,t_j),
\end{equation}
over all admissible pairs. Drawing $u \sim \mathcal{U}[0,1]$, the time increment to the next candidate event is now
\begin{equation}
    \Delta t = \frac{\ln(1/u)}{N\,\Lambda_\text{max}},
\end{equation}
which reduces to Eq.~\eqref{deltat_TN_sup} when the topology is fully connected and $N = N_A(N_A-1)/2$. A candidate interaction is then obtained by sampling a pair $(i,j)$ uniformly from $\mathcal{E}$, i.e.\ with probability $1/N$, and accepting it with probability
\begin{equation}
    p_\text{accept} = \frac{\Lambda_{ij}(t_i,t_j)}{\Lambda_\text{max}},
\end{equation}
as in Eq.~\eqref{Eq:MOSAIC_compute_pairwise_rate}. This rejection-based construction still yields $O(1)$ computational complexity per candidate event.

The choice of pairwise rate $\Lambda_{ij}$ and the associated guarantees follow the same logic as in the fully connected case. For a given node $i$ with neighbor set $\mathcal{N}(i)$, the effective survival function of its inter-event time is
\begin{equation}
    \Psi_i^\text{eff}(t)
    = \exp\!\left(-\int_0^t \sum_{j\in\mathcal{N}(i)} \Lambda_{ij}(\tau_i,\tau_j)\,d\tau_i\right).
\end{equation}
Under the same mean-field assumptions used in Sections~\ref{MOSAIC_paired_rate} and~\ref{MOSAIC_paired_rate_AA}, a multiplicative rate of the form
\begin{equation}
    \Lambda_{ij}(t_i,t_j)
    \propto \lambda_i(t_i)\,\lambda_j(t_j),
\end{equation}
with an appropriate normalization (replacing the global factor $N_A$ by the local degree $|\mathcal{N}(i)|$ and the global mean rate by a local average over neighbors), ensures that $\Psi_i^\text{eff}(t)$ converges to the intrinsic survival function $\Psi_i(t)$ for nodes with sufficiently large degree. In particular, the marginal node-level IEDs retain their imposed shape (including heavy-tailed behavior), while the constrained topology introduces heterogeneity in the rate at which nodes experience events. Thus, MOSAIC-TN preserves its mathematical guarantees on node-level inter-event distributions even when interactions are restricted to arbitrary sparse or structured topologies.

\section{Derivation of the Pairwise Interaction Rate for Poisson Processes requiring Synchronous Availability}

\label{pairwise_rate}

\noindent We consider $ N_A $ nodes, indexed by $ k = 1, 2, \ldots, N_A $, each emitting events according to independent Poisson processes with instantaneous rates $ \lambda_k $. Let $ T_k $ denote the time until the next event from node $ k $. The next pairwise interaction occurs at the minimum of these event times, $ T = \min(T_1, T_2, \ldots, T_N) $. 
The total rate of events occurring at time $ T $, $ \Lambda_{\text{total}} $, is derived from the property of the \textit{minimum of independent exponential random variables}. Specifically, if $ T_k \sim \text{Exp}(\lambda_k) $ for all $ k $, then $ T = \min(T_1, T_2, \ldots, T_N) $ is itself exponentially distributed with a rate equal to the sum of the individual rates:
\begin{equation}
    \Lambda_{\text{total}} = \sum_{k=1}^{N_A} \lambda_k.
\end{equation}
This result arises because the survival probability of the minimum is the product of the survival probabilities of the individual random variables:
\begin{equation}
    P(T > t) = \prod_{k=1}^{N_A} P(T_k > t) = \prod_{k=1}^{N_A} e^{-\lambda_k t} = e^{-\left(\sum_{k=1}^{N_A} \lambda_k\right)t}.
\end{equation}
For an interaction between two nodes $ i $ and $ j $ to occur, both nodes must contribute by aligning their events. The probability that node $ i $ contributes the next event is proportional to its rate, given by:
\begin{equation}
    P(T = T_i) = \frac{\lambda_i}{\sum_{k=1}^{N_A} \lambda_k},
\end{equation}
and similarly, the probability that node $ j $ contributes is:
\begin{equation}
    P(T = T_j) = \frac{\lambda_j}{\sum_{k=1}^{N_A} \lambda_k}.
\end{equation}
The pairwise interaction rate, $ \Lambda_{ij} $, is the product of the total rate of events, $ \Lambda_{\text{total}} $, and the joint probability of contributions from $ i $ and $ j $. This joint probability is proportional to the likelihood that both nodes emit events aligned in time:
\begin{equation}
    \Lambda_{ij} = \Lambda_{\text{total}} \cdot P(T = T_i) \cdot P(T = T_j).
\end{equation}
Substituting the expressions for $ \Lambda_{\text{total}} $, $ P(T = T_i) $, and $ P(T = T_j) $, we have:
\begin{equation}
    \Lambda_{ij} = \left( \sum_{k=1}^{N_A} \lambda_k \right) \cdot \frac{\lambda_i}{\sum_{k=1}^{N_A} \lambda_k} \cdot \frac{\lambda_j}{\sum_{k=1}^{N_A} \lambda_k}.
\end{equation}
Simplifying, this yields the effective pairwise interaction rate:
\begin{equation}
    \Lambda_{ij} = \frac{\lambda_i \cdot \lambda_j}{\sum_{k=1}^{N_A} \lambda_k}.
\end{equation}
%

 \newpage

\section{Observed backward recurrence time and instantaneous rates}

\label{mean_observed_rate}

\noindent  We consider a renewal process described by an arbitrary probability density function (PDF) $\psi_A(\tau)$, a survival density function $\Psi_A(\tau)$, and its instantaneous rate $\lambda_A(\tau) = \psi_A(\tau) / \Psi_A(\tau)$. When observing this process at a random point in time, the elapsed time $\tau$ since the last event corresponds to the \textit{backward recurrence time} in renewal theory~\cite{allison1985survival, zelen2004forward}, which represents the time elapsed since the most recent renewal event. The PDF of the backward recurrence time, $\psi_A^{\text{obs}}(\tau)$, is given by~\cite{allison1985survival}
\begin{equation}
    \psi_A^{\text{obs}}(\tau) = \frac{\Psi_A(\tau)}{\mathbb{E}[\tau]} = \lambda_0 \Psi_A(\tau),
\end{equation}
where we define $\lambda_0$ the reciprocal of the expected elapsed time $\mathbb{E}[\tau]$, defined as:
\begin{equation}
    \lambda_0 = \frac{1}{\mathbb{E}[\tau]} = \left(\int_{0}^{\infty} t \cdot \psi_A(\tau) \, d\tau\right)^{-1}.
\end{equation}
Specifically, the observed backward recurrence time does not follow the original PDF $\psi_A(t)$; instead, it is weighted by the survival of intervals, as longer intervals are more likely to be observed. This phenomenon, commonly referred to as the \textit{length-biased sampling effect}, is frequently discussed in studies related to disease screening~\cite{duffy2008correcting}.\\

In the context of MOSAIC, our focus is on the rate observed at each iteration, as it directly governs the probability of an event being accepted. The \textit{mean observed rate} can be computed by integrating the instantaneous rate $\lambda_A(\tau)$, weighted by the observed PDF $\psi_A^{\text{obs}}(\tau)$:
\begin{equation}
    \label{Eq:mean_observed_rate}
    \mathbb{E}_\text{obs}[\lambda_A(\tau)] = \int_{0}^{\infty} \lambda_A(\tau) \cdot \psi_A^{\text{obs}}(\tau) \, d\tau.
\end{equation}
Substituting $\psi_A^{\text{obs}}(\tau) = \lambda_0 \Psi_A(\tau)$, the expression becomes:
\begin{equation}
    \mathbb{E}_\text{obs}[\lambda_A(\tau)] = \lambda_0 \int_{0}^{\infty} \lambda_A(\tau) \cdot \Psi_A(\tau) \, d\tau.
\end{equation}
By noting that $\lambda_A(\tau) \cdot \Psi_A(\tau) = \psi_A(\tau)$, the integral simplifies to:
\begin{equation}
    \mathbb{E}_\text{obs}[\lambda_A(\tau)] = \lambda_0 \int_{0}^{\infty} \psi_A(\tau) \, d\tau.
\end{equation}
Since $\int_{0}^{\infty} \psi_A(\tau) \, d\tau = 1$, as $\psi_A(\tau)$ is a probability density function, the mean observed rate reduces to:
\begin{equation}
    \mathbb{E}_\text{obs}[\lambda_A(\tau)] = \lambda_0.
\end{equation}

Now, we extend this result to $N$ independent renewal processes, each characterized by its own PDF $\psi_{A_i}(\tau)$, survival function $\Psi_{A_i}(\tau)$, and mean rate $\lambda_{0_i} = 1 / \mathbb{E}[\tau_i]$. The observed PDF for randomly sampling a node and observing its rate is determined by the average contribution of all processes
\begin{equation}
    \psi_A^{\text{obs}}(\tau) =  \frac{1}{N} \sum_{i=1}^{N} \lambda_{0_i} \Psi_{A_i}(\tau).
\end{equation}
Substituting this expression into Eq.~\ref{Eq:mean_observed_rate} and following the same reasoning as in the single-process case, we find:
\begin{equation}
    \mathbb{E}_\text{obs}[\lambda_A(\tau)] = \frac{1}{N} \sum_{i=1}^{N} \lambda_{0_i}.
\end{equation}

\newpage

\section{Non uniform selection of processes in MOSAIC}

\label{MOSAICbin}

\noindent \textbf{Motivation.} The rejection approach in MOSAIC is particularly beneficial when individual rates are expensive to compute and fluctuate with each iteration. However, when handling a wide array of individual rates, this method can slow down the simulation, as processes with low rates are frequently rejected~\cite{thanh2014efficient}. This issue is notably significant in processes involving pairs of reactants. In such cases, a small number of pairs with high interaction rates can disproportionately slow down the system, while many pairs with zero propensity still undergo selection and subsequent rejection. If these zero-propensity pairs can be identified before each iteration without intensive computation, it is more efficient to exclude them from selection altogether, rather than selecting and then rejecting them as MOSAIC does.

Here, we introduce a modified version of MOSAIC that adjusts the probability of selecting certain processes over others, thereby reducing the number of rejections up to several orders of magnitudes. This modification ensures that processes with highly disparate rates are handled more efficiently, and processes with zero rates are excluded from selection entirely.\\

\noindent \textbf{Algorithm.} We denote by $t_{j}$ the time elapsed since the last event of the $j$th process $(1 \leq j \leq N)$, and by $\lambda_{j}(t_j)$ the time-dependent reaction rate of the $j$th process. At each iteration, the modified MOSAIC performs 4 steps:

\begin{enumerate}

    \item Set $\lambda_\text{max}$, the maximum reaction rate over all processes, such that:
    
    \begin{equation}
        \lambda_\text{max} \geq \max_{\{j \in [1,N]\}} \lambda_j (t_j).
    \end{equation}

    \item Compute the time increment to the next event as in SG using $\lambda_\text{max}$. Namely, a random variable is uniformly drawn from the interval $[0, 1]$, i.e. $u \in \mathcal{U}^{[0,1]}$. The time increment is computed as:  
    
    \begin{equation}
        \Delta t = \frac{\ln (1/u)}{N \cdot \lambda_\text{max}}.
        \label{eq:deltat4}
    \end{equation}

    \item Assign a weight $w_j$ to each process, ensuring that the sum of all weights $\sum_{k=1}^N w_k = N$. Then, select the process $j$ that has triggered the event, with the probability of selecting each process being
    
    \begin{equation}
        p_j = \frac{w_j}{N}.
        \label{eq:choice}    
    \end{equation}

    \item Accept the process with probability $p_\text{accept}$, given by:
    \begin{equation}
        p_\text{accept} = \frac{\lambda_j (t_j)}{\lambda_\text{max}}, 
        \label{eq:rejection2}
    \end{equation}
    and update the reactants' population accordingly. If the process is rejected, the next event is set to an empty event, i.e. the reactant populations remain unchanged. 
    
\end{enumerate}

\noindent This algorithm's output resembles the original MOSAIC, with the difference that it scales the intrinsic rate of each process by $w_j$:
    \begin{equation}
        \lambda_j^\text{obs.}(t_j) = w_j \cdot \lambda_j(t_j),
        \label{eq:rejection3}
    \end{equation}

Importantly, This modification affects the computational cost of the simulation compared to using the original MOSAIC. Denoting by $M$ the number of unique weight $M = |set_{j \leq N}\{w_j\}|$, step (iii) now has a complexity of $O(M)$ instead of $O(1)$ as in the original MOSAIC framework. Consequently, the computational complexity of simulating the system becomes O($r' M N$),  with the key observation that the ratio of attempted to accepted reactions $r'$ is now lower than the ratio when all weights were equal $r$. Therefore, employing this modified version of MOSAIC is advantageous when the number of different weights to consider is less than the computational savings from reduced rejections, i.e. $M > r/r'$.

We note that scaling the rate by a weight at each iteration will affect the obtained distribution differently depending on the rate function. In the case of the exponential distribution, this is simply equivalent to scaling the inter-event time distribution of the $j$th process by $w_j$, i.e. $\lambda_{j0}^\text{obs.} = w_j \cdot \lambda_{j0}$. Similarly, if $\lambda_j (t_j)$ follows a Weibull distribution, then this is also equivalent to scaling the distribution by $w_j$, with its shape parameter $\alpha$ remaining unchanged. Finally, if $\lambda_j (t_j)$ adheres to a Pareto distribution, the obtained distribution will also be a Pareto but the weights will affect both the rate and the shape of the distribution. For most other rate functions however, the resulting distribution does not follow a simple form.\\

\noindent \textbf{Proof that modified MOSAIC follow the same distribution as Standard Gillespie.}  To prove each processes in the weighted MOSAIC follows $\lambda_j^\text{obs.}(t_j) = w_j \cdot \lambda_j$, we take a similar approach as we did in Supplementary Section~\ref{rejection_proof}. ie. we  prove that:

\begin{itemize}[label={}]
    \item \textbf{(i) The reaction $R_j$ occurs with probability  \bm{$p = {w_j\lambda_j}/\sum{w_j}\lambda_j$}}\\

\noindent We define $p_\text{accept}(R_j)$ as the joint probability of $R_j$ being first selected and then accepted, $\lambda_\text{max} \geq \max_{\{j \in [1,N]\}} \lambda_j$ as the upper propensity bound of all reactions , and $a_{0, \text{max}} = N \lambda_{\text{max}}$. We can write
$$
p_\text{accept}(R_j) = \frac{w_j}{N} \cdot \frac{\lambda_j}{\lambda_\text{max}} = \frac{w_j  \lambda_j}{a_{0, \text{max}}}
$$
We then denote by $p_\text{accept}(R)$ the probability of any reaction being accepted:
$$p_\text{accept}(R) = \frac{1}{N} \mathlarger{\mathlarger{\sum}} w_j \cdot \frac{\lambda_j}{\lambda_\text{max}} = \frac{\sum{w_j}\lambda_j}{a_{0, \text{max}}}.$$
The conditional probability $p_{\text{accept}}(R_j \mid R)$ can be exploited to show the probability of $R_j$ being accepted given that some reaction had been accepted:
$$p_{\text{accept}}(R_j \mid R) = \frac{p_{\text{accept}}(R_j)}{p_{\text{accept}}(R)} = \left(\frac{w_i \lambda_j}{a_{0, \text{max}}}\right) / \left(\frac{\sum{w_j}\lambda_j}{a_{0, \text{max}}}\right) = \frac{w_i\lambda_j}{\sum{w_j}\lambda_j}$$

\item \textbf{(ii) The $\Delta t$ increment in time follows the same exponential distribution as in SG, i.e.}

\textbf{\bm{$\ \ \ \ \ \ f_{\Delta t}(t) \sim \sum{w_j}\lambda_j \cdot \exp( - \sum{w_j}\lambda_j \cdot t)$}}\\

\noindent Here the derivation is exactly the same as in Supplementary Section~\ref{rejection_proof}. Given that $p_\text{accept}(R) = \frac{\sum{w_j}\lambda_j}{a_{0, \text{max}}}$, we obtain $\Delta t \sim (\sum{w_j}\lambda_j) \exp({-a_{0, \text{max}} \cdot t}) \cdot \exp(x \cdot (a_{0, \text{max}} - \sum{w_j}\lambda_j)) = (\sum{w_j}\lambda_j) \cdot \exp({-\sum{w_j}\lambda_j \cdot t}).$

\end{itemize}

\section{Computational complexity of stochastic algorithms}

\label{SI_complexity}

\noindent In this section, we provide an analysis of the computational complexity of various stochastic simulation algorithms.

\begin{itemize}
    \item 	\textbf{Gillespie Algorithm:} The standard Gillespie algorithm has a known computational complexity of $O(M)$ per reaction step, where $M$ is the number of unique reaction rates. This complexity arises from the necessity of drawing a reaction based on propensity values at each iteration. Specifically, the reaction index $\mu$ is selected by searching for the smallest $\mu$ satisfying 
    \begin{equation}
        \sum_{j=1}^{\mu-1} a_j < r_2 \sum_{j=1}^{M} a_j, \leq \sum_{j=1}^{\mu} a_j.
    \end{equation}
    
    \item 	\textbf{Tree-Based Gillespie Algorithm}~\cite{gibson2000efficient}: By utilizing a binary tree structure, the selection process can be improved from $O(M)$ to $O(\log M)$. In this approach, reaction propensities are stored in a binary tree, where each node maintains a partial sum of the propensities of its subtree. Reaction selection is then performed using a binary search, reducing the selection time to $O(\log M)$. However, since the tree needs to be updated whenever reaction propensities change, its efficiency depends on the system. Specifically, each update requires modifying the relevant nodes along the path from the updated leaf to the root, leading to an update complexity of $O(k \log M)$, where $k$ is the number of updated reactions per step. For systems where propensities remain largely unchanged throughout the simulation, the tree-based approach can be highly efficient. Conversely, for systems with rapidly changing rates, maintaining the tree can be computationally expensive, sometimes outweighing the benefits over the standard Gillespie algorithm.\\
    
    \item 	\textbf{Laplace Gillespie Algorithm}~\cite{masuda2018gillespie}: The Laplace Gillespie algorithm extends the standard Gillespie algorithm to non-Markovian processes by incorporating non-exponential inter-event times. Instead of redrawing all $N$ rates at each step, it samples them initially from $ p(\lambda)$ and updates only the selected reaction’s rate thereafter. To efficiently handle selection and updates, the algorithm employs a tree-based structure where reaction propensities are stored hierarchically, enabling $O(\log N)$ selection via binary search and efficient updates. This structure maintains the algorithm’s efficiency while allowing for flexible non-Markovian dynamics.\\

    \item 	\textbf{nMGA (Non-Markovian Gillespie Algorithm)}~\cite{boguna2014simulating}: This method requires recomputing all $N$ reaction rates at each time step, making tree-based approaches inefficient. The overall complexity per iteration is thus $O(N)$, dominated by the full rate update.\\
    
    \item 	\textbf{Delayed Stochastic Simulation Algorithm (DelaySSA)}~\cite{fu2022delayssatoolkit}: This algorithm maintains a sorted list of scheduled events, allowing the next event to be extracted in $O(1)$ time. However, maintaining the sorted structure incurs an additional cost of $O(k \log N)$ per iteration, where $k$ is the number of events to be updated. New events are inserted at the correct position using binary search, followed by an efficient insertion operation in a balanced tree or linked structure~\cite{henriksen1983event}. Deletion is always performed at the front of the list in $O(1)$ time.

\end{itemize}

\noindent For stochastic simulations on temporal networks, where interactions occur among \( N \) nodes, there are \( N^2 \) potential interactions, or processes, to consider.

\begin{itemize}
    \item \textbf{Standard Gillespie Algorithm on Temporal Networks:} Since interactions take place between pairs of nodes, each with potentially different rates, the number of processes to track per time step is $O(N^2)$. Consequently, the total computational complexity for a full simulation scales as $O(N^4)$.\\
    
    \item 	\textbf{Spanning Tree Approach}~\cite{sheng2023constructing}: A spanning tree is constructed to model temporal interactions, ensuring that nodes and links follow predefined IED. The initial construction of the spanning tree incurs a computational cost of $O(N^2)$, given that each node interacts with a significant approximatly half of the other nodes at least once. During the simulation, interactions evolve over $d$ time steps, with each step requiring updates based on the IED distributions. In the worst case, all edges must be considered at each step, leading to a total computational complexity of $O(dN^2)$. \\
    
    \item 	\textbf{Activity-Driven (AD) Modeling}~\cite{perra2012activity}: At each time step, node pairs are assessed to determine potential interactions. Most activity-driven modeling studies consider complexities such as historical dependencies and heterogeneous activity levels when defining interactions~\cite{le2023modeling}, requiring iteration over all possible interactions. Given that the process unfolds over $d$ time steps, the overall computational complexity is $O(dN^2)$.
\end{itemize}

\section*{Supplementary Figures}

\begin{figure*}[h!t]
    \centering
    \captionsetup{width=1\linewidth}
    \includegraphics[width=0.9\linewidth]{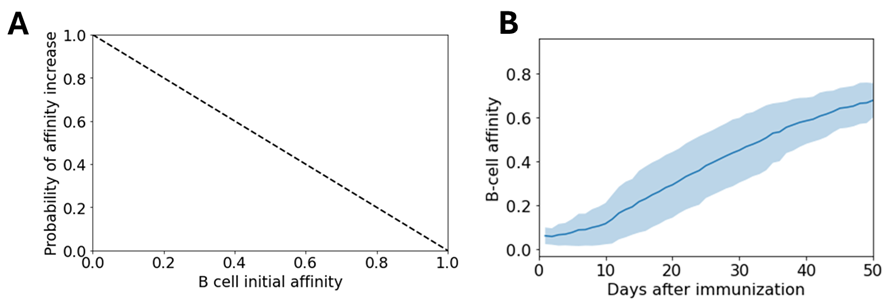}
    \caption{(A) Probability of the daughter cell increasing the affinity from the parent cell as a function of the initial affinity. (B) Affinity change of GC B cells during a germinal center simulation. The line is the average over all B~cells and the shaded area represent the mean plus one positive and one negative standard deviation.}
    \label{fig:GC_affinity}
\end{figure*}


\section*{Bibliography}

\printbibliography[title={Bibliography (Again)}, resetnumbers=true, heading=none]

@article{rabiner1986introduction,
  title={An introduction to hidden Markov models},
  author={Rabiner, Lawrence and Juang, Biinghwang},
  journal={ieee assp magazine},
  volume={3},
  number={1},
  pages={4--16},
  year={1986},
  publisher={IEEE}
}

@article{aquino_chemical_2017,
        title = {Chemical {Continuous} {Time} {Random} {Walks}},
        volume = {119},
        issn = {0031-9007, 1079-7114},
        language = {en},
        number = {23},
        journal = {Physical Review Letters},
        author = {Aquino, Tomás and Dentz, Marco},
        month = dec,
        year = {2017},
        pages = {230601}
}

@article{masuda2018gillespie,
  title={A Gillespie algorithm for non-Markovian stochastic processes},
  author={Masuda, Naoki and Rocha, Luis EC},
  journal={SIAM Review},
  volume={60},
  number={1},
  pages={95--115},
  year={2018},
  publisher={SIAM}
}

@article{thomas2019probabilistic,
  title={A probabilistic model of the germinal center reaction},
  author={Thomas, Marcel Jan and Klein, Ulf and Lygeros, John and Rodr{\'\i}guez Mart{\'\i}nez, Mar{\'\i}a},
  journal={Frontiers in immunology},
  volume={10},
  pages={689},
  year={2019},
  publisher={Frontiers}
}

@article{jiang2011study,
  title={A study of Weibull shape parameter: Properties and significance},
  author={Jiang, Renyan and Murthy, DNP},
  journal={Reliability Engineering \& System Safety},
  volume={96},
  number={12},
  pages={1619--1626},
  year={2011},
  publisher={Elsevier}
}

@article{boguna2014simulating,
  title={Simulating non-Markovian stochastic processes},
  author={Bogun{\'a}, Marian and Lafuerza, Luis F and Toral, Ra{\'u}l and Serrano, M {\'A}ngeles},
  journal={Physical Review E},
  volume={90},
  number={4},
  pages={042108},
  year={2014},
  publisher={APS}
}

@article{gillespie1977exact,
  title={Exact stochastic simulation of coupled chemical reactions},
  author={Gillespie, Daniel T},
  journal={The journal of physical chemistry},
  volume={81},
  number={25},
  pages={2340--2361},
  year={1977},
  publisher={ACS Publications}
}

@article{corral2004long,
  title={Long-term clustering, scaling, and universality in the temporal occurrence of earthquakes},
  author={Corral, Alvaro},
  journal={Physical Review Letters},
  volume={92},
  number={10},
  pages={108501},
  year={2004},
  publisher={APS}
}

@article{barabasi2005origin,
  title={The origin of bursts and heavy tails in human dynamics},
  author={Barabasi, Albert-Laszlo},
  journal={Nature},
  volume={435},
  number={7039},
  pages={207--211},
  year={2005},
  publisher={Nature Publishing Group}
}

@article{bratsun2005delay,
  title={Delay-induced stochastic oscillations in gene regulation},
  author={Bratsun, Dmitri and Volfson, Dmitri and Tsimring, Lev S and Hasty, Jeff},
  journal={Proceedings of the National Academy of Sciences},
  volume={102},
  number={41},
  pages={14593--14598},
  year={2005},
  publisher={National Acad Sciences}
}

@article{vestergaard2015temporal,
  title={Temporal gillespie algorithm: Fast simulation of contagion processes on time-varying networks},
  author={Vestergaard, Christian L and G{\'e}nois, Mathieu},
  journal={PLoS computational biology},
  volume={11},
  number={10},
  year={2015},
  publisher={Public Library of Science}
}

@article{mesin2016germinal,
  title={Germinal center B cell dynamics},
  author={Mesin, Luka and Ersching, Jonatan and Victora, Gabriel D},
  journal={Immunity},
  volume={45},
  number={3},
  pages={471--482},
  year={2016},
  publisher={Elsevier}
}

@article{gautschi1972error,
  title={Error function and Fresnel integrals},
  author={Gautschi, Walter},
  journal={Handbook of mathematical functions},
  volume={55},
  pages={297--308},
  year={1972},
  publisher={New York: Dover}
}

@book{boyce2017elementary,
  title={Elementary differential equations},
  author={Boyce, William E and DiPrima, Richard C and Meade, Douglas B},
  year={2017},
  publisher={John Wiley \& Sons}
}

@article{thanh2014efficient,
  title={Efficient rejection-based simulation of biochemical reactions with stochastic noise and delays},
  author={Thanh, Vo Hong and Priami, Corrado and Zunino, Roberto},
  journal={The Journal of chemical physics},
  volume={141},
  number={13},
  pages={10B602\_1},
  year={2014},
  publisher={American Institute of Physics}
}

@article{pelissier2020computational,
  title={Computational model reveals a stochastic mechanism behind germinal center clonal bursts},
  author={P{\'e}lissier, Aur{\'e}lien and Akrout, Youcef and Jahn, Katharina and Kuipers, Jack and Klein, Ulf and Beerenwinkel, Niko and Rodr{\'\i}guez Mart{\'\i}nez, Mar{\'\i}a},
  journal={Cells},
  volume={9},
  number={6},
  pages={1448},
  year={2020},
  publisher={Multidisciplinary Digital Publishing Institute}
}

@article{stumpf2017stem,
  title={Stem cell differentiation as a non-Markov stochastic process},
  author={Stumpf, Patrick S and Smith, Rosanna CG and Lenz, Michael and Schuppert, Andreas and M{\"u}ller, Franz-Josef and Babtie, Ann and Chan, Thalia E and Stumpf, Michael PH and Please, Colin P and Howison, Sam D and others},
  journal={Cell Systems},
  volume={5},
  number={3},
  pages={268--282},
  year={2017},
  publisher={Elsevier}
}

@article{arkin1998stochastic,
  title={Stochastic kinetic analysis of developmental pathway bifurcation in phage $\lambda$-infected Escherichia coli cells},
  author={Arkin, Adam and Ross, John and McAdams, Harley H},
  journal={Genetics},
  volume={149},
  number={4},
  pages={1633--1648},
  year={1998},
  publisher={Oxford University Press}
}

@article{figueredo2014comparing,
  title={Comparing stochastic differential equations and agent-based modelling and simulation for early-stage cancer},
  author={Figueredo, Grazziela P and Siebers, Peer-Olaf and Owen, Markus R and Reps, Jenna and Aickelin, Uwe},
  journal={PloS one},
  volume={9},
  number={4},
  pages={e95150},
  year={2014},
  publisher={Public Library of Science San Francisco, USA}
}

@article{cao2020analytical,
  title={Analytical distributions for detailed models of stochastic gene expression in eukaryotic cells},
  author={Cao, Zhixing and Grima, Ramon},
  journal={Proceedings of the National Academy of Sciences},
  volume={117},
  number={9},
  pages={4682--4692},
  year={2020},
  publisher={National Acad Sciences}
}

@article{jiang2021neural,
  title={Neural network aided approximation and parameter inference of non-Markovian models of gene expression},
  author={Jiang, Qingchao and Fu, Xiaoming and Yan, Shifu and Li, Runlai and Du, Wenli and Cao, Zhixing and Qian, Feng and Grima, Ramon},
  journal={Nature communications},
  volume={12},
  number={1},
  pages={1--12},
  year={2021},
  publisher={Nature Publishing Group}
}

@article{wang2002precision,
  title={Precision and functional specificity in mRNA decay},
  author={Wang, Yulei and Liu, Chih Long and Storey, John D and Tibshirani, Robert J and Herschlag, Daniel and Brown, Patrick O},
  journal={Proceedings of the National Academy of Sciences},
  volume={99},
  number={9},
  pages={5860--5865},
  year={2002},
  publisher={National Acad Sciences}
}

@article{herbach2017inferring,
  title={Inferring gene regulatory networks from single-cell data: a mechanistic approach},
  author={Herbach, Ulysse and Bonnaffoux, Arnaud and Espinasse, Thibault and Gandrillon, Olivier},
  journal={BMC systems biology},
  volume={11},
  number={1},
  pages={1--15},
  year={2017},
  publisher={BioMed Central}
}

@inproceedings{rubner1998metric,
  title={A metric for distributions with applications to image databases},
  author={Rubner, Yossi and Tomasi, Carlo and Guibas, Leonidas J},
  booktitle={Sixth International Conference on Computer Vision (IEEE Cat. No. 98CH36271)},
  pages={59--66},
  year={1998},
  organization={IEEE}
}

@article{sauls2019control,
  title={Control of Bacillus subtilis replication initiation during physiological transitions and perturbations},
  author={Sauls, John T and Cox, Sarah E and Do, Quynh and Castillo, Victoria and Ghulam-Jelani, Zulfar and Jun, Suckjoon},
  journal={MBio},
  volume={10},
  number={6},
  pages={e02205--19},
  year={2019},
  publisher={Am Soc Microbiol}
}

@book{ross1996stochastic,
  title={Stochastic processes},
  author={Ross, Sheldon M and Kelly, John J and Sullivan, Roger J and Perry, William James and Mercer, Donald and Davis, Ruth M and Washburn, Thomas Dell and Sager, Earl V and Boyce, Joseph B and Bristow, Vincent L},
  volume={2},
  year={1996},
  publisher={Wiley New York}
}

@article{sanft2015constant,
  title={Constant-complexity stochastic simulation algorithm with optimal binning},
  author={Sanft, Kevin R and Othmer, Hans G},
  journal={The Journal of chemical physics},
  volume={143},
  number={7},
  pages={08B609\_1},
  year={2015},
  publisher={AIP Publishing LLC}
}

@article{boxma2005off,
  title={On/off storage systems with state-dependent input, output, and switching rates},
  author={Boxma, Onno and Kaspi, Haya and Kella, Offer and Perry, David},
  journal={Probability in the Engineering and Informational Sciences},
  volume={19},
  number={1},
  pages={1--14},
  year={2005},
  publisher={Cambridge University Press}
}

@article{pakdaman2010fluid,
  title={Fluid limit theorems for stochastic hybrid systems with application to neuron models},
  author={Pakdaman, Khashayar and Thieullen, Michele and Wainrib, Gilles},
  journal={Advances in Applied Probability},
  volume={42},
  number={3},
  pages={761--794},
  year={2010},
  publisher={Cambridge University Press}
}

@article{st2019efficient,
  title={Efficient sampling of spreading processes on complex networks using a composition and rejection algorithm},
  author={St-Onge, Guillaume and Young, Jean-Gabriel and H{\'e}bert-Dufresne, Laurent and Dub{\'e}, Louis J},
  journal={Computer physics communications},
  volume={240},
  pages={30--37},
  year={2019},
  publisher={Elsevier}
}

@article{gillespie1976general,
  title={A general method for numerically simulating the stochastic time evolution of coupled chemical reactions},
  author={Gillespie, Daniel T},
  journal={Journal of computational physics},
  volume={22},
  number={4},
  pages={403--434},
  year={1976},
  publisher={Elsevier}
}

@article{bracken2014hidden,
  title={A hidden M arkov model combined with climate indices for multidecadal streamflow simulation},
  author={Bracken, C and Rajagopalan, B and Zagona, E},
  journal={Water Resources Research},
  volume={50},
  number={10},
  pages={7836--7846},
  year={2014},
  publisher={Wiley Online Library}
}

@article{nguyen2015hidden,
  title={Hidden markov model for stock selection},
  author={Nguyen, Nguyet and Nguyen, Dung},
  journal={Risks},
  volume={3},
  number={4},
  pages={455--473},
  year={2015},
  publisher={Multidisciplinary Digital Publishing Institute}
}

@article{england2010global,
  title={Global regulation of gene expression and cell differentiation in Caulobacter crescentus in response to nutrient availability},
  author={England, Jennifer C and Perchuk, Barrett S and Laub, Michael T and Gober, James W},
  journal={Journal of bacteriology},
  volume={192},
  number={3},
  pages={819--833},
  year={2010},
  publisher={Am Soc Microbiol}
}

@article{zeisel2011coupled,
  title={Coupled pre-mRNA and mRNA dynamics unveil operational strategies underlying transcriptional responses to stimuli},
  author={Zeisel, Amit and K{\"o}stler, Wolfgang J and Molotski, Natali and Tsai, Jonathan M and Krauthgamer, Rita and Jacob-Hirsch, Jasmine and Rechavi, Gideon and Soen, Yoav and Jung, Steffen and Yarden, Yosef and others},
  journal={Molecular systems biology},
  volume={7},
  number={1},
  pages={529},
  year={2011},
  publisher={John Wiley \& Sons, Ltd Chichester, UK}
}

@book{kijima2002stochastic,
  title={Stochastic processes with applications to finance},
  author={Kijima, Masaaki},
  year={2002},
  publisher={Chapman and Hall/CRC}
}

@incollection{yan2008distribution,
  title={Distribution theory, stochastic processes and infectious disease modelling},
  author={Yan, Ping},
  booktitle={Mathematical epidemiology},
  pages={229--293},
  year={2008},
  publisher={Springer}
}

@article{cao2009discrete,
  title={Discrete stochastic simulation methods for chemically reacting systems},
  author={Cao, Yang and Samuels, David C},
  journal={Methods in enzymology},
  volume={454},
  pages={115--140},
  year={2009},
  publisher={Elsevier}
}

@article{tseng2020forecasting,
  title={Forecasting the seasonal pollen index by using a hidden Markov model combining meteorological and biological factors},
  author={Tseng, Yi-Ting and Kawashima, Shigeto and Kobayashi, Satoshi and Takeuchi, Shinji and Nakamura, Kimihito},
  journal={Science of the Total Environment},
  volume={698},
  pages={134246},
  year={2020},
  publisher={Elsevier}
}

@book{van1992stochastic,
  title={Stochastic processes in physics and chemistry},
  author={Van Kampen, Nicolaas Godfried},
  volume={1},
  year={1992},
  publisher={Elsevier}
}

@article{zhang2019markovian,
  title={Markovian approaches to modeling intracellular reaction processes with molecular memory},
  author={Zhang, Jiajun and Zhou, Tianshou},
  journal={Proceedings of the National Academy of Sciences},
  volume={116},
  number={47},
  pages={23542--23550},
  year={2019},
  publisher={National Acad Sciences}
}

@article{anderson2007modified,
  title={A modified next reaction method for simulating chemical systems with time dependent propensities and delays},
  author={Anderson, David F},
  journal={The Journal of chemical physics},
  volume={127},
  number={21},
  year={2007},
  publisher={AIP Publishing}
}

@article{fu2022delayssatoolkit,
  title={DelaySSAToolkit. jl: stochastic simulation of reaction systems with time delays in Julia},
  author={Fu, Xiaoming and Zhou, Xinyi and Gu, Dongyang and Cao, Zhixing and Grima, Ramon},
  journal={Bioinformatics},
  volume={38},
  number={17},
  pages={4243--4245},
  year={2022},
  publisher={Oxford University Press}
}

@article{barrio2006oscillatory,
  title={Oscillatory regulation of Hes1: discrete stochastic delay modelling and simulation},
  author={Barrio, Manuel and Burrage, Kevin and Leier, Andr{\'e} and Tian, Tianhai},
  journal={PLoS computational biology},
  volume={2},
  number={9},
  pages={e117},
  year={2006},
  publisher={Public Library of Science San Francisco, USA}
}

@article{park2018chemical,
  title={The chemical fluctuation theorem governing gene expression},
  author={Park, Seong Jun and Song, Sanggeun and Yang, Gil-Suk and Kim, Philip M and Yoon, Sangwoon and Kim, Ji-Hyun and Sung, Jaeyoung},
  journal={Nature communications},
  volume={9},
  number={1},
  pages={297},
  year={2018},
  publisher={Nature Publishing Group UK London}
}

@article{sheng2023constructing,
  title={Constructing temporal networks with bursty activity patterns},
  author={Sheng, Anzhi and Su, Qi and Li, Aming and Wang, Long and Plotkin, Joshua B},
  journal={Nature Communications},
  volume={14},
  number={1},
  pages={7311},
  year={2023},
  publisher={Nature Publishing Group UK London}
}

@article{le2023modeling,
  title={Modeling framework unifying contact and social networks},
  author={Le Bail, Didier and G{\'e}nois, Mathieu and Barrat, Alain},
  journal={Physical Review E},
  volume={107},
  number={2},
  pages={024301},
  year={2023},
  publisher={APS}
}

@article{gibson2000efficient,
  title={Efficient exact stochastic simulation of chemical systems with many species and many channels},
  author={Gibson, Michael A and Bruck, Jehoshua},
  journal={The journal of physical chemistry A},
  volume={104},
  number={9},
  pages={1876--1889},
  year={2000},
  publisher={ACS Publications}
}

@article{genois2022combining,
  title={Combining sensors and surveys to study social contexts: Case of scientific conferences},
  author={G{\'e}nois, Mathieu and Zens, Maria and Oliveira, Marcos and Lechner, Clemens and Schaible, Johann and Strohmaier, Markus},
  journal={arXiv preprint arXiv:2206.05201},
  year={2022}
}

@article{perra2012activity,
  title={Activity driven modeling of time varying networks},
  author={Perra, Nicola and Gon{\c{c}}alves, Bruno and Pastor-Satorras, Romualdo and Vespignani, Alessandro},
  journal={Scientific reports},
  volume={2},
  number={1},
  pages={469},
  year={2012},
  publisher={Nature Publishing Group UK London}
}

@article{monk2003oscillatory,
  title={Oscillatory expression of Hes1, p53, and NF-$\kappa$B driven by transcriptional time delays},
  author={Monk, Nicholas AM},
  journal={Current biology},
  volume={13},
  number={16},
  pages={1409--1413},
  year={2003},
  publisher={Elsevier}
}

@article{hirata2002oscillatory,
  title={Oscillatory expression of the bHLH factor Hes1 regulated by a negative feedback loop},
  author={Hirata, Hiromi and Yoshiura, Shigeki and Ohtsuka, Toshiyuki and Bessho, Yasumasa and Harada, Takahiro and Yoshikawa, Kenichi and Kageyama, Ryoichiro},
  journal={Science},
  volume={298},
  number={5594},
  pages={840--843},
  year={2002},
  publisher={American Association for the Advancement of Science}
}

@article{tas2016visualizing,
  title={Visualizing antibody affinity maturation in germinal centers},
  author={Tas, Jeroen MJ and Mesin, Luka and Pasqual, Giulia and Targ, Sasha and Jacobsen, Johanne T and Mano, Yasuko M and Chen, Casie S and Weill, Jean-Claude and Reynaud, Claude-Agn{\`e}s and Browne, Edward P and others},
  journal={Science},
  volume={351},
  number={6277},
  pages={1048--1054},
  year={2016},
  publisher={American Association for the Advancement of Science}
}

@article{han2024probabilistic,
  title={Probabilistic activity driven model of temporal simplicial networks and its application on higher-order dynamics},
  author={Han, Zhihao and Liu, Longzhao and Wang, Xin and Hao, Yajing and Zheng, Hongwei and Tang, Shaoting and Zheng, Zhiming},
  journal={Chaos: An Interdisciplinary Journal of Nonlinear Science},
  volume={34},
  number={2},
  year={2024},
  publisher={AIP Publishing}
}

@article{li2006dynamics,
  title={Dynamics of opinion formation in a small-world network},
  author={Li, Ping-Ping and Zheng, Da-Fang and Hui, Pak M},
  journal={Physical Review E},
  volume={73},
  number={5},
  pages={056128},
  year={2006},
  publisher={APS}
}

@article{wang2020public,
  title={Public discourse and social network echo chambers driven by socio-cognitive biases},
  author={Wang, Xin and Sirianni, Antonio D and Tang, Shaoting and Zheng, Zhiming and Fu, Feng},
  journal={Physical Review X},
  volume={10},
  number={4},
  pages={041042},
  year={2020},
  publisher={APS}
}

@article{holme2016temporal,
  title={Temporal network structures controlling disease spreading},
  author={Holme, Petter},
  journal={Physical Review E},
  volume={94},
  number={2},
  pages={022305},
  year={2016},
  publisher={APS}
}

@article{keeling2005implications,
  title={The implications of network structure for epidemic dynamics},
  author={Keeling, Matt},
  journal={Theoretical population biology},
  volume={67},
  number={1},
  pages={1--8},
  year={2005},
  publisher={Elsevier}
}

@article{lynn2019physics,
  title={The physics of brain network structure, function and control},
  author={Lynn, Christopher W and Bassett, Danielle S},
  journal={Nature Reviews Physics},
  volume={1},
  number={5},
  pages={318--332},
  year={2019},
  publisher={Nature Publishing Group UK London}
}

@article{iacopini2018network,
  title={Network dynamics of innovation processes},
  author={Iacopini, Iacopo and Milojevi{\'c}, Sta{\v{s}}a and Latora, Vito},
  journal={Physical review letters},
  volume={120},
  number={4},
  pages={048301},
  year={2018},
  publisher={APS}
}

@article{majhi2022dynamics,
  title={Dynamics on higher-order networks: A review},
  author={Majhi, Soumen and Perc, Matja{\v{z}} and Ghosh, Dibakar},
  journal={Journal of the Royal Society Interface},
  volume={19},
  number={188},
  pages={20220043},
  year={2022},
  publisher={The Royal Society}
}

@article{koshkin2024stochastic,
  title={Stochastic modeling of a gene regulatory network driving B cell development in germinal centers},
  author={Koshkin, Alexey and Herbach, Ulysse and Mart{\'\i}nez, Mar{\'\i}a Rodr{\'\i}guez and Gandrillon, Olivier and Crauste, Fabien},
  journal={Plos one},
  volume={19},
  number={3},
  pages={e0301022},
  year={2024},
  publisher={Public Library of Science San Francisco, CA USA}
}

@article{iacopini2024temporal,
  title={The temporal dynamics of group interactions in higher-order social networks},
  author={Iacopini, Iacopo and Karsai, M{\'a}rton and Barrat, Alain},
  journal={Nature Communications},
  volume={15},
  number={1},
  pages={7391},
  year={2024},
  publisher={Nature Publishing Group UK London}
}

@article{newman2006modularity,
  title={Modularity and community structure in networks},
  author={Newman, Mark EJ},
  journal={Proceedings of the national academy of sciences},
  volume={103},
  number={23},
  pages={8577--8582},
  year={2006},
  publisher={National Acad Sciences}
}

@article{le2024flow,
  title={Flow of temporal network properties under local aggregation and time shuffling: a tool for characterizing, comparing and classifying temporal networks},
  author={Le Bail, Didier and G{\'e}nois, Mathieu and Barrat, Alain},
  journal={Journal of Physics A: Mathematical and Theoretical},
  volume={57},
  number={43},
  pages={435002},
  year={2024},
  publisher={IOP Publishing}
}

@article{ubaldi2017burstiness,
  title={Burstiness and tie activation strategies in time-varying social networks},
  author={Ubaldi, Enrico and Vezzani, Alessandro and Karsai, M{\'a}rton and Perra, Nicola and Burioni, Raffaella},
  journal={Scientific reports},
  volume={7},
  number={1},
  pages={46225},
  year={2017},
  publisher={Nature Publishing Group UK London}
}

@article{karsai2012universal,
  title={Universal features of correlated bursty behaviour},
  author={Karsai, M{\'a}rton and Kaski, Kimmo and Barab{\'a}si, Albert-L{\'a}szl{\'o} and Kert{\'e}sz, J{\'a}nos},
  journal={Scientific reports},
  volume={2},
  number={1},
  pages={397},
  year={2012},
  publisher={Nature Publishing Group UK London}
}

@article{duffy2008correcting,
  title={Correcting for lead time and length bias in estimating the effect of screen detection on cancer survival},
  author={Duffy, Stephen W and Nagtegaal, Iris D and Wallis, Matthew and Cafferty, Fay H and Houssami, Nehmat and Warwick, Jane and Allgood, Prue C and Kearins, Olive and Tappenden, Nancy and O'Sullivan, Emma and others},
  journal={American journal of epidemiology},
  volume={168},
  number={1},
  pages={98--104},
  year={2008},
  publisher={Oxford University Press}
}

@article{zelen2004forward,
  title={Forward and backward recurrence times and length biased sampling: age specific models},
  author={Zelen, Marvin},
  journal={Lifetime Data Analysis},
  volume={10},
  pages={325--334},
  year={2004},
  publisher={Springer}
}

@article{allison1985survival,
  title={Survival analysis of backward recurrence times},
  author={Allison, Paul D},
  journal={Journal of the American Statistical Association},
  volume={80},
  number={390},
  pages={315--322},
  year={1985},
  publisher={Taylor \& Francis}
}

@book{taylor1717methodus,
  title={Methodus incrementorum directa \& inversa},
  author={Taylor, Brook},
  year={1717},
  publisher={Inny}
}

@article{ku1966notes,
  title={Notes on the use of propagation of error formulas},
  author={Ku, Harry H and others},
  journal={Journal of Research of the National Bureau of Standards},
  volume={70},
  number={4},
  year={1966}
}

@article{qian2021basic,
  title={Basic mechanisms and kinetics of pause-interspersed transcript elongation},
  author={Qian, Jin and Dunlap, David and Finzi, Laura},
  journal={Nucleic Acids Research},
  volume={49},
  number={1},
  pages={15--24},
  year={2021},
  publisher={Oxford University Press}
}

@article{rajala2010effects,
  title={Effects of transcriptional pausing on gene expression dynamics},
  author={Rajala, Tiina and H{\"a}kkinen, Antti and Healy, Shannon and Yli-Harja, Olli and Ribeiro, Andre S},
  journal={PLoS computational biology},
  volume={6},
  number={3},
  pages={e1000704},
  year={2010},
  publisher={Public Library of Science San Francisco, USA}
}

@article{kelly2012framework,
  title={A framework for evaluating pairwise and multiway synchrony among stimulus-driven neurons},
  author={Kelly, Ryan C and Kass, Robert E},
  journal={Neural computation},
  volume={24},
  number={8},
  pages={2007--2032},
  year={2012},
  publisher={MIT Press}
}

@article{guevara2014use,
  title={Use of the harmonic mean to the determination of dissociation constants of stereoisomeric mixtures of biologically active compounds},
  author={Guevara-Salazar, J Alberto and Quintana-Zavala, Delia and Jim{\'e}nez-V{\'a}zquez, Hugo A and Trujillo-Ferrara, Jos{\'e}},
  journal={Journal of Enzyme Inhibition and Medicinal Chemistry},
  volume={29},
  number={6},
  pages={884--894},
  year={2014},
  publisher={Taylor \& Francis}
}

@article{newman2002assortative,
  title={Assortative mixing in networks},
  author={Newman, Mark EJ},
  journal={Physical review letters},
  volume={89},
  number={20},
  pages={208701},
  year={2002},
  publisher={APS}
}

@article{watts1998collective,
  title={Collective dynamics of ‘small-world’networks},
  author={Watts, Duncan J and Strogatz, Steven H},
  journal={nature},
  volume={393},
  number={6684},
  pages={440--442},
  year={1998},
  publisher={Nature Publishing Group}
}

@article{newman2003structure,
  title={The structure and function of complex networks},
  author={Newman, Mark EJ},
  journal={SIAM review},
  volume={45},
  number={2},
  pages={167--256},
  year={2003},
  publisher={SIAM}
}

@book{posfai2016network,
  title={Network science},
  author={P{\'o}sfai, M{\'a}rton and Barab{\'a}si, Albert-L{\'a}szl{\'o}},
  year={2016},
  publisher={Citeseer}
}

@article{albert2002statistical,
  title={Statistical mechanics of complex networks},
  author={Albert, R{\'e}ka and Barab{\'a}si, Albert-L{\'a}szl{\'o}},
  journal={Reviews of modern physics},
  volume={74},
  number={1},
  pages={47},
  year={2002},
  publisher={APS}
}

@article{blondel2008fast,
  title={Fast unfolding of communities in large networks},
  author={Blondel, Vincent D and Guillaume, Jean-Loup and Lambiotte, Renaud and Lefebvre, Etienne},
  journal={Journal of statistical mechanics: theory and experiment},
  volume={2008},
  number={10},
  pages={P10008},
  year={2008},
  publisher={IOP Publishing}
}

@article{unicomb2021dynamics,
  title={Dynamics of cascades on burstiness-controlled temporal networks},
  author={Unicomb, Samuel and I{\~n}iguez, Gerardo and Gleeson, James P and Karsai, M{\'a}rton},
  journal={Nature communications},
  volume={12},
  number={1},
  pages={133},
  year={2021},
  publisher={Nature Publishing Group UK London}
}

@article{lin2024higher,
  title={Higher-order non-Markovian social contagions in simplicial complexes},
  author={Lin, Zhaohua and Han, Lilei and Feng, Mi and Liu, Ying and Tang, Ming},
  journal={Communications Physics},
  volume={7},
  number={1},
  pages={175},
  year={2024},
  publisher={Nature Publishing Group UK London}
}

@inproceedings{henriksen1983event,
  title={Event list management-a tutorial},
  author={Henriksen, James O},
  booktitle={Proceedings of the 15th conference on Winter Simulation-Volume 2},
  pages={543--551},
  year={1983}
}

@article{le2025generalizing,
  title={Generalizing Egocentric Temporal Neighborhoods to probe for spatial correlations in temporal networks and infer their topology},
  author={Bail, Didier Le},
  journal={Journal of Physics A: Mathematical and Theoretical},
  year={2025},
  publisher={IOP Publishing}
}

@article{han2023non,
  title={Non-Markovian epidemic spreading on temporal networks},
  author={Han, Lilei and Lin, Zhaohua and Yin, Qingqing and Tang, Ming and Guan, Shuguang and Bogu{\~n}{\'a}, Mari{\'a}n},
  journal={Chaos, Solitons \& Fractals},
  volume={173},
  pages={113664},
  year={2023},
  publisher={Elsevier}
}

@article{scholtes2014causality,
  title={Causality-driven slow-down and speed-up of diffusion in non-Markovian temporal networks},
  author={Scholtes, Ingo and Wider, Nicolas and Pfitzner, Ren{\'e} and Garas, Antonios and Tessone, Claudio J and Schweitzer, Frank},
  journal={Nature communications},
  volume={5},
  number={1},
  pages={5024},
  year={2014},
  publisher={Nature Publishing Group UK London}
}

@article{williams2019effects,
  title={Effects of memory on spreading processes in non-Markovian temporal networks},
  author={Williams, Oliver E and Lillo, Fabrizio and Latora, Vito},
  journal={New Journal of Physics},
  volume={21},
  number={4},
  pages={043028},
  year={2019},
  publisher={IOP Publishing}
}

@article{grossmann2021heterogeneity,
  title={Heterogeneity matters: Contact structure and individual variation shape epidemic dynamics},
  author={Gro{\ss}mann, Gerrit and Backenk{\"o}hler, Michael and Wolf, Verena},
  journal={Plos one},
  volume={16},
  number={7},
  pages={e0250050},
  year={2021},
  publisher={Public Library of Science San Francisco, CA USA}
}

@article{gregg2021agent,
  title={Agent-based modeling reveals benefits of heterogeneous and stochastic cell populations during cGAS-mediated IFN$\beta$ production},
  author={Gregg, Robert W and Shabnam, Fathima and Shoemaker, Jason E},
  journal={Bioinformatics},
  volume={37},
  number={10},
  pages={1428--1434},
  year={2021},
  publisher={Oxford University Press}
}

@article{garcia2023understanding,
  title={Understanding repertoire sequencing data through a multiscale computational model of the germinal center},
  author={Garcia-Valiente, Rodrigo and Merino Tejero, Elena and Stratigopoulou, Maria and Balashova, Daria and Jongejan, Aldo and Lashgari, Danial and P{\'e}lissier, Aur{\'e}lien and Caniels, Tom G and Claireaux, Mathieu AF and Musters, Anne and others},
  journal={npj Systems Biology and Applications},
  volume={9},
  number={1},
  pages={8},
  year={2023},
  publisher={Nature Publishing Group UK London}
}

@article{sahneh2017gemfsim,
  title={GEMFsim: A stochastic simulator for the generalized epidemic modeling framework},
  author={Sahneh, Faryad Darabi and Vajdi, Aram and Shakeri, Heman and Fan, Futing and Scoglio, Caterina},
  journal={Journal of computational science},
  volume={22},
  pages={36--44},
  year={2017},
  publisher={Elsevier}
}

@inproceedings{bisset2009epifast,
  title={EpiFast: a fast algorithm for large scale realistic epidemic simulations on distributed memory systems},
  author={Bisset, Keith R and Chen, Jiangzhuo and Feng, Xizhou and Kumar, VS Anil and Marathe, Madhav V},
  booktitle={Proceedings of the 23rd international conference on Supercomputing},
  pages={430--439},
  year={2009}
}

@inproceedings{barrett2008episimdemics,
  title={Episimdemics: an efficient algorithm for simulating the spread of infectious disease over large realistic social networks},
  author={Barrett, Christopher L and Bisset, Keith R and Eubank, Stephen G and Feng, Xizhou and Marathe, Madhav V},
  booktitle={SC'08: Proceedings of the 2008 ACM/IEEE Conference on Supercomputing},
  pages={1--12},
  year={2008},
  organization={IEEE}
}

@article{sheng2024strategy,
  title={Strategy evolution on higher-order networks},
  author={Sheng, Anzhi and Su, Qi and Wang, Long and Plotkin, Joshua B},
  journal={Nature Computational Science},
  volume={4},
  number={4},
  pages={274--284},
  year={2024},
  publisher={Nature Publishing Group US New York}
}

@article{chen2025epihiper,
  title={Epihiper—A high performance computational modeling framework to support epidemic science},
  author={Chen, Jiangzhuo and Hoops, Stefan and Mortveit, Henning S and Lewis, Bryan L and Machi, Dustin and Bhattacharya, Parantapa and Venkatramanan, Srinivasan and Wilson, Mandy L and Barrett, Chris L and Marathe, Madhav V},
  journal={PNAS nexus},
  volume={4},
  number={1},
  pages={pgae557},
  year={2025},
  publisher={Oxford University Press US}
}

@article{hoops2006copasi,
  title={COPASI—a complex pathway simulator},
  author={Hoops, Stefan and Sahle, Sven and Gauges, Ralph and Lee, Christine and Pahle, J{\"u}rgen and Simus, Natalia and Singhal, Mudita and Xu, Liang and Mendes, Pedro and Kummer, Ursula},
  journal={Bioinformatics},
  volume={22},
  number={24},
  pages={3067--3074},
  year={2006},
  publisher={Oxford University Press}
}

@article{koher2019contact,
  title={Contact-based model for epidemic spreading on temporal networks},
  author={Koher, Andreas and Lentz, Hartmut HK and Gleeson, James P and H{\"o}vel, Philipp},
  journal={Physical Review X},
  volume={9},
  number={3},
  pages={031017},
  year={2019},
  publisher={APS}
}

@article{valdano2015analytical,
  title={Analytical computation of the epidemic threshold on temporal networks},
  author={Valdano, Eugenio and Ferreri, Luca and Poletto, Chiara and Colizza, Vittoria},
  journal={Physical Review X},
  volume={5},
  number={2},
  pages={021005},
  year={2015},
  publisher={APS}
}

@article{cure2025fast,
  title={Fast and exact stochastic simulations of epidemics on static and temporal networks},
  author={Cure, Samuel and Pflug, Florian G and Pigolotti, Simone},
  journal={PLOS Computational Biology},
  volume={21},
  number={9},
  pages={e1013490},
  year={2025},
  publisher={Public Library of Science San Francisco, CA USA}
}

\end{refsection} 

\end{document}